\documentclass[twocolumn]{aastex631}

\usepackage{upgreek}

\usepackage{threeparttablex}
\usepackage{comment}

\begin{document}

\title{Morphology of 137 Fast Radio Bursts down to Microseconds Timescales from The First CHIME/FRB Baseband Catalog}
\shorttitle{CHIME/FRB Baseband Catalog 1 -- Morphology}

\correspondingauthor{Ketan R. Sand}
\email{ketan.sand@mail.mcgill.ca}

\author[0000-0003-3154-3676]{Ketan R. Sand}
\affiliation{Department of Physics, McGill University, 3600 rue University, Montr\'eal, QC H3A 2T8, Canada}
\affiliation{Trottier Space Institute, McGill University, 3550 rue University, Montr\'eal, QC H3A 2A7, Canada}

\shortauthors{Sand et al.}
\newcommand{\allacks}{}

\author[0000-0002-8376-1563]{Alice P.Curtin}
\affiliation{Department of Physics, McGill University, 3600 rue University, Montr\'eal, QC H3A 2T8, Canada}
\affiliation{Trottier Space Institute, McGill University, 3550 rue University, Montr\'eal, QC H3A 2A7, Canada}

\author[0000-0002-2551-7554]{Daniele Michilli}
\affiliation{MIT Kavli Institute for Astrophysics and Space Research, Massachusetts Institute of Technology, 77 Massachusetts Ave, Cambridge, MA 02139, USA}
\affiliation{Department of Physics, Massachusetts Institute of Technology, 77 Massachusetts Ave, Cambridge, MA 02139, USA}

\author[0000-0001-9345-0307]{Victoria M. Kaspi}
\affiliation{Department of Physics, McGill University, 3600 rue University, Montr\'eal, QC H3A 2T8, Canada}
\affiliation{Trottier Space Institute, McGill University, 3550 rue University, Montr\'eal, QC H3A 2A7, Canada}

\author[0000-0001-8384-5049]{Emmanuel Fonseca}
\affiliation{Department of Physics and Astronomy, West Virginia University, PO Box 6315, Morgantown, WV 26506, USA }
\affiliation{Center for Gravitational Waves and Cosmology, West Virginia University, Chestnut Ridge Research Building, Morgantown, WV 26505, USA}

\author[0000-0003-0510-0740]{Kenzie Nimmo}
\affiliation{MIT Kavli Institute for Astrophysics and Space Research, Massachusetts Institute of Technology, 77 Massachusetts Ave, Cambridge, MA 02139, USA}
\affiliation{Department of Physics, Massachusetts Institute of Technology, 77 Massachusetts Ave, Cambridge, MA 02139, USA}

\author[0000-0002-4795-697X]{Ziggy Pleunis}
\affiliation{Anton Pannekoek Institute for Astronomy, University of Amsterdam, Science Park 904, 1098 XH Amsterdam, The Netherlands}
\affiliation{ASTRON, Netherlands Institute for Radio Astronomy, Oude Hoogeveensedijk 4, 7991 PD Dwingeloo, The Netherlands}

\author[0000-0002-6823-2073]{Kaitlyn Shin}
\affiliation{MIT Kavli Institute for Astrophysics and Space Research, Massachusetts Institute of Technology, 77 Massachusetts Ave, Cambridge, MA 02139, USA}
\affiliation{Department of Physics, Massachusetts Institute of Technology, 77 Massachusetts Ave, Cambridge, MA 02139, USA}

\author[0000-0002-3615-3514]{Mohit Bhardwaj}
\affiliation{McWilliams Center for Cosmology, Department of Physics, Carnegie Mellon University, Pittsburgh, PA 15213, USA}

\author[0000-0002-1800-8233]{Charanjot Brar}
\affiliation{Department of Physics, McGill University, 3600 rue University, Montr\'eal, QC H3A 2T8, Canada}
\affiliation{Trottier Space Institute, McGill University, 3550 rue University, Montr\'eal, QC H3A 2A7, Canada}

\author[0000-0001-7166-6422]{Matt Dobbs}
\affiliation{Department of Physics, McGill University, 3600 rue University, Montr\'eal, QC H3A 2T8, Canada}
\affiliation{Trottier Space Institute, McGill University, 3550 rue University, Montr\'eal, QC H3A 2A7, Canada}

\author[0000-0003-3734-8177]{Gwendolyn M. Eadie}
\affiliation{David A. Dunlap Department of Astronomy \& Astrophysics, University of Toronto, 50 St.~George Street, Toronto, ON M5S 3H4, Canada}
\affiliation{Department of Statistical Sciences, University of Toronto, 700 University Ave 9th Floor, Toronto, ON M5G 1X6, Canada}
\affiliation{Data Sciences Institute, University of Toronto, 700 University Avenue, 10th Floor, Toronto, ON M5G 1Z5, Canada}

\author[0000-0002-3382-9558]{B.M. Gaensler}
\affiliation{Department of Astronomy and Astrophysics, University of California Santa Cruz, 1156 High Street, Santa Cruz, CA 95064, USA}
\affiliation{Dunlap Institute for Astronomy \& Astrophysics, University of Toronto, 50 St.~George Street, Toronto, ON M5S 3H4, Canada}
\affiliation{David A. Dunlap Department of Astronomy \& Astrophysics, University of Toronto, 50 St.~George Street, Toronto, ON M5S 3H4, Canada}

\author[0000-0003-3457-4670]{Ronniy C. Joseph}
\affiliation{Department of Physics, McGill University, 3600 rue University, Montr\'eal, QC H3A 2T8, Canada}
\affiliation{Trottier Space Institute, McGill University, 3550 rue University, Montr\'eal, QC H3A 2A7, Canada}

\author[0000-0002-4209-7408]{Calvin Leung}
\affiliation{Department of Astronomy, University of California Berkeley, Berkeley, CA 94720, USA}
\affiliation{NASA Hubble Fellowship Program~(NHFP) Einstein Fellow}

\author[0000-0002-7164-9507]{Robert Main}
\affiliation{Department of Physics, McGill University, 3600 rue University, Montr\'eal, QC H3A 2T8, Canada}
\affiliation{Trottier Space Institute, McGill University, 3550 rue University, Montr\'eal, QC H3A 2A7, Canada}

\author[0000-0002-4279-6946]{Kiyoshi W. Masui}
\affiliation{MIT Kavli Institute for Astrophysics and Space Research, Massachusetts Institute of Technology, 77 Massachusetts Ave, Cambridge, MA 02139, USA}
\affiliation{Department of Physics, Massachusetts Institute of Technology, 77 Massachusetts Ave, Cambridge, MA 02139, USA}

\author[0000-0001-7348-6900]{Ryan Mckinven}
\affiliation{Department of Physics, McGill University, 3600 rue University, Montr\'eal, QC H3A 2T8, Canada}
\affiliation{Trottier Space Institute, McGill University, 3550 rue University, Montr\'eal, QC H3A 2A7, Canada}

\author[0000-0002-8897-1973]{Ayush Pandhi}
\affiliation{Dunlap Institute for Astronomy \& Astrophysics, University of Toronto, 50 St.~George Street, Toronto, ON M5S 3H4, Canada}
\affiliation{David A. Dunlap Department of Astronomy \& Astrophysics, University of Toronto, 50 St.~George Street, Toronto, ON M5S 3H4, Canada}

\author[0000-0002-8912-0732]{Aaron B. Pearlman}
\affiliation{Department of Physics, McGill University, 3600 rue University, Montr\'eal, QC H3A 2T8, Canada}
\affiliation{Trottier Space Institute, McGill University, 3550 rue University, Montr\'eal, QC H3A 2A7, Canada}
\affiliation{Banting Fellow}
\affiliation{McGill Space Institute Fellow}
\affiliation{FRQNT Postdoctoral Fellow}

\author[0000-0001-7694-6650]{Masoud Rafiei-Ravandi}
\affiliation{Department of Physics, McGill University, 3600 rue University, Montr\'eal, QC H3A 2T8, Canada}
\affiliation{Trottier Space Institute, McGill University, 3550 rue University, Montr\'eal, QC H3A 2A7, Canada}

\author[0000-0002-4623-5329]{Mawson W. Sammons}
\affiliation{Department of Physics, McGill University, 3600 rue University, Montr\'eal, QC H3A 2T8, Canada}
\affiliation{Trottier Space Institute, McGill University, 3550 rue University, Montr\'eal, QC H3A 2A7, Canada}

\author[0000-0002-2088-3125]{Kendrick Smith}
\affiliation{Perimeter Institute for Theoretical Physics, 31 Caroline Street N, Waterloo, ON N25 2YL, Canada}

\author[0000-0001-9784-8670]{Ingrid H. Stairs}
\affiliation{Department of Physics and Astronomy, University of British Columbia, 6224 Agricultural Road, Vancouver, BC V6T 1Z1 Canada}

\begin{abstract}

We present a spectro-temporal analysis of 137 fast radio bursts (FRBs) from the first CHIME/FRB baseband catalog, including 125 one-off bursts and 12 repeat bursts, down to microsecond resolution  using the least-squares optimization fitting routine: \texttt{fitburst}. Our measured values are compared with those in the first CHIME/FRB intensity catalog, revealing that nearly one-third of our sample exhibits additional burst components at higher time resolutions. We measure sub-burst components within burst envelopes as narrow as $\sim$23 $\upmu$s (FWHM), with 20\% of the sample displaying sub-structures narrower than 100 $\upmu$s, offering constraints on emission mechanisms. Scattering timescales in the sample range from 30 $\upmu$s to 13 ms at 600 MHz. We observe no correlations between scattering time and dispersion measure, rotation measure, or linear polarization fraction, with the latter suggesting that depolarization due to multipath propagation is negligible in our sample. Bursts with narrower envelopes ($\leq$ 1 ms) in our sample exhibit higher flux densities, indicating the potential presence of sub-ms FRBs that are being missed by our real-time system below a brightness threshold. Most multicomponent bursts in our sample exhibit sub-burst separations of $\leq$ 1 ms, with no bursts showing separations $<$41 $\upmu$s, even at a time resolution of 2.56 $\upmu$s, but both scattering and low signal-to-noise ratio can hinder detection of additional components. Lastly, given the morphological diversity of our sample, we suggest that one-off and repeating FRBs can come from different classes but have overlapping property distributions.

\end{abstract}

\section{Introduction} \label{sec:intro}

Fast radio bursts (FRBs) are $\upmu$s - ms duration radio bursts originating from cosmological distances (see \citealp{petroff_dawn_2022, cordes2019fast, Bailes_review} for reviews).  Numerous theories have been proposed to elucidate their origins \citep{platts2019living}, with magnetars emerging as primary candidates following the detection of FRB-like bursts from the Galactic magnetar SGR 1935+2154 \citep{bochenek2020fast, andersen2020bright}. Multiple bursts have been detected from a subset of these sources, which are classified as repeaters \citep{chimefrb_RN1_2019, Fonseca2020_RN2, RN3}. Presently, there are approximately 790 published FRB sources, with around 50 known to exhibit repetition, although the intrinsic fraction of repeating FRB sources may be much higher \citep{2023PASA...40...57J, ygl+2024}.

Burst morphology, defined both in time and in radio frequency, plays a pivotal role in unraveling the FRB emission mechanisms and  the distribution of matter lying between us and the source. To elaborate, the intrinsic morphologies of FRBs are distorted by multiple propagation effects due to the interaction of radio waves with free electrons along their path to Earth, thereby giving us information about the media traveled by the FRB signal \citep{macquart2020census}. Dispersion measure (DM) is the integrated column density of free electrons, which cause a frequency-dependent delay of the radio waves. In addition to the dispersion of the radio waves, scattering can broaden the pulses due to multipath propagation caused by plasma inhomogeneities along the line of sight. 

It is imperative to disentangle FRB propagation effects to understand the local environments of FRBs. Scattering timescales are a great probe for mapping turbulence or any spatial gradient in electron density in the intervening medium. Studies such as that of \cite{chawla_scattering_dm} propose that the majority of the scattering contribution to FRBs originates from the source's local environment, with possible additional contributions from intervening halos and circumgalactic medium (CGM : \citealp{vedantham_phinney_2019}). Measurements by \cite{Gupta2022} detected FRB scattering down to $\sim$ 20 $\upmu$s at 835 MHz. This supports the notion that the Intergalactic medium (IGM) is not the primary source of scattering as the high DM of the FRB (715 pc\,cm$^{-3}$) did not necessitate to a higher scattering timescale. \cite{occ_2021} inferred minimal contributions to scattering from halos. \cite{fro2024} detected a heavily scattered event from FRB\,20221219A, where the dominant scattering screen was inferred to be in the CGM, corresponding to two intervening galaxy halos.

It has been shown that FRBs with high Faraday rotation measures (RMs) inhabit dense, magnetized environments \citep{Masui2015,mic2018}, which might lead to significant scattering. However, this effect -- a correlation between scattering and RM -- has not been observed across a large sample of FRBs. Additionally, frequency-dependent depolarization due to multipath propagation has been suggested in some repeating sources \citep{feng_2022_depolarization}. Thus, it is conceivable that sources with low linear polarization fractions may exhibit longer scattering timescales. Scattering time estimates can however, be hindered by hidden subcomponents in the burst envelope as well as incorrect DM measurements, if measured at a coarser time resolution. 

Various origins of FRB emission have been proposed. The two major hypotheses are synchrotron maser emission due to shock wave interaction with a relativistic plasma ejected during a magnetar flare (e.g., \citealp{mms19}), or magnetic perturbation in proximity to the magnetar surface, leading to both X-ray and subsequent FRB emission (e.g., \citealp{lu2018radiation}). The temporal and spectral properties of a large population of FRBs can provide useful constraints on the two models. 

Microsecond-duration sub-bursts have been observed in FRBs. Other telescopes have seen 3--4 $\upmu$s bursts for FRB\,20180916B \citep{nimmo2021_micro} and individual bursts $\leq$ 10 $\upmu$s for FRB\,20121102A \citep{snelders_2023}. Durations a further order of magnitude shorter --  a 60 ns sub-burst -- were observed from the repeating FRB\,20200120E at 1.4 GHz \citep{nimmo2022burst} and at 2.3 GHz \citep{majid_nano_2021}.  20--30 $\upmu$s structures have also been seen in apparent one-off bursts \citep{ffb18, cms+20}. The observation of micro- and even nanosecond-duration bursts argues that the emission originates close to the source surface, based on light travel time arguments. However, population-wide observations of such substructures would provide stronger evidence to support this hypothesis and  could constrain other emission scenarios. 

In addition to microstructure, a variety of other complex burst structures have been observed in both repeaters and one-off FRB sources. For instance, microshots have been detected from FRB\,20220912A \citep{Hewitt_2023}, while FRB\,20201124A has exhibited a polychotomy of morphological archetypes, including both upward and downward drifting sub-bursts \citep{zhz_2022}. \cite{dds+20} conducted a high-time resolution study of five bursts from one-off sources, concluding that there might be an emerging archetype between repeaters and non-repeaters. Recently, \cite{fmm+23} highlighted 12 one-off sources which included additional drifting archetypes. However, an analysis of a large sample of one-off sources at high time resolution can provide more insight on whether there exists a continuum in morphology between repeaters and one-off FRBs.

\cite{pleunismorph} conducted the largest morphological analysis of the FRB population to date, leveraging data from 536 bursts published in the first Canadian Hydrogen Intensity Mapping Experiment/Fast Radio Burst (CHIME/FRB) catalog \citep{chimefrbcatalog1}. They identified four major archetypes in FRB morphology, with a key finding being that repeaters exhibit narrower frequency ranges and broader band-averaged temporal profiles compared to one-off events. This observation hints at potentially distinct emission mechanisms or separate FRB populations. Their analysis was based on total-intensity data (`intensity data' hereafter), with a time resolution of 0.983 ms and a frequency resolution of 24.4 kHz \citep[see][for an overview of the CHIME/FRB observing system]{abb+18}. The limited time resolution prevented them to accurately estimate DM, scattering timescales, and inhibited their ability to study sub-ms components within burst envelopes \citep{aaa+23_basecat}. Additionally, their bandwidth estimates were influenced by the off-axis spectral response arising from the CHIME/FRB formed beams \citep{2017arXiv170204728N, apb+23}. 

Complex voltage data (hereafter ``baseband'' data) from the CHIME antennas provide significant advantages \citep{lk04}. Baseband data retain information on the phase of the detected radio waves, enabling more precise localization for CHIME/FRB, and allowing us to study the burst properties at microsecond timescales \citep{mmm+21}. Furthermore, thanks to the phase information, the effect of dispersion can be coherently corrected \citep{1971ApJ...169..487H}. This effectively mitigates intra-channel temporal smearing, allowing for the investigation of narrower components within burst envelopes. Additionally, these data store full Stokes parameter information, enabling the study of polarization properties \citep{mmm+21b}.

We here present the largest analysis to date of FRB morphologies down to microsecond timescales, using data from the CHIME/FRB baseband catalog \citep{aaa+23_basecat}. In Section \ref{sec:obs}, we detail our data collection and analysis methodology, including our burst fitting routine and inherent sample biases. In Section \ref{sec:results}, we compare our measurements with the intensity data of these bursts as reported by \cite{chimefrbcatalog1}, and we then explore correlations among various measured properties and identify complex morphologies in our sample. In Section \ref{sec:discuss} we discuss the implications of our findings for the FRB population, emission mechanisms, and propagation effects. We conclude in Section \ref{sec:conc}. 

\section{Observations and data analysis} \label{sec:obs}

CHIME is an array of four cylindrical parabolic reflectors, each with dimensions of 100 m × 20 m. Within each cylinder, 256 equispaced feeds record dual-polarization data in the 400--800 MHz frequency range \citep{abb+22}. 

The CHIME/FRB backend utilizes the CHIME telescope to conduct real-time searches for FRB signals as the sky transits overhead \citep{abb+18}. It has a triggered baseband system that archives raw voltage data for FRBs above a specified signal-to-noise (S/N) threshold. Details of this pipeline are outlined in \cite{mmm+21}.  Briefly, this pipeline forms multiple tied-array beams around an initial position and estimates the best position by applying a 2-D Gaussian model to the S/N distribution in these beams. Once localized, a `singlebeam' beamformed file is generated, resolved at a time resolution of 2.56 $\upmu$s and frequency resolution of $\sim$390 kHz, which can be analysed further. The baseband system has a data buffer of 20 s, permitting the recording of CHIME’s full bandwidth for a DM of $\sim$ 1000 pc\,cm$^{-3}$, above which we start loosing channels from the top portion of the band i.e., starting 800 MHz and going towards lower frequencies.

Among the 536 bursts reported in the first CHIME/FRB catalog \citep{chimefrbcatalog1}, 140 had baseband data collected. An analysis of these, including updated localization, exposure, fluence, and DM values, is presented in the first CHIME/FRB baseband catalog \citep{aaa+23_basecat}. Moreover, the investigation of polarization properties for 128 bursts from this subset is detailed by \cite{ppm+24}. 

Here we present a morphological analysis of 137 bursts (125 one-off and 12 from repeaters) from the first CHIME/FRB baseband catalog. Out of the full sample of 140 bursts, two bursts from FRB\,20190612A and FRB\,20190628C were at the edge of the baseband buffer, resulting in incomplete datasets. FRB\,20190627D had less than 80 channels stored and hence was too faint for reliable analysis. Therefore, results from these three bursts are omitted. Updated morphological parameters derived as detailed below are provided in the Appendix in Table \ref{table}, while Figure \ref{fig:burst_wfalls} illustrates a sub-sample of the waterfall plots of these bursts along with their respective best fits. 

We have not performed an extensive repeater vs one-off analysis in this study. Results from which will be published in future work utilizing high-time resolution morphological properties of a larger repeater burst sample \citep{curtin2024}.

\subsection{Baseband FRB Morphology Pipeline} \label{sec:Morphology_pipe}

Our software pipeline dedicated to the study of FRB burst morphologies systematically analyzes the Stokes intensity (I) data within the singlebeam files to infer the spectro-temporal characteristics of the burst. The methodology has been outlined in detail by \cite{sbm+23}. 

\textbf{Step 1:} The data is coherently dedispersed to a preliminary DM which is usually the signal-to-noise (S/N) maximizing DM determined by the baseband pipeline. Then radio frequency interference (RFI) masking is done using the functionality described in \cite{mmm+21}.

\textbf{Step 2:} The pipeline determines the appropriate downsampling factor. Given that not all bursts within our dataset are bright enough to be studied at 2.56 $\upmu$s time resolution, we systematically downsample or reduce time samples, by averaging out adjacent time bins of the data until a sufficient S/N is achieved to reliably extract morphological parameters. Specifically, we iteratively downsample the data until the burst profile generated by summing across all non-masked channels reaches an S/N of 15, with a maximum downsampling factor of 256, corresponding to a temporal resolution of 0.65 ms. Details regarding the temporal resolution for each burst are provided in Table \ref{table}. 

\textbf{Step 3:} Using the \texttt{DM phase} routine \citep{dm_phase}, we estimate the structure-maximizing DM for each burst \citep{Hes2019}. We then rereduce the singlebeam data by coherently dedispersing at the optimized DM, and then extract the burst at the determined downsampling factor for subsequent analysis.

\textbf{Step 4:} We derive the integrated temporal profile of each burst by summing across all the non-masked frequency channels. We then smooth the profile using the locally weighted scatter plot smoothing method (LOWESS) implemented by \texttt{statsmodel} package \citep{cleveland1979robust,seabold2010statsmodels}. This smoothed profile is then used to estimate the number of burst components using \texttt{find\_peaks} implemented by \texttt{SciPy} \citep{scipy_Virtanen_2020}. A visual assessment is conducted to validate the accuracy of the algorithm's component estimation. In cases where the algorithm inaccurately estimates the number of components, the process is repeated by manually providing our best estimate regarding the number of components present within the burst.

\textbf{Step 5:} We proceed to estimate initial temporal profile parameters by fitting a sum of exponentially modified Gaussians (EMGs) to the pulse profile \citep{mckinnon_2014}, with one EMG term per sub-burst, while maintaining a fixed scattering timescale for all sub-bursts. Initially, we employ a nonlinear least-squares algorithm, \texttt{curve\_fit}, from \texttt{Scipy}, to fit the data. In cases where the fit does not converge sufficiently, indicated by a high reduced $\upchi^2$ value ($\geq$ 2), we utilize the best-fit model parameters as initial conditions for a Markov Chain Monte Carlo (MCMC) sampling algorithm. This algorithm is implemented using the \texttt{emcee} routine \citep{emcee}, employing independent wide uniform prior distributions for all parameters. In case of further ambiguity in number of components the model with reduced $\upchi^2$ closest to unity is chosen.

The resulting fit yields essential parameters such as the time of arrival (\textit{$t_{\text{arr}}$}), intrinsic width ($\sigma$ of the Gaussian) for each sub-burst, and scattering timescale ($\tau$) for the burst. These parameters are stored as ``profile" parameters and subsequently provided to \texttt{fitburst} (see Section \ref{sec:fit}) as an initial guess to perform a two-dimensional spectro-temporal fit to the data.

\subsection{Fitburst}\label{sec:fit}

FRBs exhibit complex structures often comprising multiple sub-bursts, each characterized by distinct temporal shapes (\textit{T$_l$}) and spectral energy distributions (SEDs, \textit{F$_l$}) for each \textit{l$^{th}$} sub-component. Furthermore, propagation effects such as dispersion and multi-path scattering introduce frequency-dependent variations in the burst morphology. 

In order to efficiently characterize these features, CHIME/FRB has developed a least-squares optimization routine, \texttt{fitburst}\footnote{\url{https://github.com/CHIMEFRB/fitburst}} \citep{fpb+23, chimefrbcatalog1}. This routine optimizes for 4+5N parameters for an N component burst. Among these, four parameters remain global across all sub-bursts : DM, scattering timescale ($\tau$) at a reference frequency ($\nu_r$), scattering index ($\delta \rightarrow \nu^{-\delta}$), and the DM index ($\epsilon \rightarrow \nu^{-\epsilon}$). The remaining five parameters are individually fit for each component, specifically amplitude (\textit{A}), the intrinsic width or standard deviation of Gaussian ($\sigma$), the time of arrival (\textit{t$_{arr}$}), the power-law spectral index ($\gamma$) and ``running" of the power-law ($\beta$) \citep{pleunismorph}.

Each of these parameters are estimated using two formalisms: sum of EMGs \citep{mckinnon_2014} for \textit{T$_l$} and a running power law for \textit{F$_l$}. Mathematical formulations for these methodologies have been shown in detail in prior work by \cite{fpb+23}. 

For this work, we fixed $\epsilon$ = 2 assuming cold plasma dispersion and $\delta$ = 4 assuming Gaussian density fluctuations in the scattering medium for scatter broadening. The input bursts were already dedispersed to their structure-maximizing DM while allowing \texttt{fitburst} to refine this value further. These optimized DM values are reported in Table \ref{table}. Our reference frequency for scattering is chosen to be 600 MHz, the central frequency of the CHIME band. 

Following this, we derived four distinct models for each burst using initial estimates from our pipeline. These models were:

\begin{enumerate}
    \item Model 1: The temporal profile fit generated by the initial guess pipeline (Section \ref{sec:Morphology_pipe}) using EMGs, with no SED parameter estimates.
    \item Model 2: Fixing $\tau$ = 0, i.e., no scattering. 5N parameter \texttt{fitburst} model.
    \item Model 3: Allowing $\tau$ to be a free parameter, enabling its optimization. 1+5N parameter \texttt{fitburst} model.
    \item Model 4: Employing a detailed formalism described in equation (10) of \cite{fpb+23}, where amplitude is individually fit across channels alongside non-zero scattering, without SED parameter estimates. 1+3N parameter \texttt{fitburst} model.
\end{enumerate}

We assessed the goodness of fit for each model by scrutinizing the residuals, ultimately identifying the most suitable model for each burst. Results of these fits are presented in Table \ref{table}. Notably, Model 4 demonstrated exceptional efficacy, particularly in fitting bright bursts and accommodating data with masked channels or bursts exhibiting stochastic brightness variation across frequency. Consequently, this model was predominantly favored in fitting the majority of bursts in our sample. 

We chose Model 2 fits for bursts where visually the decay time dominated the intrinsic width and there was no obvious scatter broadening with frequency. For these we provide an upper limit on scattering equivalent to the minimum component width (see Table \ref{tab:burst}). The implications due to this are discussed in Section \ref{sec:corr}. 


In cases where none of the last three models converged, parameter values from the profile fit (Model 1) are reported. Note that for these bursts, there is no frequency dependence taken into consideration, since we fit the frequency-summed temporal profile. 

For the bursts lacking SED parameters, we averaged across time to generate a combined spectrum and then did a running power-law fit \citep{pleunismorph} to this spectrum to get the burst model. We then estimated the bandwidth by computing the full width at tenth maximum (FWTM) of the peak of the time-averaged burst models. These estimates for each burst were bounded by the lower end of the CHIME band (400 MHz) and the highest frequency channel for which data were available for that particular burst. The bandwidth values are shown in Table \ref{table}.

\subsection{Simulations}\label{sec:sim}

Our investigation revealed that measurement uncertainties, particularly within the Model 4 framework, were underestimated. To estimate these uncertainties and examine potential measurement biases, we conducted simulations using the \texttt{simpulse}\footnote{\url{https://github.com/kmsmith137/simpulse}} routine \citep{marcus_injection}. These simulations are done using the similar approach as presented by \cite{sbm+23}.

For each burst, uncertainty estimation involved simulating a broadband signal ($\sim$ 400 MHz) with a width matching the measured minimum width in the case of multicomponent bursts and measured scattering timescale. The S/N was set to match that reported by \texttt{fitburst}. Subsequently, we inserted this simulated burst into 100 iterations of random noise at a time resolution of 40 microseconds (our median time resolution). These 100 iterations were then processed through \texttt{fitburst}. The resulting standard deviation across these measurements was converted into relative percentage uncertainties, with final uncertainty estimates scaled accordingly.

In addition to our observed distribution, we conducted supplementary simulations to further understand the measurement capabilities of \texttt{fitburst}. These simulations encompassed scenarios involving narrow yet highly scattered bursts ($\sigma$ = 20 $\upmu$s, $\tau$= 20 ms) and wide bursts with negligible scattering ($\sigma$ = 10 ms, $\tau$= 100 $\upmu$s). The implications from these are discussed in Section \ref{sec:corr}.

\subsection{Biases}\label{sec:bias}
Considerations of instrumental and measurement biases are essential for accurately interpreting results. While baseband data mitigates many biases inherent in intensity data for CHIME (see Section \ref{sec:Int vs Base}), such as due to coarser time resolution and formed beam effects \citep{pleunismorph}, they are not without their own limitations. A primary concern is that baseband data storage is restricted to events above a certain realtime S/N cutoff ($\sim$ 10--12 for this dataset), leading to a lack of representation of low-fluence events in our dataset. Furthermore, we are unable to retain full bandwidth information for events with DMs exceeding 1000 pc\,cm$^{-3}$. Additionally, a subset of events lack data for a significant fraction of channels due to system issues during the initial development of the baseband infrastructure. Consequently, our bandwidth estimates may not accurately reflect true bandwidth values even within the CHIME band. For events with sparse channel coverage, estimations of DM, scattering, and other parameters maybe biased. 

Notably, no primary beam corrections have been applied to our dataset, which will be investigated in a future work to estimate spectral indices of the FRB population. The lack of these corrections could introduce biases in bandwidth, particularly for events detected off the meridian. This is attributed to the chromatic nature of the beam, which exhibits a drop in sensitivity at higher frequencies when the source is more than 1 degree off the meridian, with the effect becoming increasingly significant at greater offsets \citep{chime_cosmo_overview, apb+23}. However, fewer than 10 events in our sample meet this criterion. Lastly, inherent selection biases of the instrument during the first catalog period, discussed by \cite{marcus_injection}, contribute to the overall bias considerations, notably bias against wide bursts and highly scattered events.

\section{Results} \label{sec:results}
Below, we discuss the results of our analysis. First we conduct a comparison of the measured morphological properties in baseband data with corresponding measurements obtained from intensity data \citep{chimefrbcatalog1}. 

We then perform correlation studies among various measured parameters. To account for Galactic disk contributions in DM we utilize the NE2001 model \citep{ne2001}. For the Milky Way halo contribution to DM, we adopt a value of 30 pc\,cm$^{-3}$ \citep{cbg+23}. We remove events for which the measured scattering is less than 3 times the expected scattering from the NE2001 model; this amounted to only 3 events, namely FRB\,20190110A, FRB\,20190517C and FRB\,20190609C.

Polarization properties utilized in this analysis are sourced from \cite{ppm+24}. The flux and fluence values are taken from \cite{aaa+23_basecat}. Lastly, we identify bursts exhibiting intriguing morphological features warranting further analysis in order to investigate emission and propagation effects.

\subsection{Comparison with Intensity data} \label{sec:Int vs Base}

The CHIME/FRB intensity catalog was published at a time resolution of $\sim$ 0.983 ms \citep{chimefrbcatalog1}. Here, we present a comparative analysis of this catalog with the higher time resolution baseband data.

\begin{figure*}
\centering
\gridline{\fig{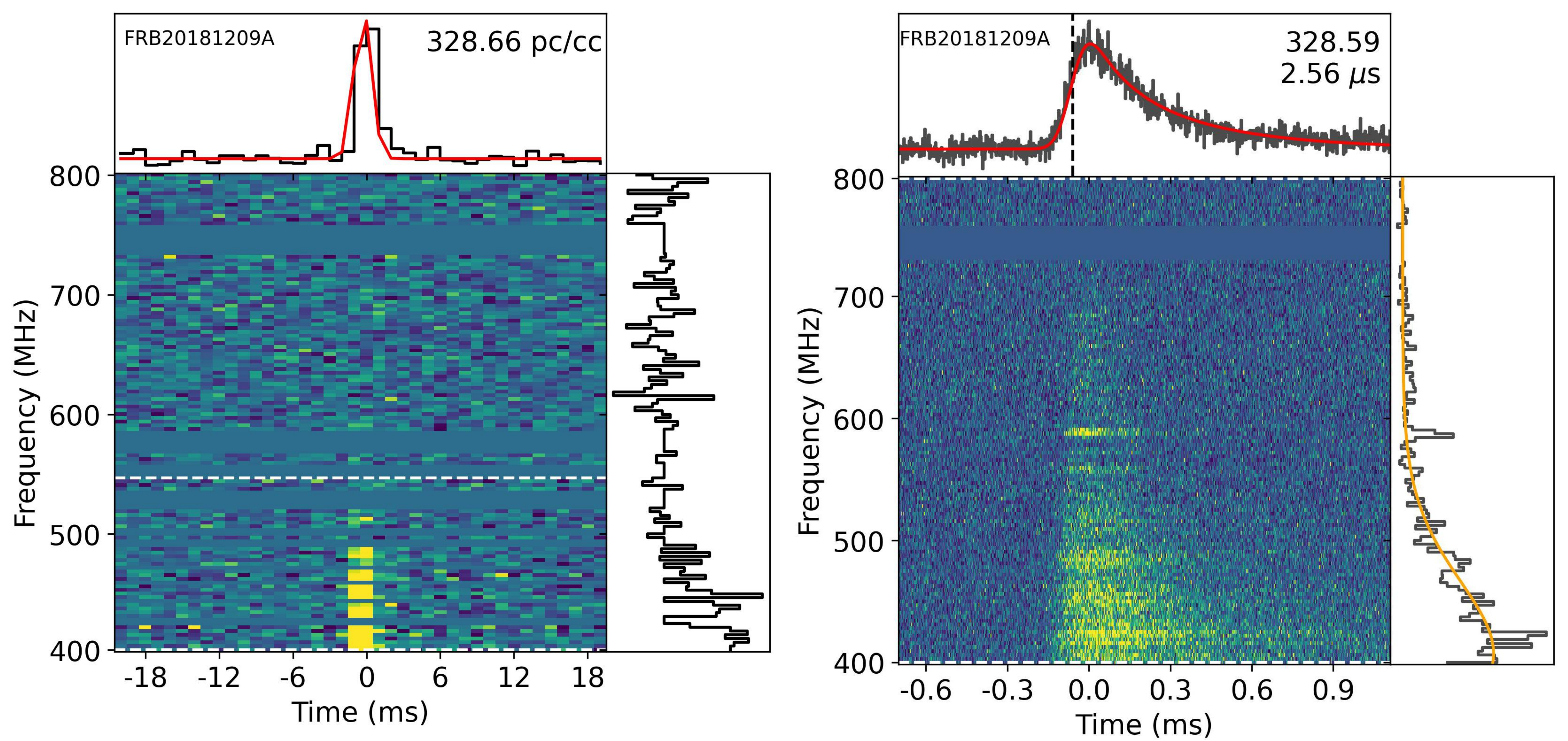}{1.0\textwidth}{}}
\gridline{\fig{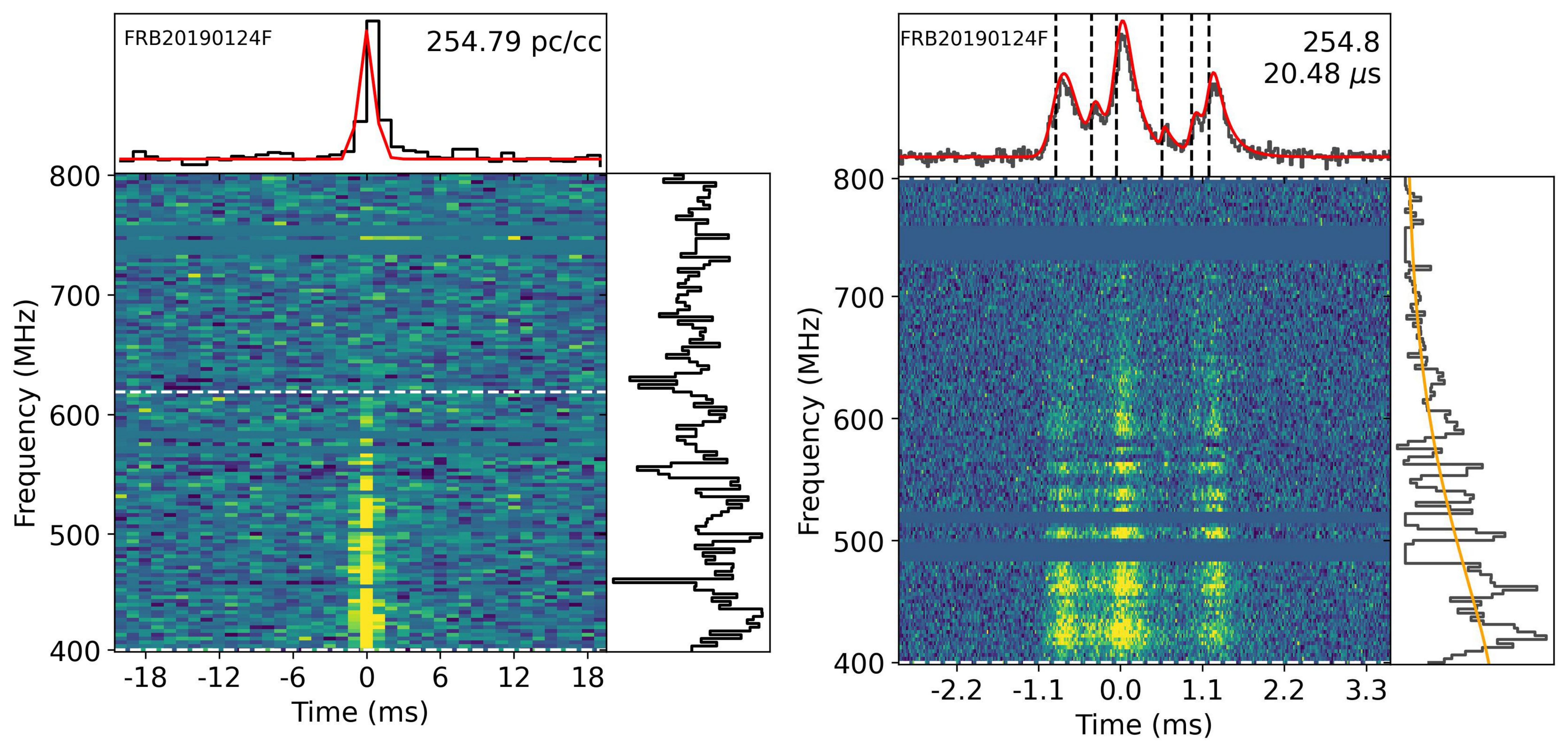}{1.0\textwidth}{}}
    \caption{Comparisons between bursts at intensity data resolution of 0.983 ms (\textit{Left}) and their corresponding representations at baseband resolution (\textit{Right}). Each subfigure depicts the burst's intensity as a function of time in the top panel, with our best-fit model represented by a red line. Below this, the burst's dynamic spectrum is displayed alongside the frequency distribution of power, with the best-fit spectrum denoted by an orange line. In the top left corner of each time series panel is the Transient Name Server (TNS) name, and in the top right corner is the DM to which it has been dedispersed. For baseband bursts, the time resolution is also indicated. For the top burst, we see its narrowband nature attributed to secondary beam effects, and its lack of scattering due to limited time resolution in intensity data, both of which are resolved in baseband data. For the bottom burst, we again observe the resolution of narrowband characteristics, along with the identification of 6 components, rendering the burst more complex. White dashed lines show the extent of emitting bandwidth. The number of channels is 128.}
    \label{fig:bb_vs_int}
\end{figure*}

Figure \ref{fig:bb_vs_int} illustrates the dynamic spectra, along with summed intensity, for FRB\,20180912A and FRB\,20190124F. We show both the intensity and the baseband data for these bursts. For FRB\,20180912A (shown in the top panel), \cite{chimefrbcatalog1} initially provided an upper limit on scattering at intensity time resolution. However, upon further examination at a resolution of 2.56 $\upmu$s, we observe evidence of scattering with a timescale of approximately 100 $\upmu$s at 600 MHz. Additionally, this burst appears broadband in the baseband data, as we eliminate all formed beam effects in this domain. Similarly, the burst in the bottom row of Figure \ref{fig:bb_vs_int}(FRB\,20190124F) covers the whole CHIME bandwidth and reveals 6 components when observed at a resolution of 20 $\upmu$s. These examples are representative of the broader trend observed in many other bursts within our sample. Quantitatively, nearly one-third of the bursts in the baseband sample exhibit additional components at the microsecond timescale resolution. 

The bias in baseband DM compared to intensity data has been discussed by \cite{aaa+23_basecat}. Our DM values were further refined from that presented by \cite{aaa+23_basecat} using the \texttt{fitburst} algorithm. Notably, baseband DMs are generally lower compared to those measured using intensity data. This difference can be attributed to our enhanced ability to characterize scattering timescales and sub-burst drifting with baseband data, both of which are somewhat covariant with DM in low resolution data. 

\begin{figure}
    \centering
    \includegraphics[scale=0.35]{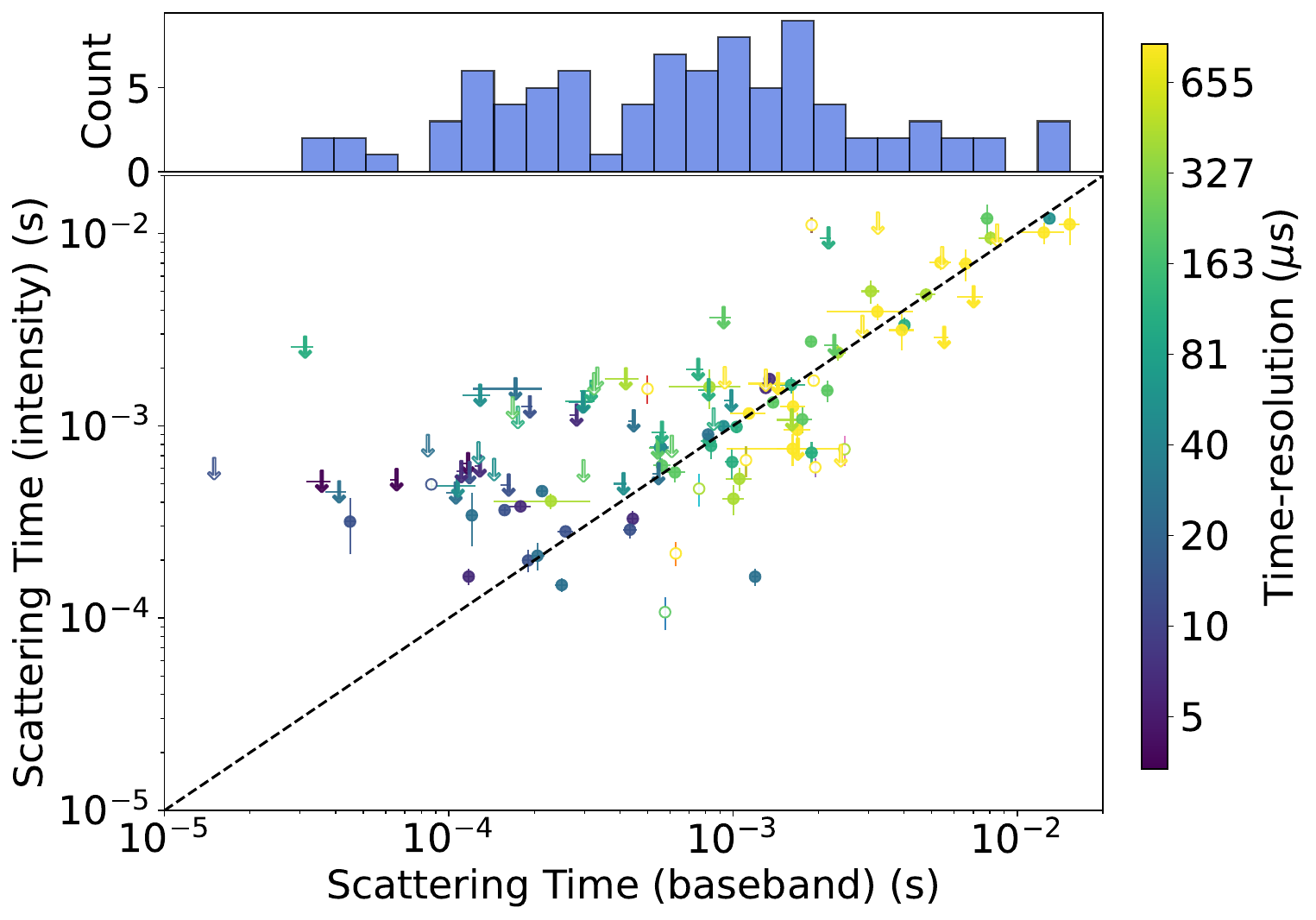}
    \caption{
Top panel: distribution of measured scattering times at 600 MHz obtained from baseband data, excluding upper limits and profile measurements. Lower panel: scatter plot of measured scattering timescales for the same burst in intensity data vs baseband data, including limits. The colorbar represents the time resolution of the baseband measurement (refer to Section \ref{sec:Morphology_pipe}). Downward arrows denote upper limits on scattering in intensity data. Filled arrows indicate measurements obtained in baseband data, while empty arrows signify upper limits in baseband data as well. Empty circles indicate that scattering was measured solely using the profile in baseband data (see Section \ref{sec:fit}). The dashed line represents equal values on both axes. Enhanced time resolution yields more stringent scattering estimates, particularly for cases previously reported as limits in intensity data. }
    \label{fig:scat_bb_vs_int}
\end{figure}

Comparing the measurements of scattering timescales with intensity data, as depicted in Figure \ref{fig:scat_bb_vs_int}, we observe more robust estimates with baseband data. The time resolution at which each measurement was conducted is indicated for each burst, with finer resolutions yielding more precise values. This improvement can be attributed to our ability to resolve more components and better characterize DMs. Solid downward arrows denote upper limits in intensity data for which robust scattering estimates are now achievable. Our lowest measured value is approximately 30 $\upmu$s at 600 MHz. However, some events exhibit increased scattering in baseband data, which can be attributed to the correction of scattering tails as part of DM smearing at coarser resolutions. With improved DM values, accurate measurements of scattering become feasible.

\begin{figure}
    \centering
    \includegraphics[scale=0.35]{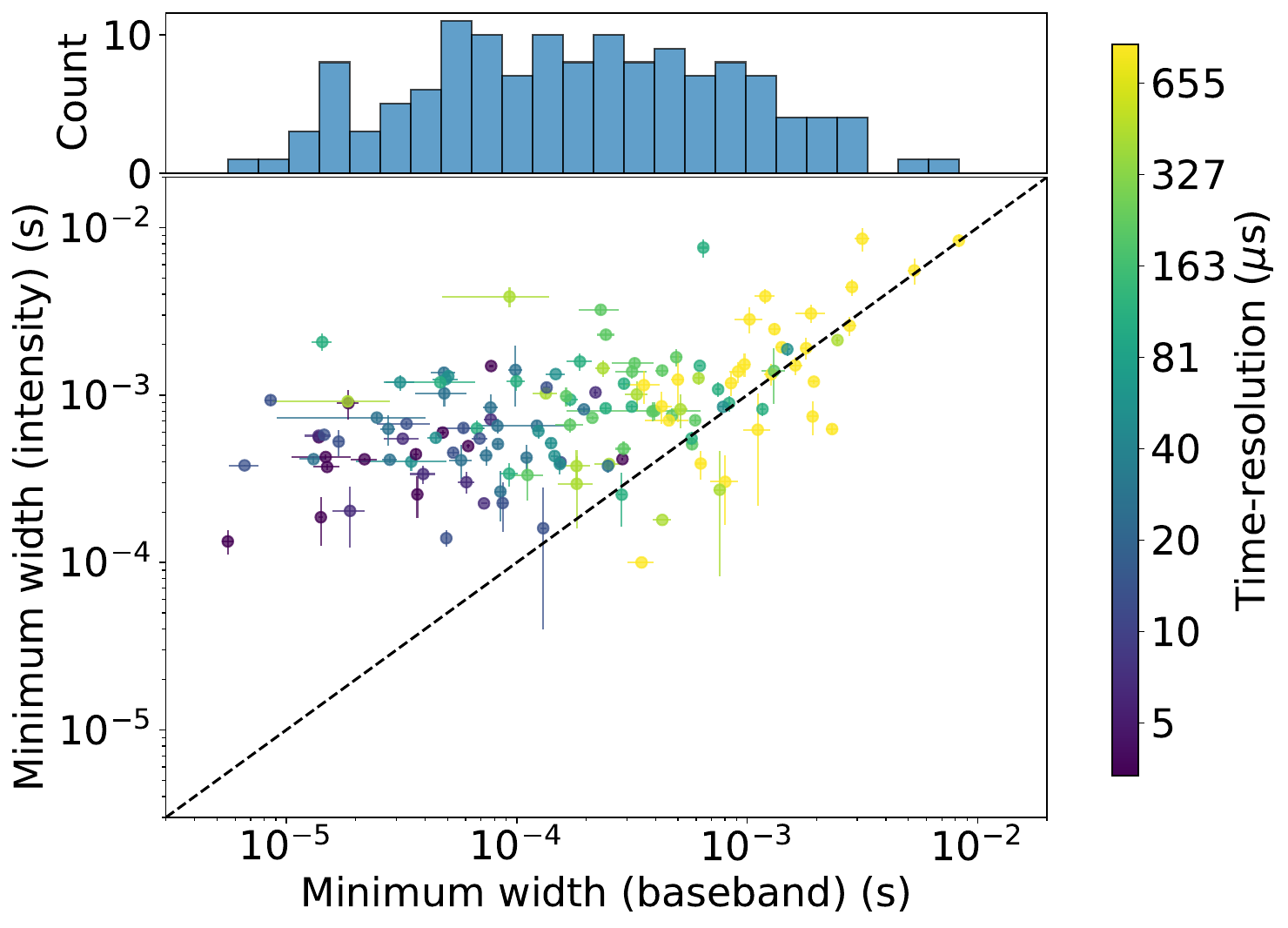}
    \caption{ Figure is similar to Figure \ref{fig:scat_bb_vs_int}, but here we show the minimum width measured for each FRB, in case it has multiple components. As expected, we measure narrower components with baseband data. }
    \label{fig:width_bb_vs_int}
\end{figure}

The impact of time resolution is particularly evident in the measurement of minimum width for multicomponent bursts as shown in Figure \ref{fig:width_bb_vs_int}. In the baseband sample, where we identify more components and implement coherent dedispersion, we discern narrower components within the dataset. In our current sample, we have successfully measured intrinsic minimum widths ($\sigma$) as low as 10 $\upmu$s.

\begin{figure}
    \centering
    \includegraphics[scale=0.32]{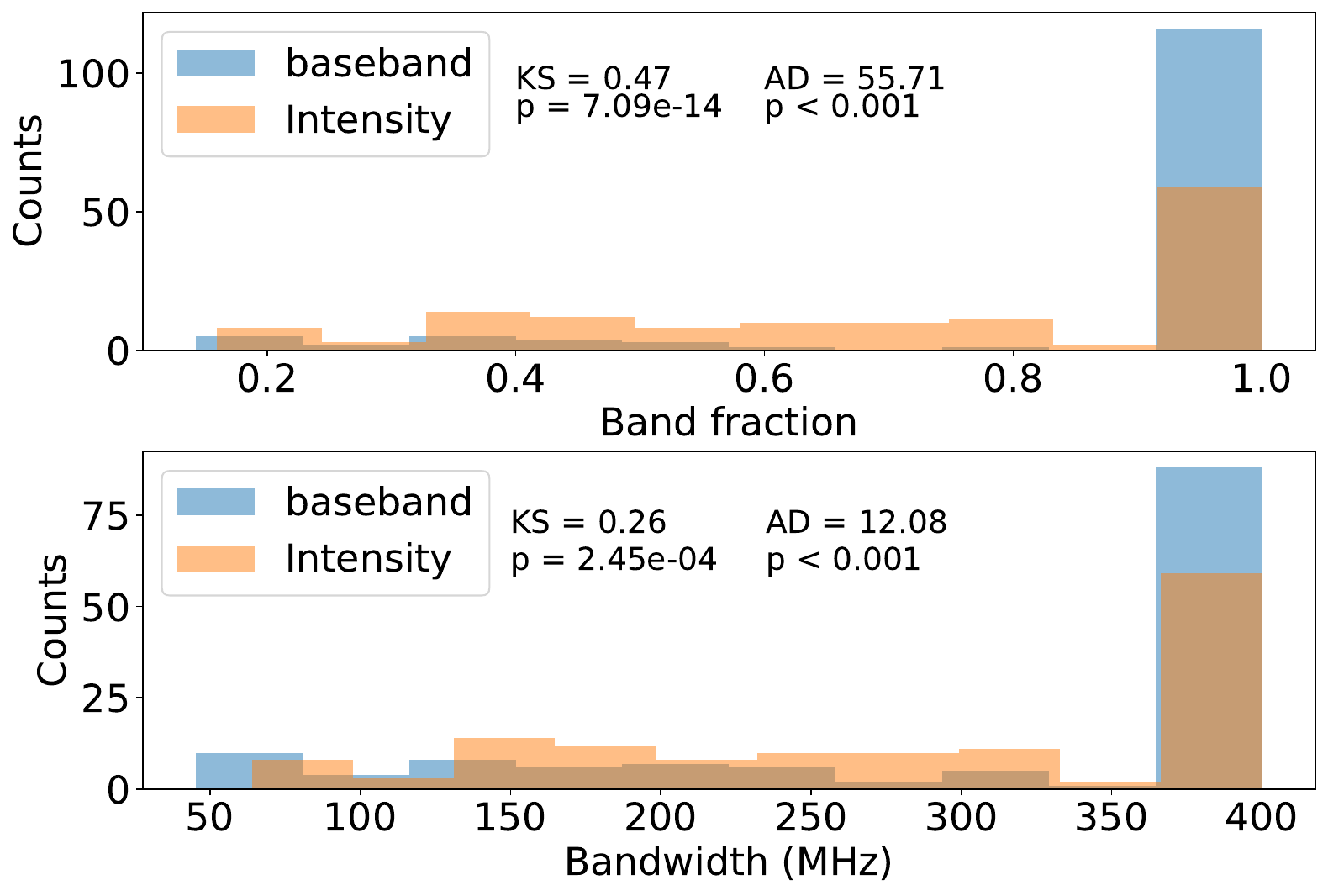}
    \caption{ Top panel: comparison of band fraction between intensity and baseband data. Band fraction represents the ratio of the bandwidth occupied by a burst to the total available observing band. Both KS and AD tests show significant statistical dissimilarity. Bottom panel: same comparison using actual measured bandwidths. Again, bursts appear more broadband in baseband data. This is because baseband data permit beamforming at the most accurately known source position, effectively mitigating secondary beam effects. }
    \label{fig:bw_bb_vs_int}
\end{figure}

Baseband data acquisition enables precise beamforming towards the most accurately known sky position of an FRB \citep{mmm+21}, effectively circumventing secondary beam effects and facilitating robust measurements of bandwidths. In Figure \ref{fig:bw_bb_vs_int}, the top panel illustrates a comparison of band fractions between intensity and baseband data. Band fraction represents the ratio of the burst bandwidth to the available observing band. Despite potential missing channels (see Section \ref{sec:bias}), bursts observed in baseband data exhibit higher band fractions, with nearly 85\% of the bursts in our sample having band fraction equal to 1. We conducted a Kolmogorov-Smirnoff (KS) test \citep{massey1951kolmogorov} and a Anderson-Darling (AD) test \citep{scholz1987k} between the two distributions using \texttt{scipy.stats.ks\_2samp} and \texttt{scipy.stats.anderson\_ksamp} package respectively.  We find them to be statistically distinct (p-value $<$ 0.001 in both cases)i.e., the two samples are drawn from different populations, highlighting the large effect of beam response in intensity data. This disparity is further underscored by the measured bandwidths of the bursts, as depicted in the bottom panel of the Figure \ref{fig:bw_bb_vs_int}. Thus, baseband data, as expected, enhance our comprehension of the spectral property distribution within our sample.

\subsection{Searching for Correlations} \label{sec:corr}

We investigated correlations among various FRB properties using the Spearman rank correlation \citep{spearman1961proof} via the \texttt{scipy.stats.spearmanr} function, which assesses the strength and direction of association between two variables. However, this method only considers robust measured values, overlooking censored data (e.g., upper limits) in our scattering estimates from Model 1 and Model 2 fits.

To address this limitation, we employed a censored Kendall $\tau$ test using the \texttt{cenken} function in R's NADA \citep{NADApackage, Rsoftware} package. The censored Kendall $\tau$ test adapts the traditional Kendall $\tau$ test to handle censored data, adjusting rankings and pairwise comparisons to accurately measure correlations despite incomplete data \citep{helsel2005nondetects}.

Results from the correlation analysis are presented in Table \ref{Tab:corr}. The correlation coefficient indicates the strength of the relationship e.g., +1 for perfect correlation, $-$1 for perfect anti-correlation. The p-value assesses the statistical significance of the observed correlation. 

In the figures discussed below, the filled circles represent measured scattering values derived from Model 3 and Model 4 fits, while the downward arrows indicate upper limits from Model 2 fits. The open circles correspond to profile fits from Model 1. For the Spearman rank correlation analysis, only the measured scattering values (filled circles) are included. In contrast, the Kendall $\tau$ test incorporates all data points, including both measured values, upper limits, and profile fits.

\begin{figure}
    \centering
    \includegraphics[scale=0.30]{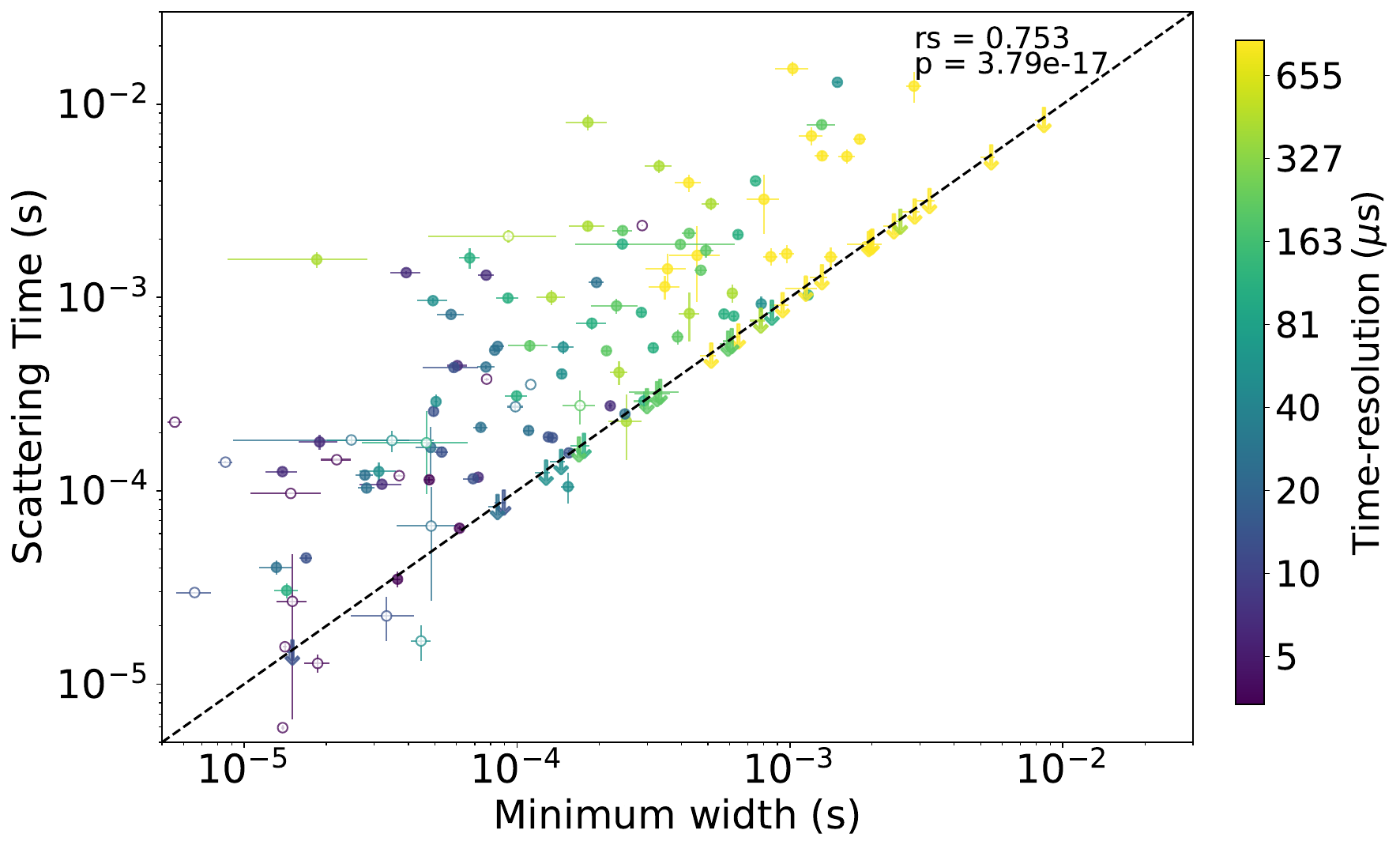}
    \caption{ Scattering timescale at 600 MHz vs minimum intrinsic component width, with the color bar representing the time resolution of the baseband measurements. Downward arrows denote upper limits on scattering, while empty circles represent scattering measurements derived solely from the burst profile. We observe a positive correlation between the scattering timescale and minimum width, as corroborated by the Spearman rank correlation coefficient (rs) value displayed in the top right corner, along with its corresponding p-value. The rs value was calculated solely using bursts with robust scattering measurements (filled circles). However, this correlation is likely an observational bias, as discussed in the text (\S\ref{sec:corr}).}
    \label{fig:corr_width_scat}
\end{figure}

We observe a positive correlation between the scattering timescale and the measured minimum width, as depicted in Figure \ref{fig:corr_width_scat}. However, this correlation likely arises from observational bias. Simulations reveal that accurately estimating scattering timescales becomes challenging for broader bursts. Notably, when the scattering timescale ($\tau$) at 600 MHz matches the intrinsic width ($\sigma$), \texttt{fitburst} consistently retrieves both parameters with high precision. For cases where $\tau < \sigma$, the estimation accuracy depends on the burst's S/N. But in extreme cases if the scattering is an order of magnitude smaller than the intrinsic width, \texttt{fitburst} produces scattering estimates that fluctuate around the initial guess. Hence to be conservative, for all Model 2 events, we report an upper limit on the scattering time, equivalent to the minimum intrinsic width of the burst components, this explains the lack of measured data points in the bottom right corner of the figure. Furthermore, our search pipeline is known to be biased against highly scattered events, as identified through injections in our initial catalog \citep{marcus_injection, chimefrbcatalog1}.  

Additionally, we assumed a single thin scattering screen, resulting in a scattering kernel with a very low rise time. In reality there may be multiple scattering screens along the line of sight or a single thick screen resulting in a scattering kernel with a longer rise time \citep{kbm+2019,gsa+21}, which we have interpreted as a larger intrinsic width, giving rise to the observed correlation. This may explain the absence of extremely narrow but highly scattered events in the top left corner of the figure. The biases discussed here regarding our scattering time measurements should be taken into consideration while interpreting the correlations below as they can potentially mask weak correlations. 

\begin{figure}
    \centering
    \includegraphics[scale=0.32]{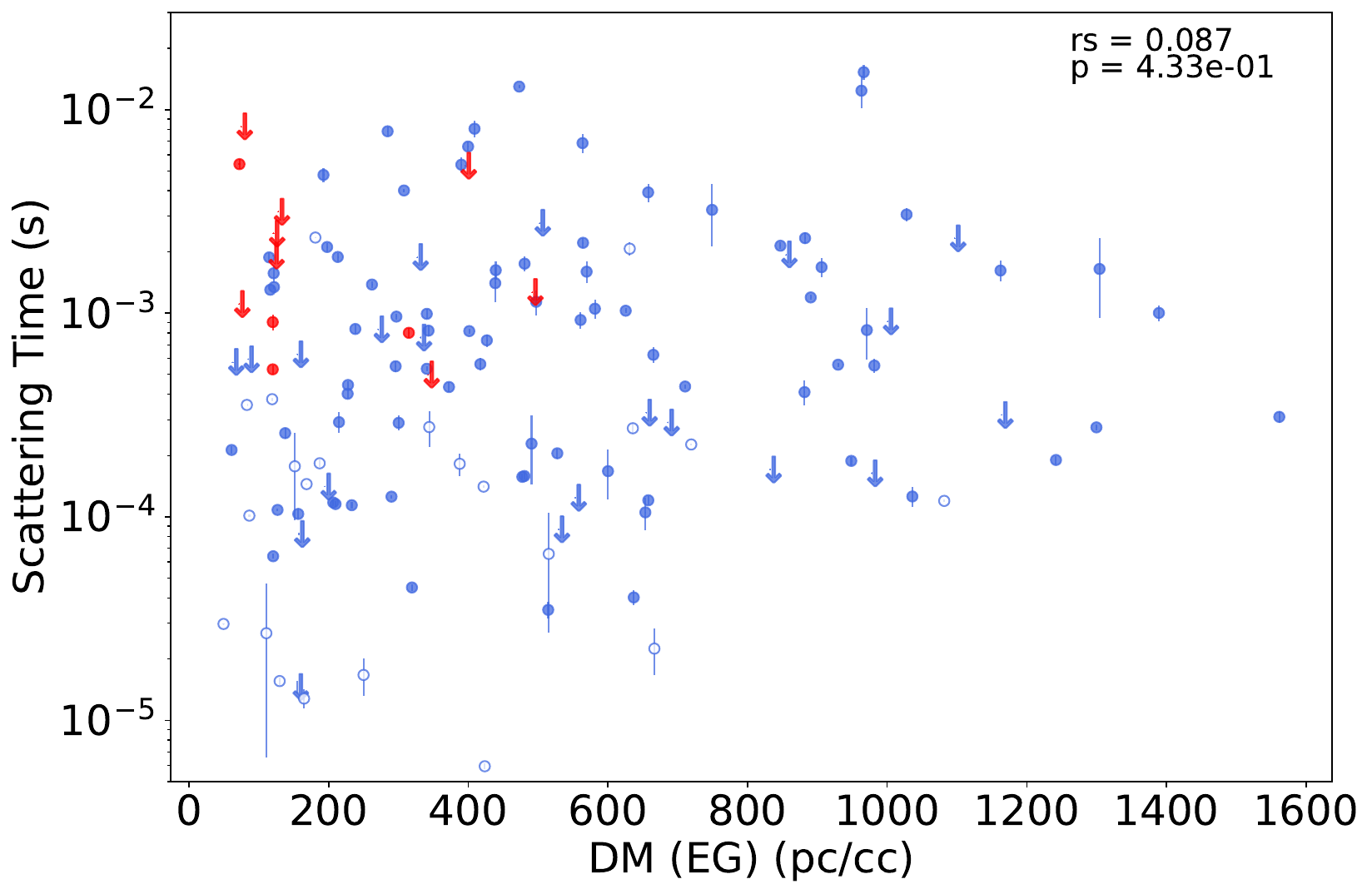}
    \caption{ Extragalactic DM vs scattering timescale at 600 MHz. The Galactic contribution is subtracted using estimates from the NE2001 model \citep{ne2001}. The red points are bursts from repeaters and blue show the one-off sources. The downward arrows are upper limits on scattering times and empty circles signify measurements done by fitting EMG models to just the temporal profile. We observe no significant correlation between the two values, suggesting that the sources responsible for the majority of the contributions to the scattering and DM are distinct (see Section \ref{sec:corr}). All the values used here and in following figures are in observers frame, i.e., not corrected for redshift. Spearman rank correlation coefficient (rs) value displayed in the top right corner, along with its corresponding p-value.}
    \label{fig:corr_DM_scat}
\end{figure}

We performed correlation studies between scattering and other measured properties. Given that scattering is a manifestation of line-of-sight propagation effects, it offers valuable insights into both the local environment and the intervening medium. We observed no significant correlation between scattering and extragalactic DM as shown in Figure \ref{fig:corr_DM_scat}, suggesting that the majority of contributions to DM and scattering originate from distinct sources. Alternatively, our observations may lack the requisite redshift range to discern any potential correlations. 

\begin{figure}
    \centering
    \includegraphics[scale=0.32]{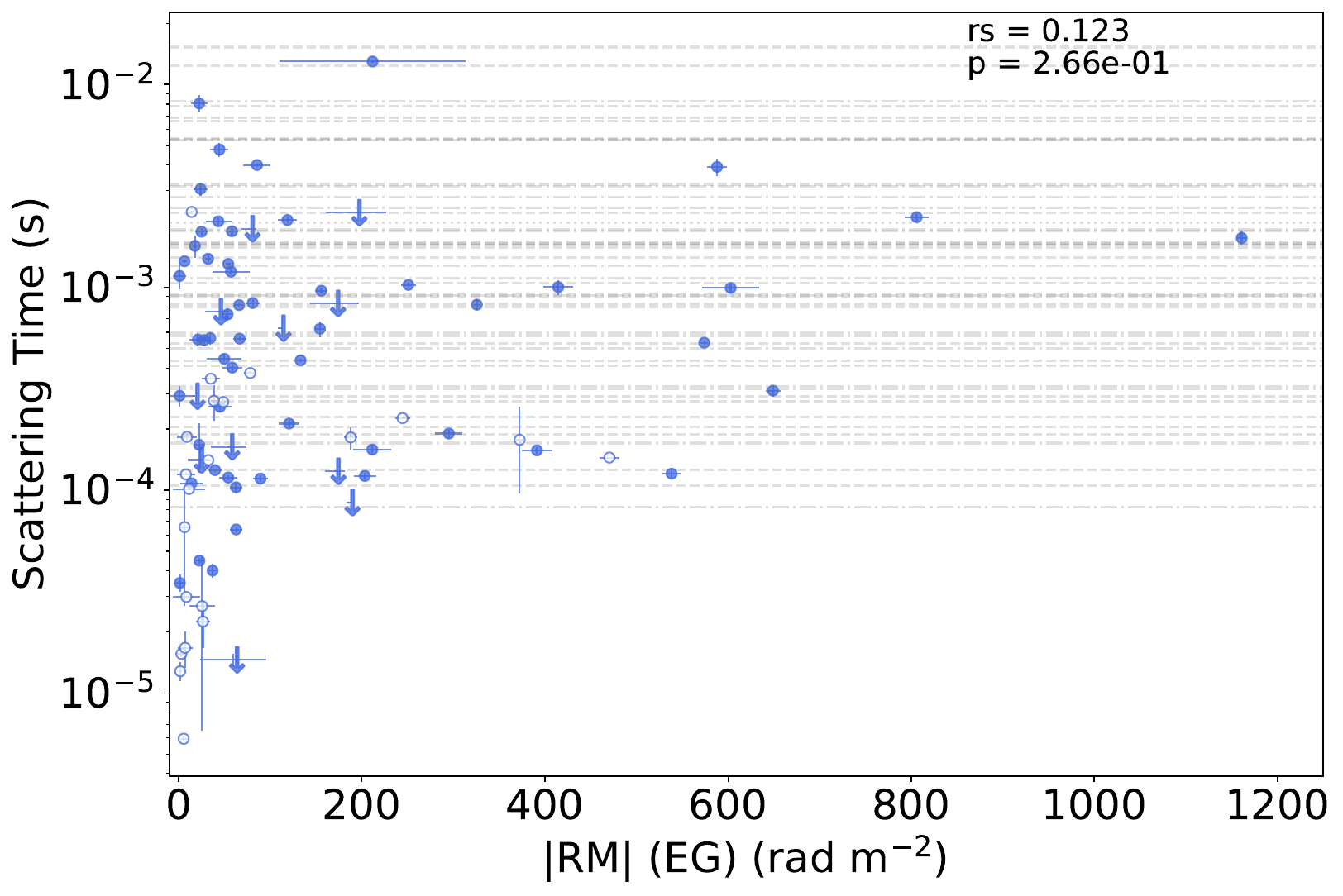}
    \caption{ Extragalactic RM vs scattering timescale at 600 MHz. Dashed lines represent the scattering timescale of sources with no measured RM value. Our analysis reveals no significant correlation between the two parameters. Its important to note that our sample is limited at high RM values ($>$ 500 rad m$^{-2}$) hence devoid of sources residing in extreme magneto-ionic environments that can potentially affect the correlation. Spearman rank correlation coefficient (rs) value displayed in the top right corner, along with its corresponding p-value.} 
    \label{fig:corr_RM_scat}
\end{figure}

\begin{figure}
    \centering
    \includegraphics[scale=0.32]{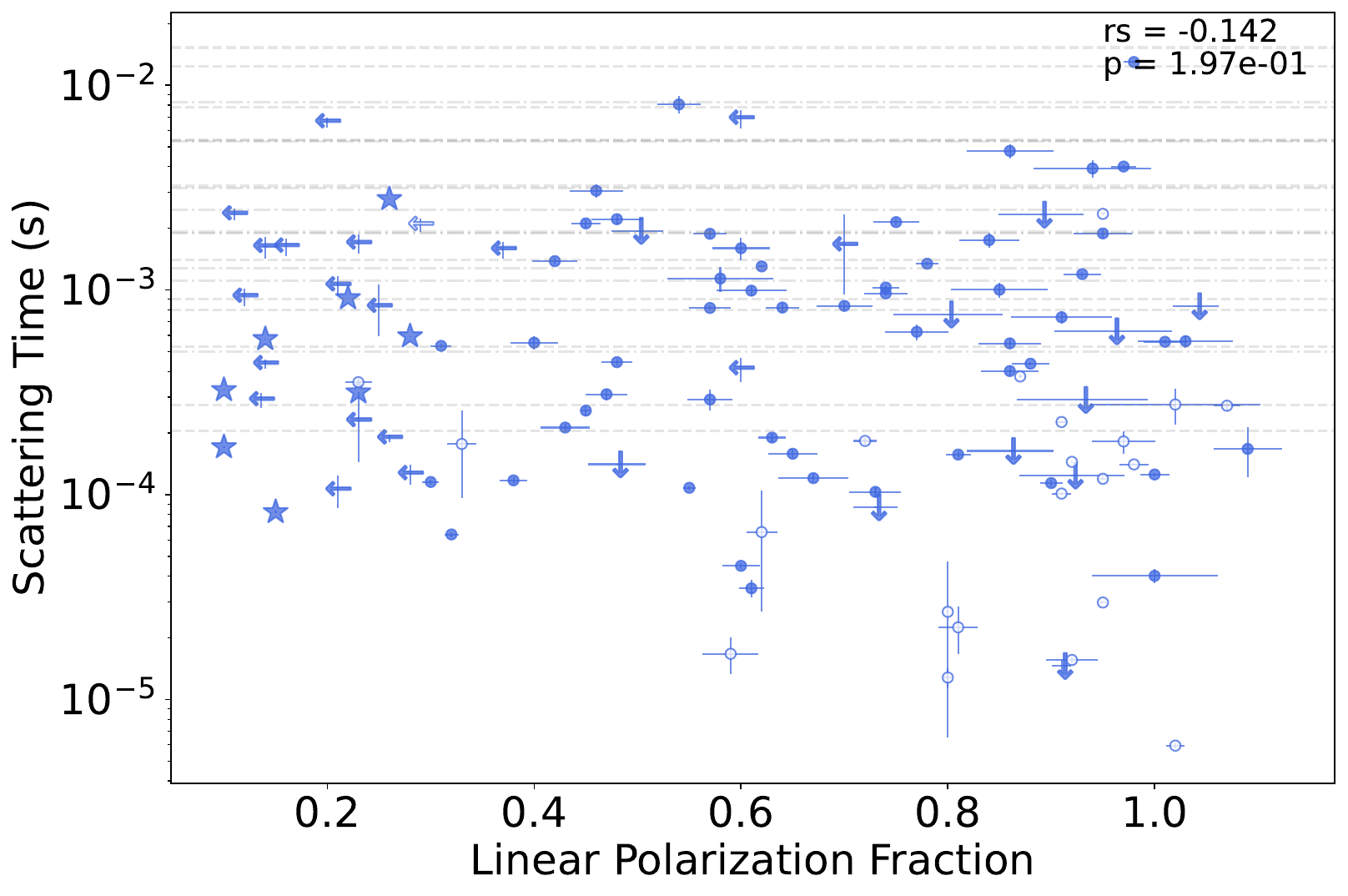}
    \caption{ Linear polarization fraction in the CHIME/FRB band vs scattering timescale at 600 MHz. Downward filled arrows are sources with upper limits on scattering but with measured polarization fractions. Left filled arrows are sources with measured scattering but with upper limits on the polarization fraction. Stars are sources with upper limits on both scattering and polarization fraction. Empty circle indicate scattering estimates from profiles only. Dashed lines represent sources with no measured linear polarization fractions. The dots indicate sources with both measured scattering and polarization fractions. We see no correlation between these two parameters. This finding suggests that depolarization might not be the result of multipath propagation in our sample. Spearman rank correlation coefficient (rs) value displayed in the top right corner, along with its corresponding p-value.}
    \label{fig:corr_polfrac_scat}
\end{figure}

Similarly, we found no strong correlation between scattering and RM, as shown in Figure \ref{fig:corr_RM_scat}. This suggests that the scattering screen may not coincide with the Faraday active medium responsible for the majority of the RM contribution. The absence of correlation between scattering and polarization fraction, as depicted in Figure \ref{fig:corr_polfrac_scat}, shows lack of frequency-dependent depolarization due to multipath propagation in our sample. Notably lack of high RM sources in our sample might bias our overall correlations. 

To further assess the absence of correlations between scattering and DM, RM, and L/I , we conducted a median absolute deviation from the median (MADFM) analysis utilizing the \texttt{scipy.stats} package. Specifically, we calculated the MADFM for the measured scattering data points (represented by filled circles) in groups of 10 across all correlation plots, tracking its evolution. As illustrated in Figure \ref{fig:madfm_corr}, the resulting trend remains consistent with a horizontal line, thereby reinforcing the conclusion of no statistically significant correlations within the dataset.

\begin{figure}
    \centering
    \includegraphics[scale=0.45]{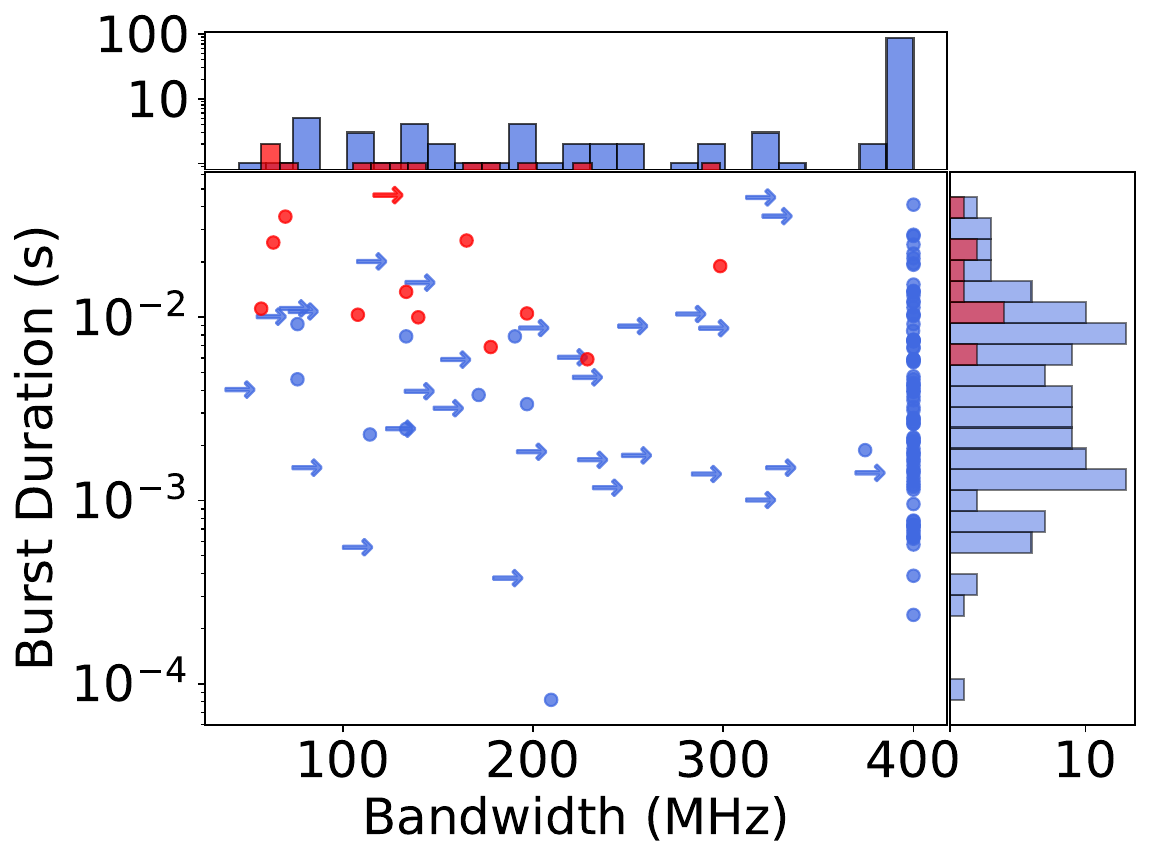}
    \caption{ Burst duration vs bandwidth for repeaters (red) and one-off events (blue). Top panel: distribution of bandwidth. Right panel: distribution of burst duration. In the scatter plot, right-facing arrows denote events for which not all channels were saved (refer to Section \ref{sec:bias}), providing a lower limit on the bandwidth within the CHIME band. Repeaters exhibit wider temporal profiles and narrower bandwidths overall compared to one-off events, consistent with previous findings by \cite{pleunismorph}.}
    \label{fig:corr_width_vs_bw}
\end{figure}

We define burst duration or burst envelope as the time interval between the 10\% fluence level at the rise of the initial component and the 90\% fluence level at the decay of the final component. The blue shaded region in the top panels of Figure \ref{fig:burst_wfalls} show the extent of the burst duration. In Figure \ref{fig:corr_width_vs_bw}, we examine the bandwidth and burst duration distribution of repeaters (in red) and one-off events (in blue). The right-facing arrows denote events for which we lacked data from all channels. Our findings bolster the hypothesis proposed in \cite{pleunismorph} that repeaters exhibit narrower frequency bandwidths and wider temporal durations compared to one-off events. However, a more extensive analysis with a substantially larger repeater sample will be presented elsewhere \citep{curtin2024}.

\begin{figure}
    \centering
    \includegraphics[scale=0.34]{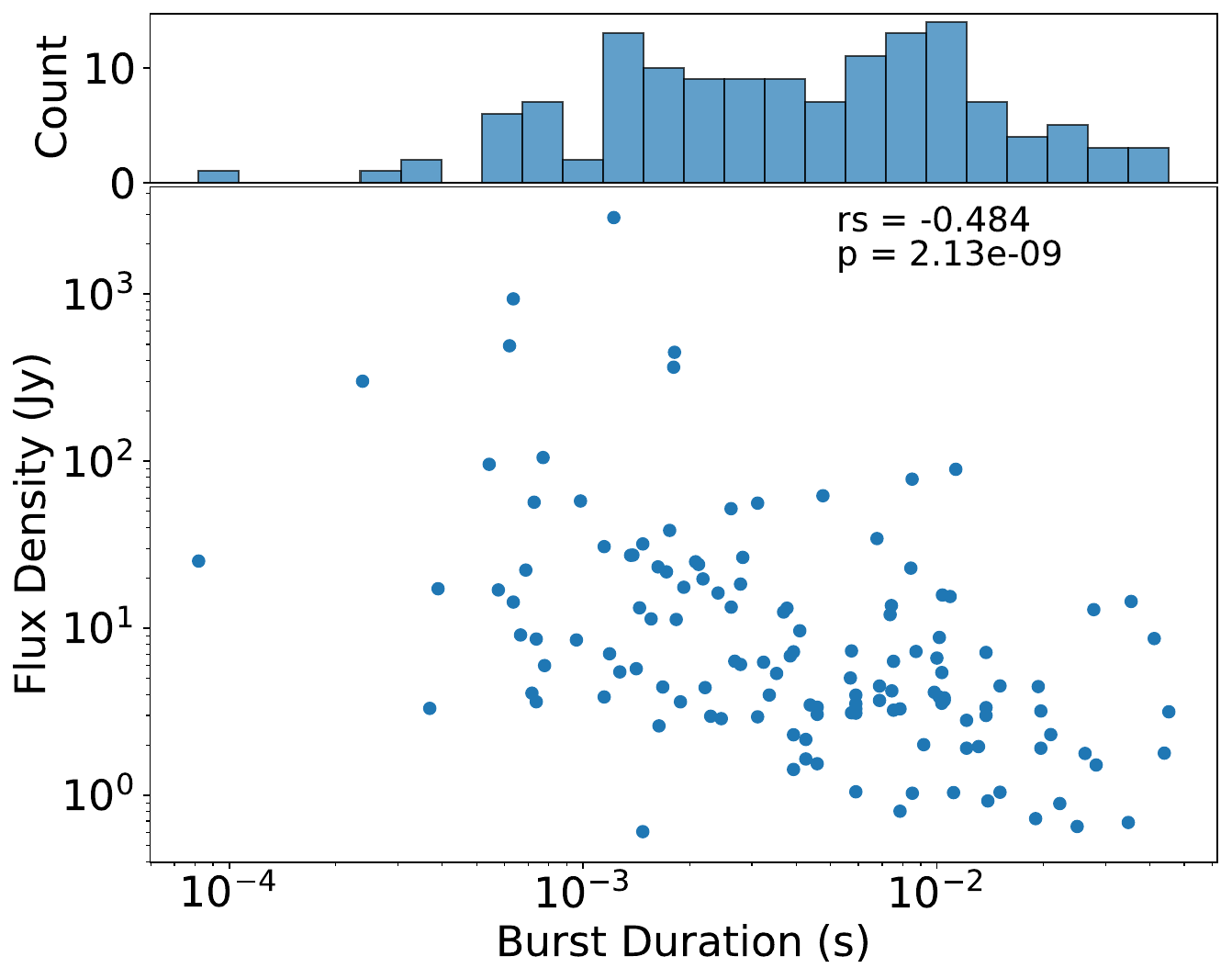}
    \caption{ Top panel shows the distribution of burst duration in our sample. The bottom panel shows correlation between peak flux density and the burst duration. We found brighter bursts have narrower durations.}
    \label{fig:corr_boxcar_flux}
\end{figure}

In Figure \ref{fig:corr_boxcar_flux}, we compare the burst duration with their measured peak flux density \citep{aaa+23_basecat}. The analysis reveals that bursts characterized by narrower temporal envelopes exhibit higher peak flux densities.

We observed no significant correlations among other parameters like Fluence vs Scattering, Fluence vs DM and Width vs DM. The plots for all of these are shown in the Appendix in Figure \ref{fig:additional_corr}.

\begin{table}[!ht]
    \centering
    \caption{The table provides both the Spearman rank correlation coefficient ($r_s$) and censored Kendall’s $\tau$ coefficient ($\tau_{coeff}$), along with their respective p-values, to understand the correlation between the properties. We performed the $\tau$ test only for datasets where we had censored values (see Section \ref{sec:corr}).}  \label{Tab:corr}
    \begin{tabular}{ccccc}
    \hline
        Property & $r_s$ & $p_{rs}$ & $\tau_{coeff}$ & $p_{\tau}$ \\ \hline
        Min. Width vs Scattering & 0.76 & $10^{-17}$ & 0.36 & $10^{-10}$ \\ 
        DM vs Scattering & 0.09 & 0.433 & 0.11 & 0.051  \\ 
        RM vs Scattering & 0.12 & 0.266  & 0.15 & 0.045 \\ 
        Pol. Fraction vs Scattering & $-$0.14 & 0.197 & $-$0.11 & 0.095 \\ 
        Fluence vs Scattering & -0.03& 0.791 & 0.02 & 0.709  \\ 
        Duration vs Flux & $-$0.48 & $10^{-9}$ & - & - \\
        Fluence vs DM & $-$0.19 & 0.195 & - & - \\ 
        Width vs DM & 0.15 & 0.080 & - & - \\ \hline
    \end{tabular}
\end{table}

\subsection{Complex Morphologies}\label{sec:complex}

In the first catalog, limited time resolution hindered our ability to explore the full morphological diversity of the FRB sample. \cite{pleunismorph} identified four archetypes within the first catalog. While we still observe similar behavior in our dataset, we note a higher fraction of bursts categorized into the two archetypes involving complex multicomponent structures, approximately 40\% compared to 10\% reported in the intensity data catalog. 

We identified 33 sources in our sample with component intrinsic widths of $\leq$ 50 $\upmu$s. It is plausible that even narrower microstructures ($\leq$ 20 $\upmu$s) may be present in these bursts; however, robustly claiming such detections necessitates careful DM estimation alongside other measurement constraints.

\begin{figure}
    \centering
    \includegraphics[scale=0.45]{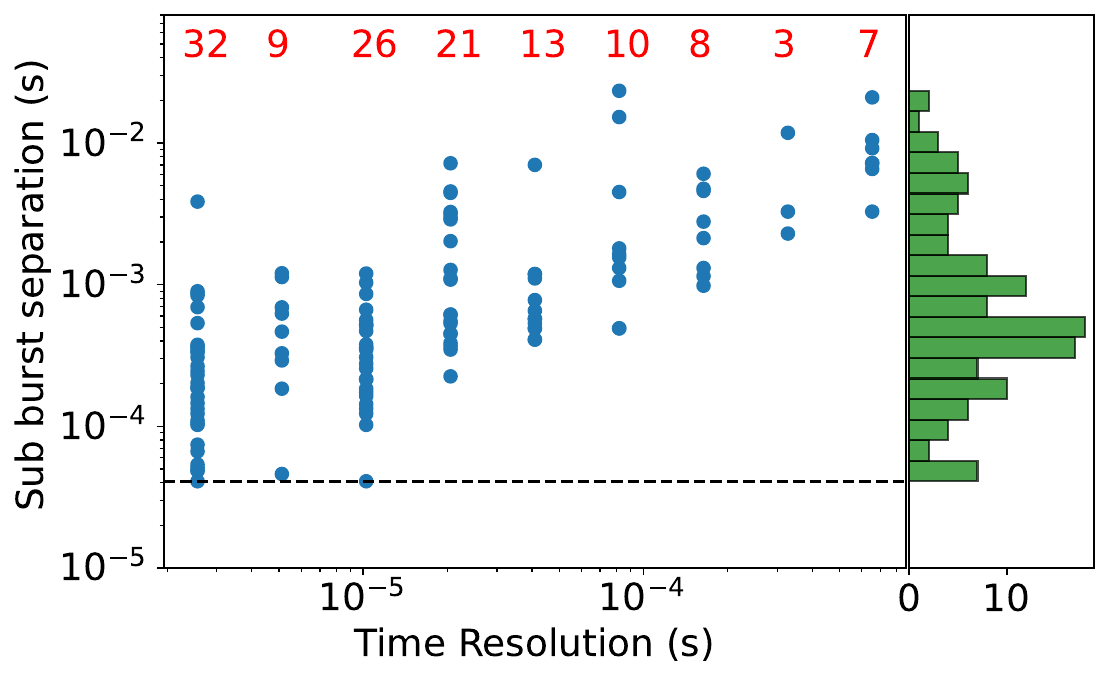}
    \caption{ Sub-burst separation vs time resolution. The numbers in red quantify sub-burst arrival times measured at each time resolution. The right panel of the figure illustrates the separation between the arrival times of sub-bursts for multicomponent bursts. The scatter plot shows the sub-burst separation measured at the respective time resolution of a burst. Our analysis reveals no sub-burst separation below 41 $\upmu$s, as indicated by the black dashed line. }
    \label{fig:arrival_time}
\end{figure}

We investigated the separation between the arrival times of sub-pulses in multicomponent bursts. As illustrated in Figure \ref{fig:arrival_time}, we did not observe any characteristic sub-burst separation or bimodality in our sample. Notably, we found no separations below 41 $\upmu$s, despite analyzing bursts at a finer time resolution (2.56 $\upmu$s), as depicted by the numbers at the top of the figure. The majority of our multicomponent bursts exhibited separations of less than 1 ms, indicating that studying FRBs at finer resolutions will yield a more detailed morphological dataset. This approach also holds promise for elucidating quasi-periodicity in FRBs, as previously observed by \cite{abb+21b} in FRB\,20191221A.  In our sample, we identified approximately 10 bursts with 5 or more components, making them candidates for quasi-periodicity searches noteworthy examples, such as FRBs 20190122C, 20190411C, and 20190617A, are depicted in Figure \ref{fig:burst_wfalls}. The search results from one of these bursts (FRB\,20190425A) has been published by \cite{fmm+23}; they found no quasi-periodicity. 

Upon visual examination of the waterfall plots, we observed sub-bursts displaying drifting that is best described by a linear function along with that best described with a power law \citep{fmm+23}. Notably, FRB\,20181224E, FRB\,20190224D, FRB\,20190411C, FRB\,20190502C, FRB\,20190612B and FRB\,20190630D exhibited likely instances of such behavior. This phenomenon could be attributed to plasma lensing along the line of sight \citep{cwh+17}. To validate this hypothesis, a phase-coherence search among sub-bursts utilizing raw voltage data can be employed \citep{kader2024}, but is beyond the scope of this paper.

\section{Discussion}\label{sec:discuss}

\subsection{Morphology : Repeaters vs One-off sources}

This study presents the largest sample of one-off FRBs analyzed at microsecond timescales to date. Nearly 40\% of our sample exhibits multiple components, compared to 10\% in the previous study at lower time resolution \citep{pleunismorph}. This sample includes 12 bursts from repeaters within the baseband catalog timeframe, detailed in Table \ref{tab:burst}.

In Figure \ref{fig:corr_width_vs_bw}, a distinction is observed in the temporal and bandwidth distributions of repeaters and one-off sources, consistent with findings by \cite{pleunismorph}. This suggests that bursts with wider durations and narrower bandwidths are predominantly attributed to repeating sources. In our reference sample apart from known repeat bursts (Figure \ref{fig:burst_wfalls}), FRB\,20181221A, FRB\,20190130B, FRB\,20190430C, FRB\,20190609C, FRB\,20190612B, and FRB\,20190630D exhibit such morphology. Notably, FRB\,20190430C and FRB\,20190609C eventually repeated \citep{RN3}, while others remain fruitful targets for follow-up observations.

Sub-burst drifting, a key morphological feature associated with repeaters \citep{Hes2019}, has also been identified in extremely bright bursts from apparent one-off sources \citep{fmm+23}. Qualitative analysis of our sample reveals bidirectional drifting in the central emission frequency across sub-bursts in FRB\,20190411C and FRB\,20190502C (Figure \ref{fig:burst_wfalls}). In sources such as FRB\,20181224E, FRB\,20190106B, and FRB\,20190630D, sub-bursts are identified where a single DM does not align all components. This phenomenon is reminiscent of the residual drift observed in wider components of FRB\,20220912A by \cite{Hewitt_2023}, despite narrower ``microshot" components within the same burst envelope aligning perfectly after dedispersion. Additionally, they suggested such diverse morphology is analogous to what have been observed in solar radio bursts (SRBs). Type I SRBs consist of short-duration narrowband emissions, whereas Type II and III SRBs exhibit drifting structures and are associated with coronal mass ejections and emission along open magnetic field lines, respectively. A similar scenario might be at play for FRBs, where a single central engine could produce morphologically diverse emissions through multiple physical mechanisms.

Baseband data allowed for an accurate estimation of the true band fraction by mitigating formed beam effects present in the measurements from the first CHIME/FRB catalog \citep{chimefrbcatalog1}. Although primary beam effects might still cause underestimation, our analysis shows that nearly 85\% of sources have a band fraction of unity, indicating that the FRB population is generally broadband, extending at least over the full CHIME bandwidth.

However, to explain the volumetric rate of FRBs, it has been aruged that a significant fraction must be repeaters \citep{rav19b}. Repeaters are usually (though not exclusively) narrowband \citep{pleunismorph}. It is possible that apparent one-off bursts represent the brighter end of the luminosity distribution of a repeating source, while weaker repeat bursts are narrowband. Such a dichotomy in bandwidth distribution has been observed for FRB\,20121102A \citep{hsh+22}. Furthermore, \cite{koh+24} demonstrated that FRB\,20201124A exhibits a broken power law in its energy distribution of bursts, with brighter bursts showing a much flatter index, consistent with what has been observed for one-off sources \citep{smb+23}. Additionally, in their sample they found bursts from the low and high ends of the energy distribution to be statistically indistinguishable in morphology. This suggests that bright, broadband bursts with morphology similar to those of apparently one-off sources may also be emitted by repeating sources, albeit rarely.

Comparing the morphological properties of a large sample of repeat bursts studied at high time resolution to the sample presented here will provide further insights into whether there is a continuum in emission between repeaters and one-off sources or a clear dichotomy. Such a study is underway using the CHIME/FRB repeater sample baseband data \citep{curtin2024}. Access to such a large dataset at high time resolution could also enhance the exploration of modeling capabilities using neural networks, potentially aiding in the identification and prediction of repeating sources, as initially suggested by \cite{pleunismorph}.

\subsection{Constraints on FRB emission}

We achieve component measurements two orders of magnitude finer than those reported by \cite{chimefrbcatalog1}, with 26 bursts, or roughly 20\% of the sample, displaying components with FWHM $\leq$ 100 $\upmu$s. There might be even narrower components, but we may be limited by scattering and low S/N in some cases. Investigating components smaller than 10 $\upmu$s will require careful optimization of DM to remove intrachannel smearing. Regardless, our analysis suggests that structures tens of microseconds wide might be common in FRBs, provided that they are not smeared out by propagation effects such as scattering. Structures down to tens of nanoseconds have been observed for the repeating source FRB\,20200120E at higher frequencies (1.4 GHz; \citealp{nimmo2022burst}, 2.3 GHz \citealp{majid_nano_2021}), and other repeaters have also shown structures narrower than 10~$\upmu$s \citep{nimmo2021_micro, Hewitt_2023}.

Such narrow bursts favor an emission model acting within or near the magnetosphere of a compact object, likely a neutron star \citep{lu2018radiation,2020MNRAS.498..651B}. This is consistent with the sub-second periodicity observed in FRB\,20191221A \citep{abb+21b}, and the recent report of a S-shaped linear polarization angle sweep in FRB\,20221022A \citep{mckinven24}.  An alternative scenario involving shock wave interaction with past flare ejecta at greater distances \citep{mms19} predicts longer-duration pulses, but may still be plausible, by concentrating the maser emission to a small patch of shocked plasma. 

We observe no sub-components separated by $\leq$ 41 $\upmu$s despite having sufficient time resolution to detect much shorter timescales (see Fig.\ref{fig:arrival_time}). Sub-component separations less than 10$\upmu$s have been observed from FRB\,20201220E \citep{majid_nano_2021, nimmo2022burst} and FRB\,20220912A \citep{Hewitt_2023}, notably in extremely bright bursts at higher frequencies (both at 1.4 GHz). Hence, in the CHIME band, scattering and low S/N ratios may be biasing us against detecting smaller separations. A similar analysis with a larger sample may eventually identify shorter separations in CHIME/FRB bursts.

\subsection{Correlations and FRB propagation}

We observe a diverse range of scattering timescales, spanning three orders of magnitude, from 30 $\upmu$s to 13 ms. These scattering estimates greatly exceed those predicted by the NE2001 model for the Milky Way \citep{ne2001} which is expected because of the extragalactic origins of FRBs. The dominant scattering screen can be located anywhere from the circumburst environment, as seen for FRB\,20190520B \citep{ocker_20190520B}, to the circumgalactic medium along the line of sight, as observed for FRB\,20221219A \citep{fro2024}. Scintillation estimates can further constrain the position of the screen, as demonstrated for FRB\,20221022A \citep{npb2024}. A scintillation analysis for our sample is underway and will be presented in a future work.

The lack of strong correlation (see Section \ref{sec:corr}) presented here provides valuable insights into the origins of morphological and polarization propagation effects. However, it is important to interpret these correlations in light of the strong bias CHIME/FRB has against detecting events with very long scattering times \citep{chimefrbcatalog1}. 

\subsubsection{Extragalactic DM vs Scattering}

A correlation between DM and scattering is seen among Galactic radio pulsars \citep{Cor16b}, which are generally strongly confined to the Galactic Plane. For FRBs we find no correlation between extragalactic DM and scattering time in our study (see Section ref{sec:corr}). The lack of any apparent correlation argues that the bulk of the extragalactic component of FRB DMs is not in the ISM of the host (as it is for Galactic pulsars).  This is not surprising because extragalactic DM correlates strongly with redshift \citep{macquart2020census}, arguing the former lies in the intergalactic medium, which itself does not cause much scattering \citep{occ_2022, coc22}.  With a well calibrated DM vs redshift relation and a large sample of host-localized FRBs, it may be possible in the future to discern a DM/scattering correlation after removing the average IGM DM contribution from the extragalactic DM.  This would be an interesting line of investigation for the upcoming CHIME/FRB Outriggers \citep{lanman_kko}. 

Additionally, with availability of more constraining measurements with this dataset one can forward model the DM and scattering budget for the observed population as presented by \cite{chawla_scattering_dm}. We can then constraint where the dominant scattering screen lies for most FRBs, whether its the CGM \citep{vedantham_phinney_2019} or do most FRBs indeed inhabit local environments with extreme properties.

\subsubsection{Extragalactic $|RM|$ vs Scattering}

We find no obvious extragalactic $|RM|$ vs scattering correlation, which would have been expected if the local highly magneto-ionic FRB environments that cause high RMs can also be turbulent and cause high scattering. As shown in Figure \ref{fig:corr_RM_scat} the correlation is not statistically significant. This suggests that the Faraday active medium responsible for the majority of the RM contribution might not coincide with the dominant scattering screen. Supporting this claim are RM, scattering measurements from repeating sources such as FRB\,20180916B which demonstrate intervals of significant secular RM variations but no comparable behavior in scattering \citep{mckinven_2022_R3, sbm+23}.

The magnetar near the center of the Milky Way, SGR J1745$-$2900, has both very high RM local to the source \citep{Desvignes_2018} and very long scattering time (e.g., $\sim$130 ms at 2 GHz; \citealt{pennucci2015, spitler+2014, pmp+18}). It has been shown using angular broadening measurements that the majority of the scattering contribution arises from a single screen $\sim$ 5 kpc towards the Galactic centre \citep{Bower_GC,Wucknitz_GC}, instead of a screen in proximity to the source.

FRB\,20121102A has shown something similar. It has an extremely high RM of $\sim$ 10$^{5}$ rad m$^{-2}$ \citep{mic2018}. But the only CHIME/FRB detection from the source had an upper limit on scattering of $\sim$ 2 ms at 600 MHz \citep{josephy2019chime}, which is nominal compared to our sample and in agreement with what has been measured from the source at higher frequencies \citep{Hes2019} using its scintillation bandwidth. This suggests that an extreme magneto-ionic environment does not also cause highly scattered bursts, unless the scattering screen happens to coincide with it. 

\subsubsection{Linear Polarization Fraction vs Scattering}

\cite{feng_2022_depolarization}, in their analysis of active repeaters across a wide frequency range, argued that the linear polarization fraction decreases with frequency because of enhanced scattering at low frequencies, where multi-path propagation also depolarizes. Assuming all FRBs are intrinsically 100\% linearly polarized, then sources with long scattering times should exhibit a decrease in their linear polarization fraction (L/I). We do not find any statistically significant correlation for our sample (see Fig.~\ref{fig:corr_polfrac_scat}). This suggests that depolarization does not seems to be the rule for one-off FRBs. Indeed, repeating FRBs may occupy magneto-ionic environments where the Faraday active medium simultaneously produces extreme scattering, thereby depolarization (e.g., FRB\,20190520B \citealp{ocker_20190520B}), however, this scenario appears far less common for one-off FRBs and has interesting implications for the nature and evolution FRB circumburst environments. Additionally CHIME/FRB is heavily biased against bursts having scattering times $>$ 10 ms \citep{marcus_injection} and it is these that may well have lower L/I. 

Our results further support the L/I measurements presented by \cite{ppm+24} for 128 CHIME FRBs at 600 MHz, which statistically agree with the results published by \cite{DSA_pol_2024} for 25 Deep Synoptic Array-110 (DSA-110) FRBs at 1.4 GHz. Additionally, \cite{USG+2024} do not observe any spectral depolarization in their analysis of the ASKAP FRB sample.

\subsubsection{Other Possible Correlations}

We do not observe any correlation between DM and fluence in our dataset, unlike the correlation observed in the ASKAP and Parkes FRB samples \citep{2018Natur.562..386S}. Given that we probe a similar DM and fluence range, this suggests a broad luminosity distribution among the FRB population, allowing for a range of fluences from sources at similar distances. This variability is shown by the repeater FRB\,20201124A, which exhibits bursts spanning five orders of magnitude in fluence \citep{xnc+22, koh+24}. 

Additionally, we do not observe any correlation between DM and burst width. The relationship between DM and redshift  \citep{macquart2020census} implies a $1+z$ factor increment in the measured width for sources at higher DMs. The absence of such a relationship suggests that we may not be probing a large redshift range in this sample, especially considering the unknown host galaxy DM contribution, which can be substantial, as seen in FRB\,20190520B, with contributions reaching several hundred DM units \citep{ocker_20190520B}.
Furthermore, we have observed a wide range of intrinsic widths even from the same source, ranging from microseconds to tens of milliseconds. Notable examples include FRB\,20180916B \citep{nimmo2021_micro,sbm+23} and FRB\,20121102A \citep{Gaj18,snelders_2023}. It is also challenging to disentangle all propagation effects particularly in the case of multiple screen or a thick screen to accurately estimate the true intrinsic width of the burst.

\subsection{Population of sub-ms FRBs}

In Figure \ref{fig:corr_boxcar_flux}, bursts with durations $\leq$ 1 ms exhibit higher peak flux densities. CHIME/FRB conducts searches with a sampling rate of 1 ms or greater \citep{chime18_overview}. This suggests that bursts with sub-ms duration may be indistinguishable from noise in our searches below a certain brightness threshold. A similar phenomenon was observed in FRB\,20121102A by \cite{snelders_2023}, where upon re-analysis of only 30 mins of the archival GBT data (4 -- 8 GHz) 8 pulses were found with burst envelope $\leq$ 15 $\upmu$s, which were missed by searches done at coarser temporal sampling and S/N cutoff \citep{Gaj18,zhang2018fast}. Although scattering effects and intra-channel smearing at CHIME frequencies limit such detections, our real-time system may be missing a population of such ultra-FRBs by capturing only the brighter end of the distribution. A more careful analysis, accounting for search algorithms and system biases, is needed to quantify the implications this might have on the overall FRB population.

Furthermore, weaker burst components are maybe overlooked at ms durations due to low S/N and lack of coherent dedispersion. For instance in our sample, FRB\,20190425A displays much weaker pre-cursor components followed by a bright burst, which might have triggered the baseband system. Similarly FRB\,20190624B has weaker post cursor components. We have seen bursts from repeaters to be clustered in their arrival times \citep{Gaj18, FAST_121102A, FAST_R67}, this might imply that lowering the search S/N threshold near a bright component could reveal additional sub-bursts.

\section{Conclusion}\label{sec:conc}

We studied the morphological properties of 137 FRBs from the first CHIME/FRB baseband catalog, enabling exploration of temporal and spectral properties down to microsecond timescales. Our analysis extends to three orders of magnitude finer time resolution, with the brightest burst studied at 2.56 $\upmu$s, compared to $\sim$ 0.983 ms in the first CHIME/FRB intensity catalog \citep{chimefrbcatalog1}. This increased time resolution available in baseband data allowed us to mitigate contamination from underlying substructures, better estimate DM, and accurately measure scattering timescales ($\tau$). We found 85\% of the bursts in our sample span their entire available bandwidth.  We measure $\tau$ values ranging from 30 $\upmu$s to 13 ms. 

We compared our results with values published in the first CHIME/FRB catalog. We identified additional components in one-third of the bursts in our sample. Nearly 20\% of the bursts exhibited intrinsic widths ($\sigma$) less than 50 $\upmu$s, with the narrowest being 10 $\upmu$s. The observation of such microsecond features lends support to models invoking magnetospheric emission but a small patch of shocked plasma at large distances may still be a possible emission mechanism. We also find that our bandwidth measurements are more robust than those published in the intensity catalog as the former are not influenced by formed beam effects.

We studied the distribution of arrival times between sub-bursts and found no source with components separated by less than 41 $\upmu$s with nearly 50 bursts in the sample exhibiting multiple components.

We have searched for correlations between scattering time and other measured parameters, while considering our observational biases against FRBs with long scattering times. We find no correlation between DM and scattering, suggesting distinct primary contributors to these effects.  One possibility is that the majority of the DM contribution comes from the IGM, whereas scattering can originate from a screen local to the source or in the intervening CGM. 

Similarly, we find no correlation between RM and scattering, indicating that the Faraday active medium may not coincide with the scattering screen. Our lack of correlation between scattering and linear polarization fraction suggests no significant depolarization due to multipath propagation in our FRB sample, unlike what has been observed for active repeaters. Though constraints on both of these correlation are limited by lack of high RM FRBs in our sample. 

We also do not observe any correlation between DM and fluence, suggesting that FRBs exhibit a wide luminosity distribution. Similarly no correlation was found between DM and width, suggesting FRB sources emit bursts across a range of intrinsic widths.  

This study reveals that apparent one-off FRBs exhibit immense morphological diversity. We identify narrowband bursts with wider envelopes, bursts with bidirectional drifting and bursts with sub-components exhibiting a residual drift even after dedsipersion.  All this is reminiscent of features we have seen in repeating sources. This could mean either they both come from similar sources or there are two different classes but with overlapping property distributions. 

We also identify that bursts with narrower durations tend to be brighter in our sample. This might imply that the CHIME/FRB realtime system is missing a population of sub-ms FRBs and only catching the brighter end of the distribution. 

This study underscores the importance of high time resolution, as studying FRBs at microsecond timescales is essential for accurate measurements of their properties and for constraining implications related to their emission mechanisms and progenitor scenarios. 
\newline
 
\begin{acknowledgments}

We acknowledge that CHIME is located on the traditional, ancestral, and unceded territory of the Syilx (Okanagan) people.

We thank Amanda Cook for comments that have improved the quality of this manuscript.

We thank the Dominion Radio Astrophysical Observatory, operated by the National
Research Council Canada, for gracious hospitality and expertise. 
CHIME is funded by a grant from the Canada Foundation for Innovation (CFI) 2012 Leading Edge Fund (Project 31170) and by contributions from the provinces of British Columbia, Qu\'ebec and Ontario. The CHIME/FRB Project is funded by a grant from the CFI 2015 Innovation Fund (Project 33213) and by contributions from the provinces of British Columbia and Qu\'ebec, and by the Dunlap Institute for Astronomy and Astrophysics at the University of Toronto. Additional support was provided by the Canadian Institute for Advanced Research (CIFAR), McGill University and the Trottier Space Institute via the Trottier Family Foundation, and the University of British Columbia. The Dunlap Institute is funded through an endowment established by the David Dunlap family and the University of Toronto. Research at Perimeter Institute is supported by the Government of Canada through Industry Canada and by the Province of Ontario through the Ministry of Research \& Innovation. The National Radio Astronomy Observatory is a facility of the National Science Foundation (NSF) operated under cooperative agreement by Associated Universities, Inc. FRB Reseach at UBC is supported by an NSERC Discovery Grant and by the Canadian Institute for Advanced Research.  The CHIME/FRB basband system is funded in part by a CFI John R. Evans Leaders Fund award to IHS.

\allacks
\end{acknowledgments}

\begin{acknowledgments}
K.R.S. acknowledges support from Fonds de Recherche du Quebec – Nature et Technologies (FRQNT) Doctoral Research Award. A.P.C is a Vanier Canada Graduate Scholar. V.M.K. holds the Lorne Trottier Chair in Astrophysics \& Cosmology, a Distinguished James McGill Professorship, and receives support from an NSERC Discovery grant (RGPIN 228738-13), from an R. Howard Webster Foundation Fellowship from CIFAR. K.N. is an MIT Kavli fellow. Z.P. is supported by an NWO Veni fellowship (VI.Veni.222.295). K.S. is supported by the NSF Graduate Research Fellowship Program. M.B is a McWilliams fellow and an International Astronomical Union Gruber fellow. M.B. also receives support from the McWilliams seed grant. M.D. is supported by a CRC Chair, NSERC Discovery Grant, CIFAR, and by the FRQNT Centre de Recherche en Astrophysique du Qu\'ebec (CRAQ). G.M.E. acknowledges support from a Canadian Statistical Sciences Institute (CANSSI) Collaborative Research Team Grant. CANSSI is supported under the Discovery Institutes Support program of NSERC. B.M.G. acknowledges the support of the Natural Sciences and Engineering Research Council of Canada (NSERC) through grant RGPIN-2022-03163. C. L. is supported by NASA through the NASA Hubble Fellowship grant HST-HF2-51536.001-A awarded by the Space Telescope Science Institute, which is operated by the Association of Universities for Research in Astronomy, Inc., under NASA contract NAS5-26555. K.W.M. holds the Adam J. Burgasser Chair in Astrophysics. A.P. is funded by the NSERC Canada Graduate Scholarships -- Doctoral program. A.B.P. is a Banting Fellow, a McGill Space Institute~(MSI) Fellow, and a FRQNT postdoctoral fellow. M.W.S. acknowledges support from the Trottier Space Institute Fellowship program. 
\allacks
\end{acknowledgments}

\vspace{5mm}
\facilities{CHIME}

\software{astropy \citep{astropy1,astropy2,astropy3}, bitshuffle \citep{mas17_bitshuffle}, DM Phase \citep{dm_phase}, emcee \citep{emcee}, fitburst \citep{fonseca_fitburst}, hdf5 \citep{hdf5}, matplotlib \citep{matplotlib}, NADA \citep{helsel2005nondetects, Rsoftware, NADApackage}, numpy \citep{numpy}, pandas \citep{pandas}, scipy \citep{scipy_Virtanen_2020}, statsmodel \citep{seabold2010statsmodels}.}

\appendix

\setcounter{table}{0}
\renewcommand{\thetable}{A\arabic{table}}

\setcounter{figure}{0}
\renewcommand{\thefigure}{A\arabic{figure}}

\section{Data Table and Burst Waterfalls}
\begin{figure*}
\centering
\gridline{\fig{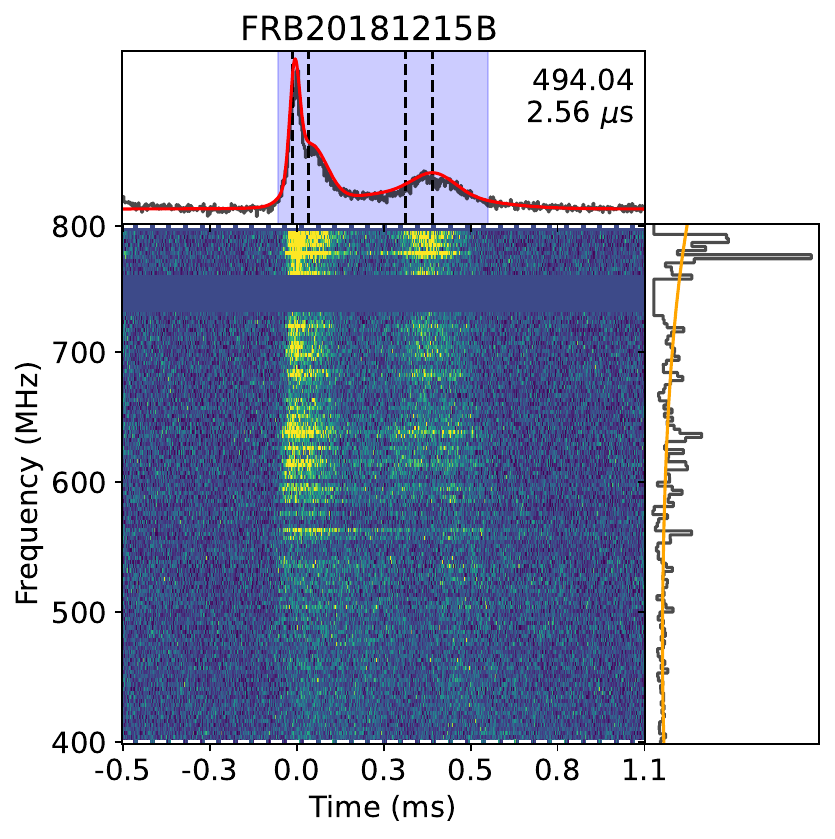}{0.18\textwidth}{}
          \fig{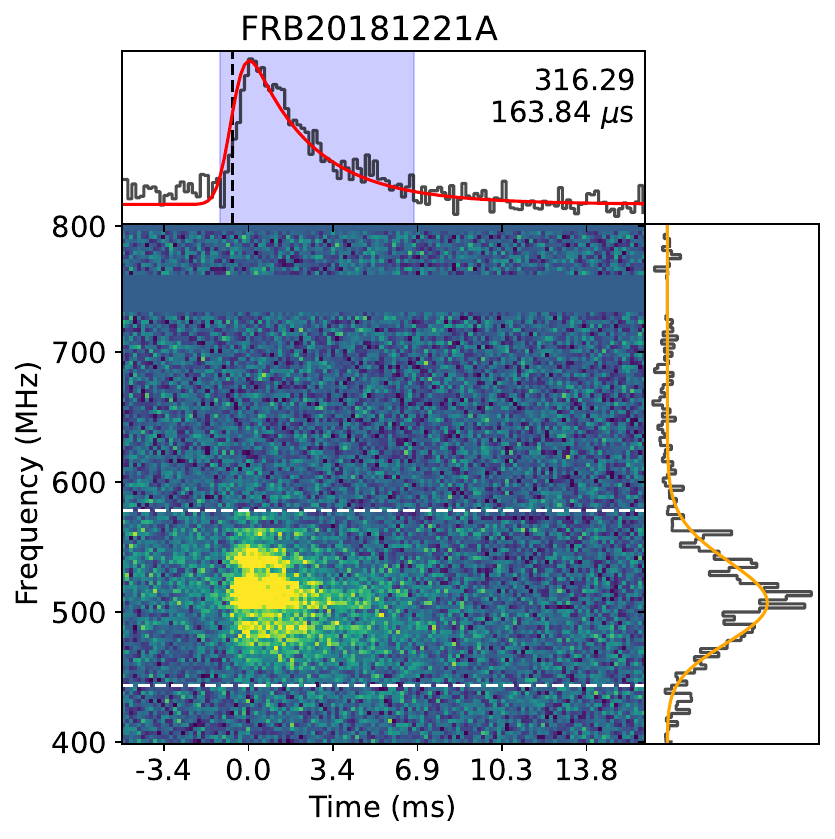}{0.18\textwidth}{}
          \fig{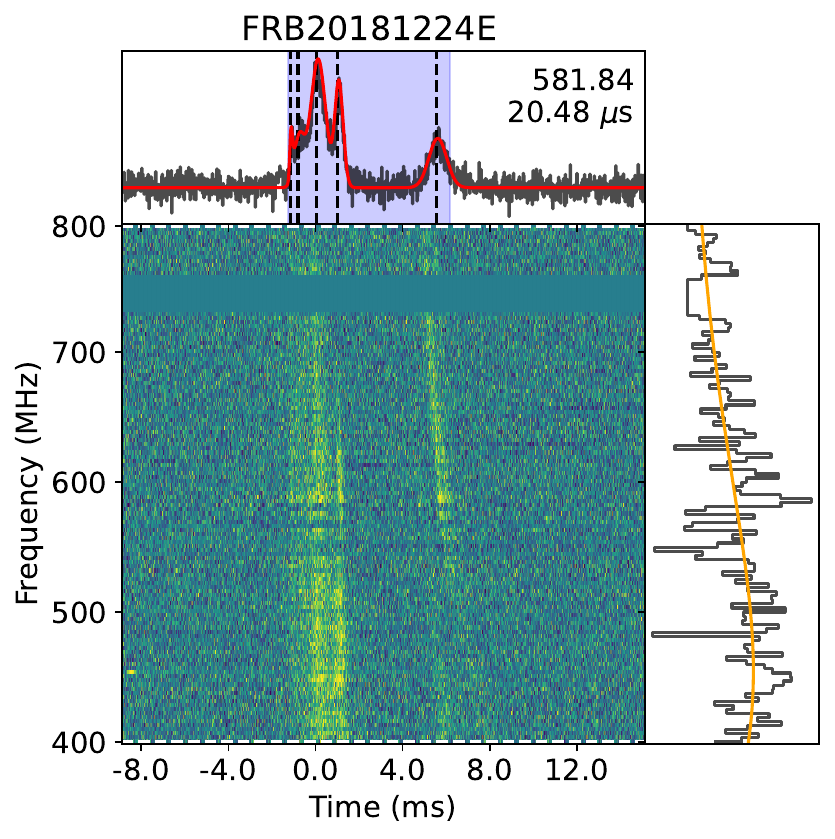}{0.18\textwidth}{}
          \fig{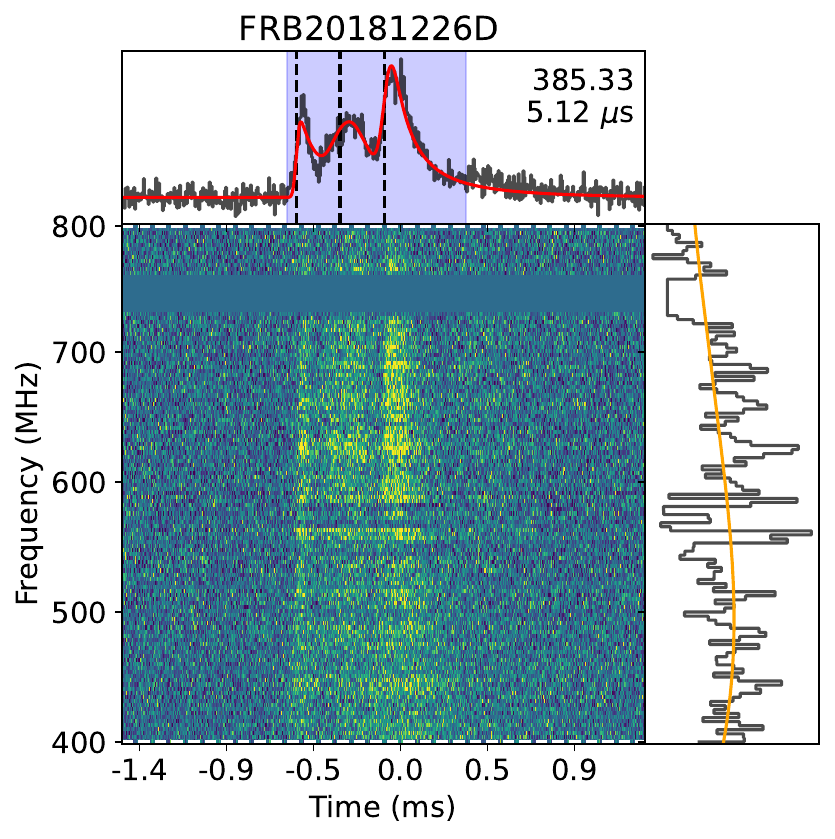}{0.18\textwidth}{}
          \fig{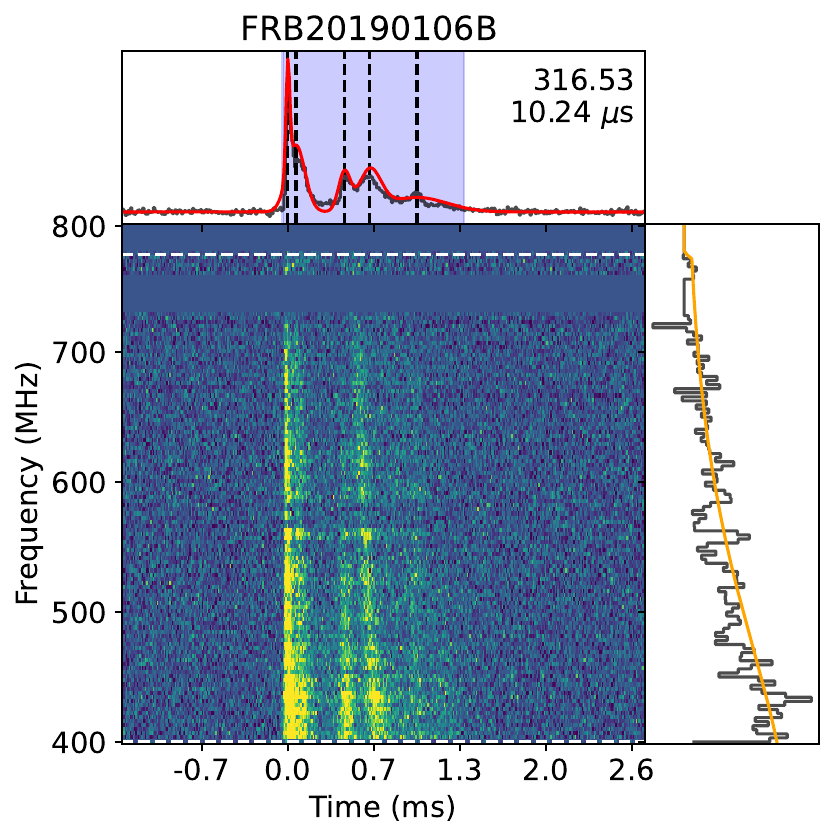}{0.18\textwidth}{}
          }
\gridline{\fig{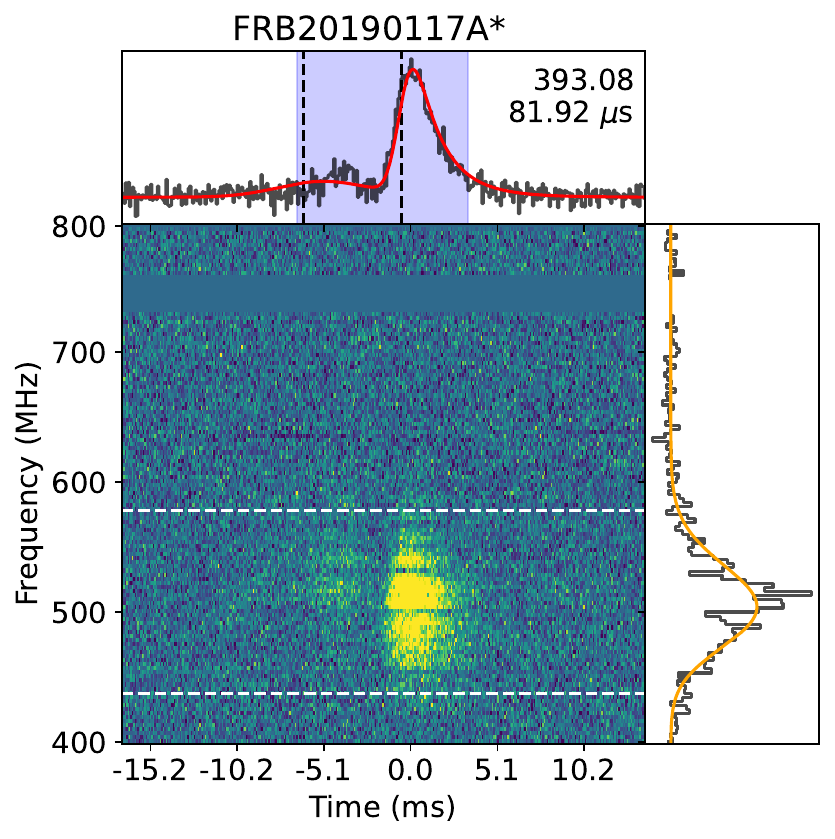}{0.18\textwidth}{}
          \fig{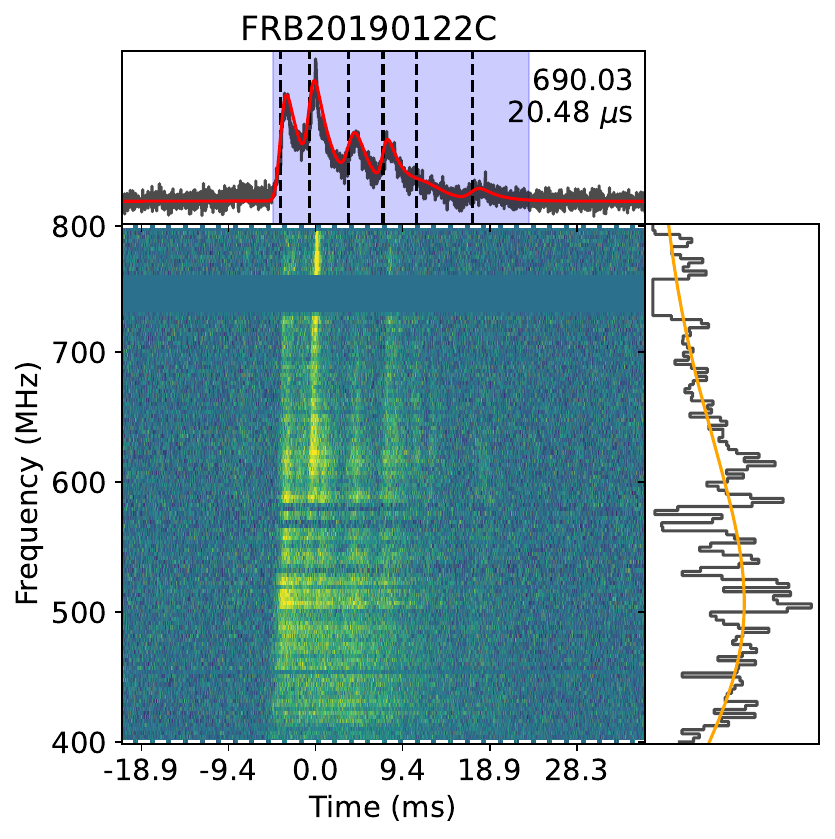}{0.18\textwidth}{}
          \fig{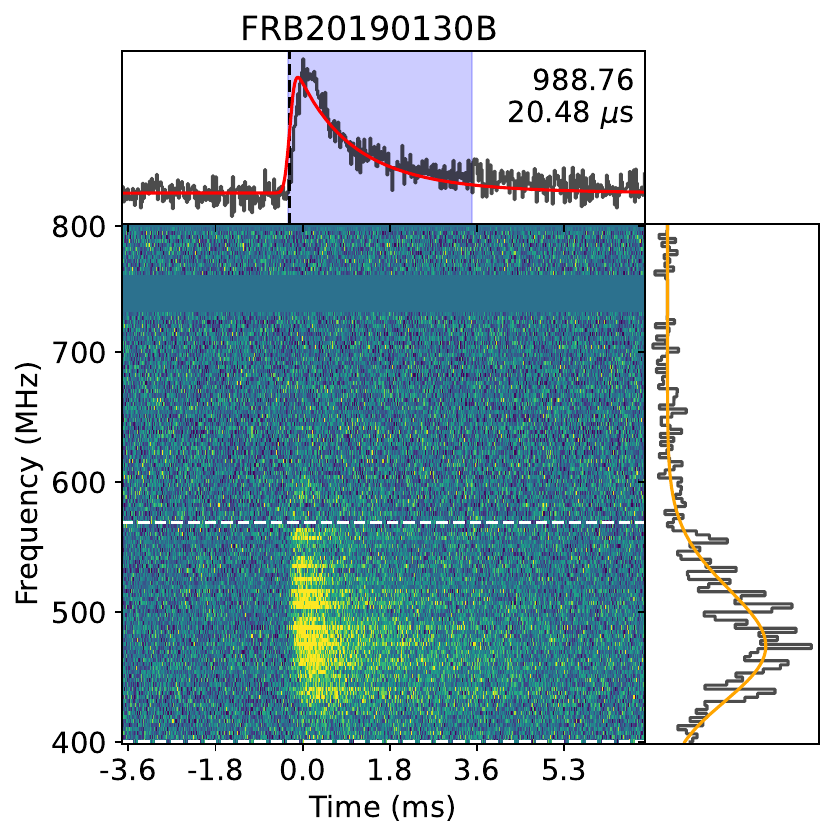}{0.18\textwidth}{}
          \fig{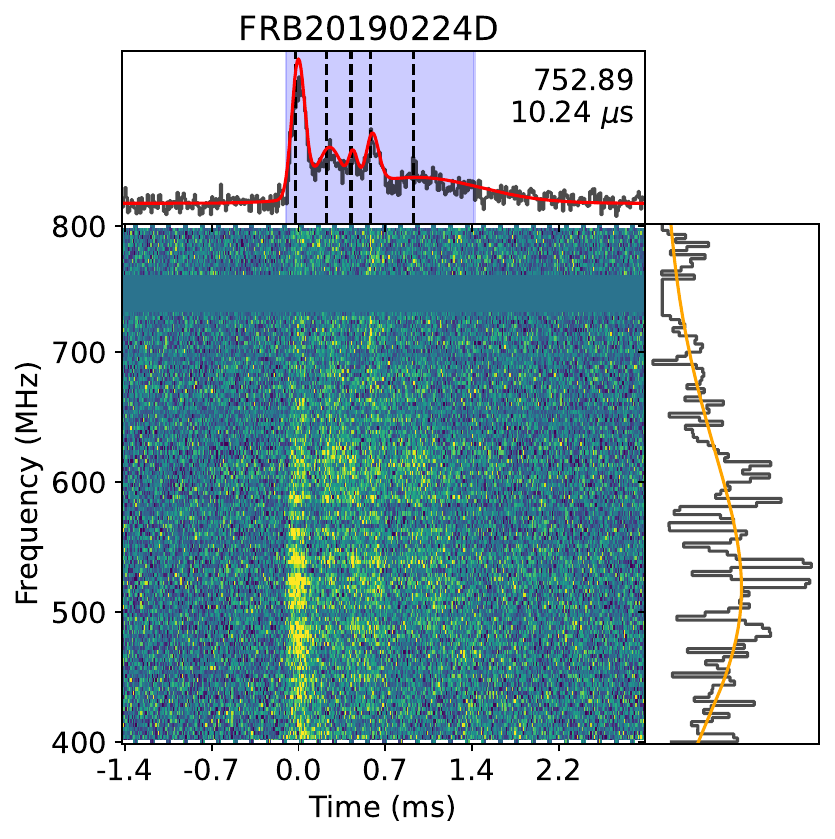}{0.18\textwidth}{}
          \fig{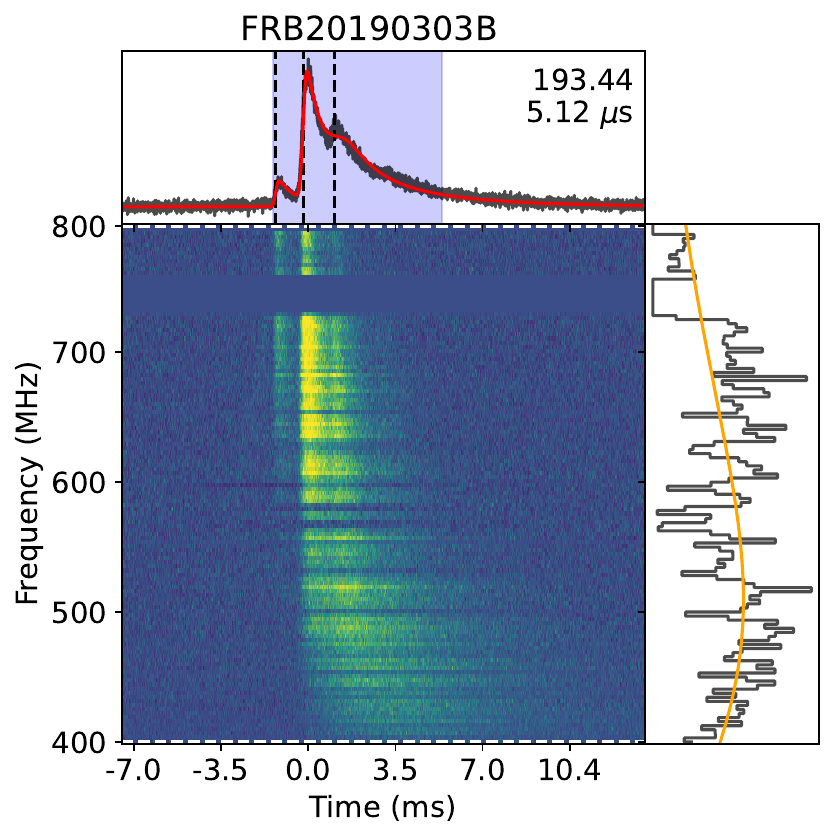}{0.18\textwidth}{}
          }
\gridline{\fig{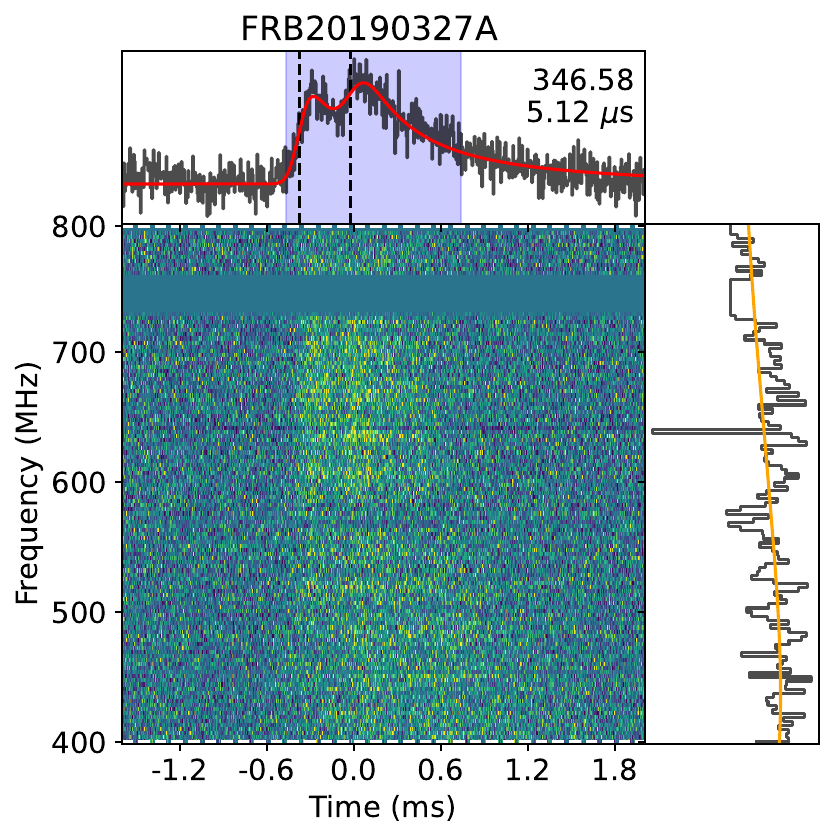}{0.18\textwidth}{}
          \fig{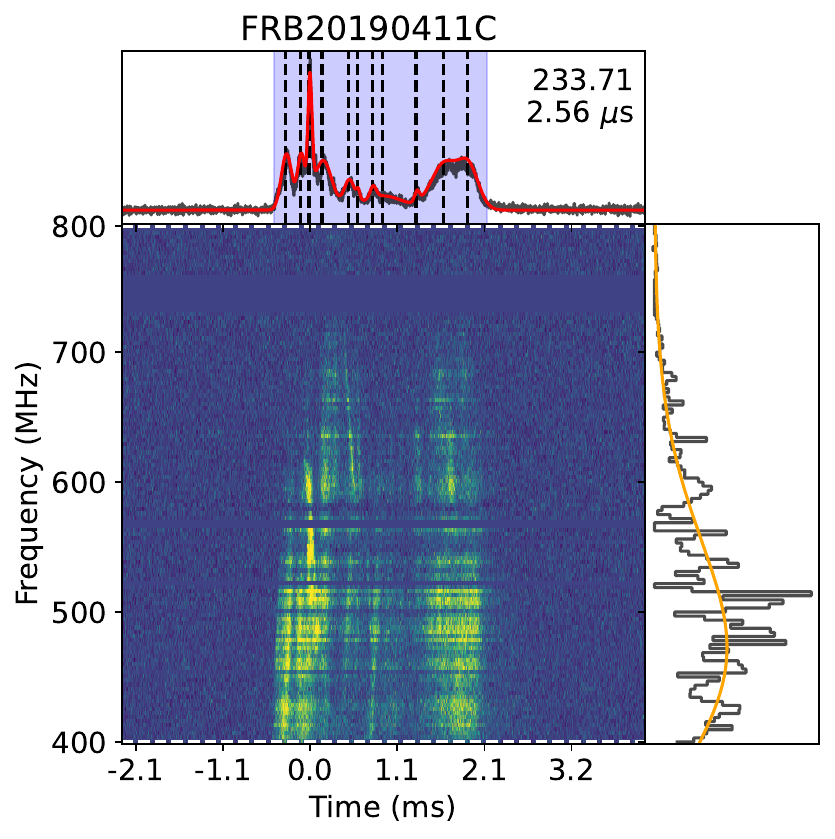}{0.18\textwidth}{}
          \fig{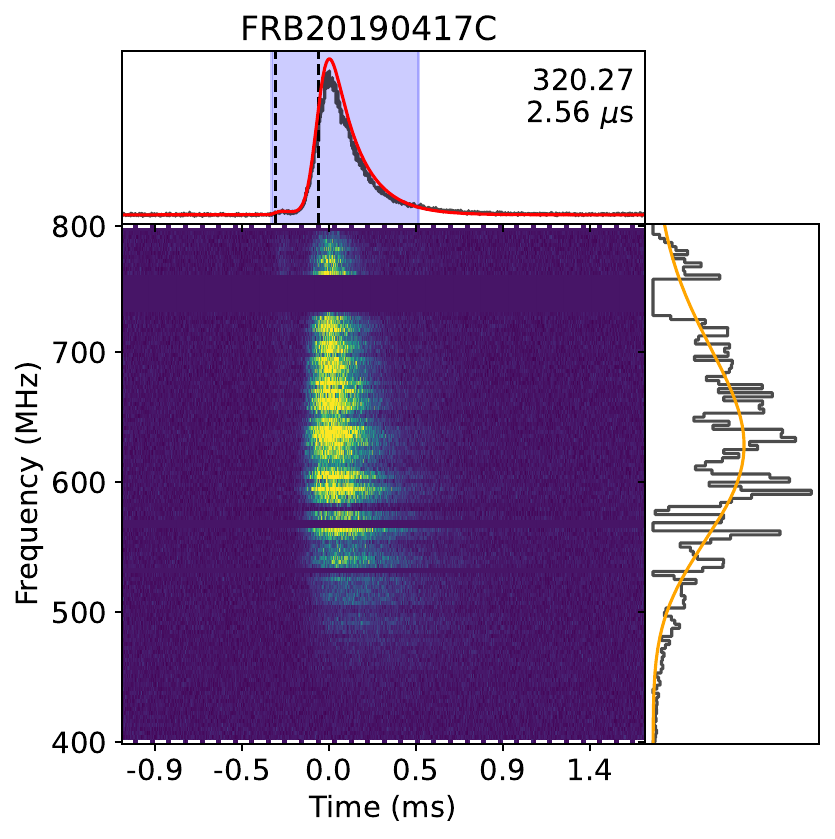}{0.18\textwidth}{}
          \fig{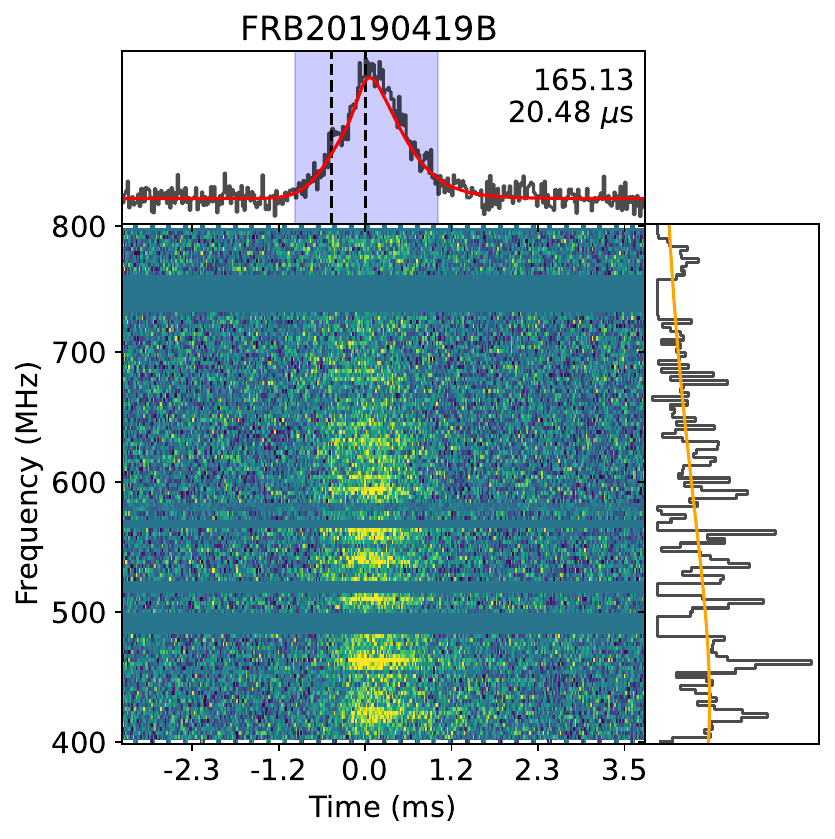}{0.18\textwidth}{}
          \fig{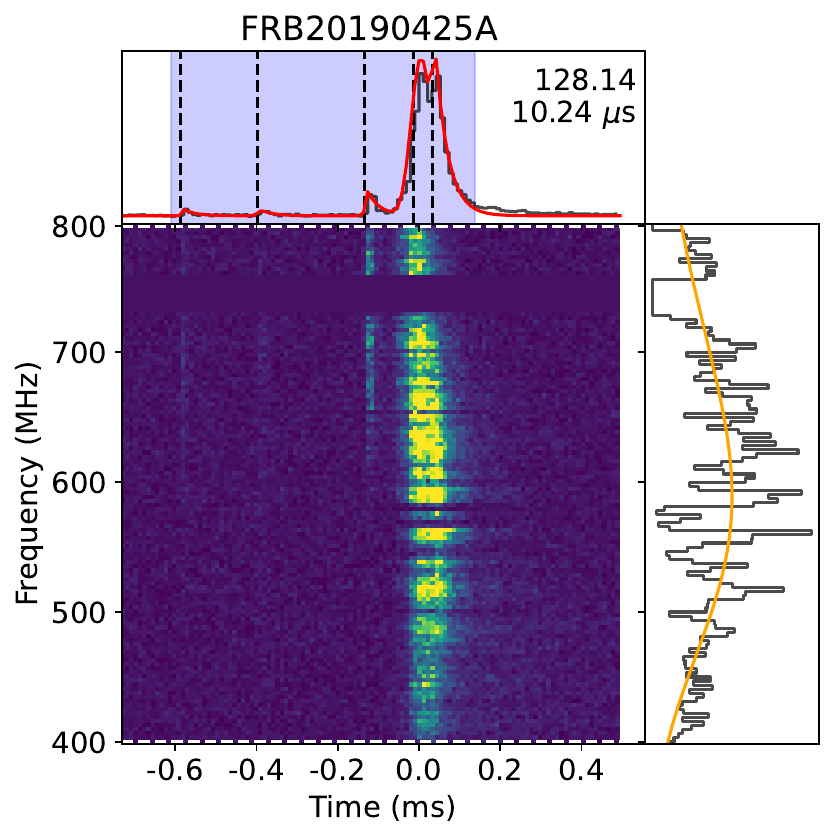}{0.18\textwidth}{}
          }
\gridline{\fig{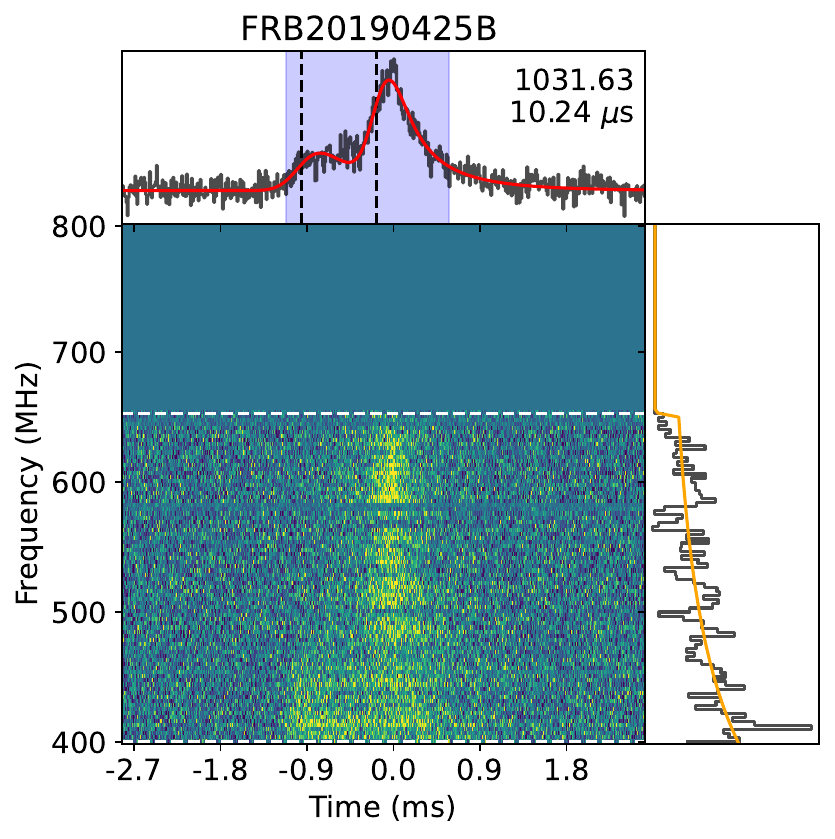}{0.18\textwidth}{}
          \fig{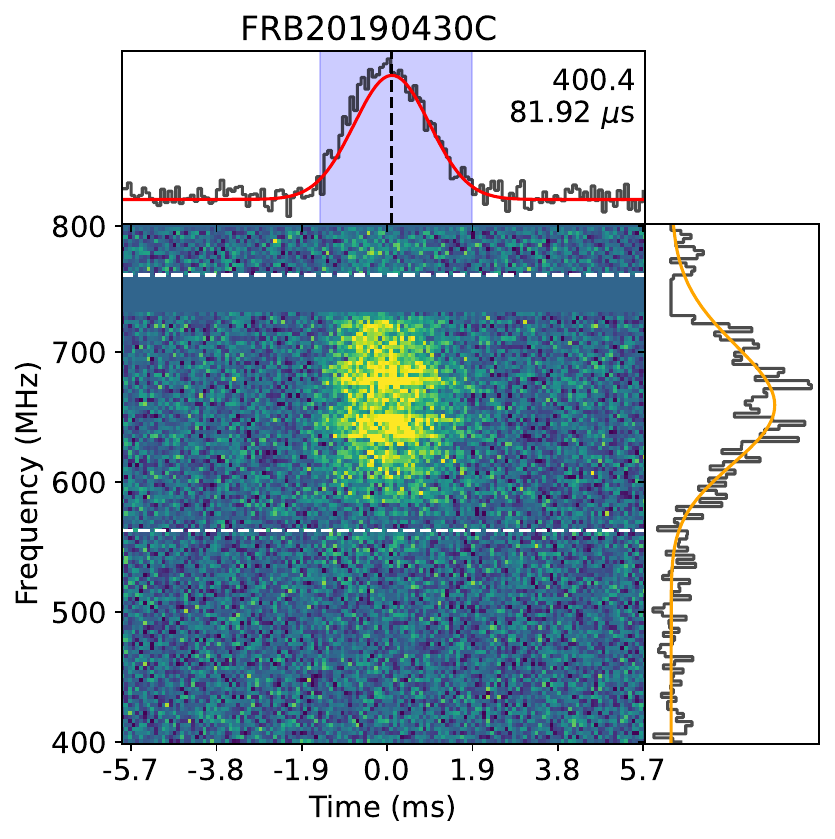}{0.18\textwidth}{}
          \fig{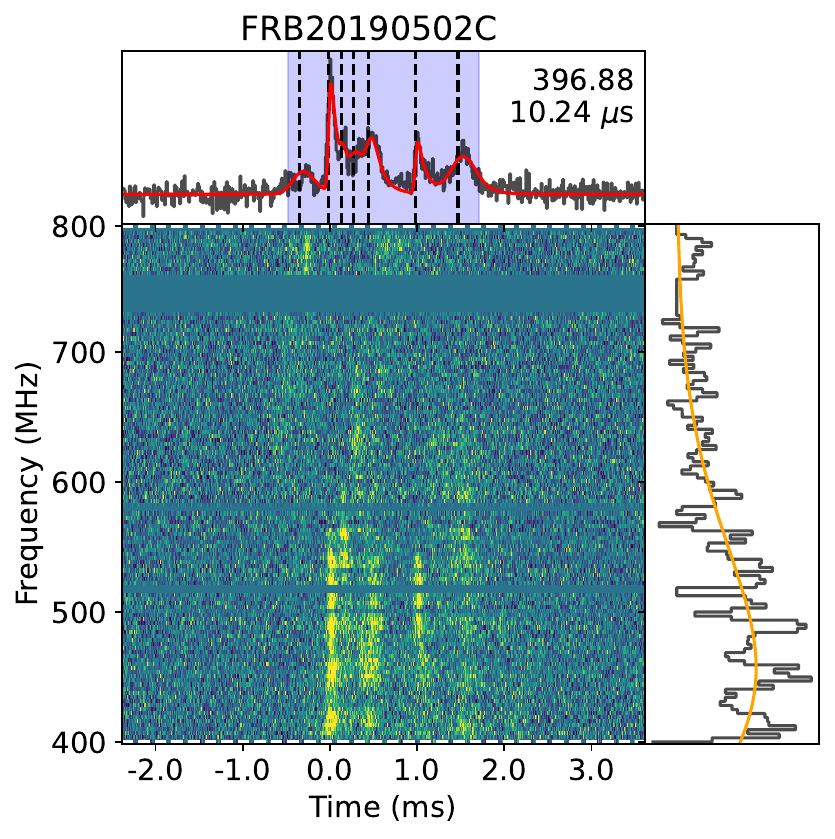}{0.18\textwidth}{}
          \fig{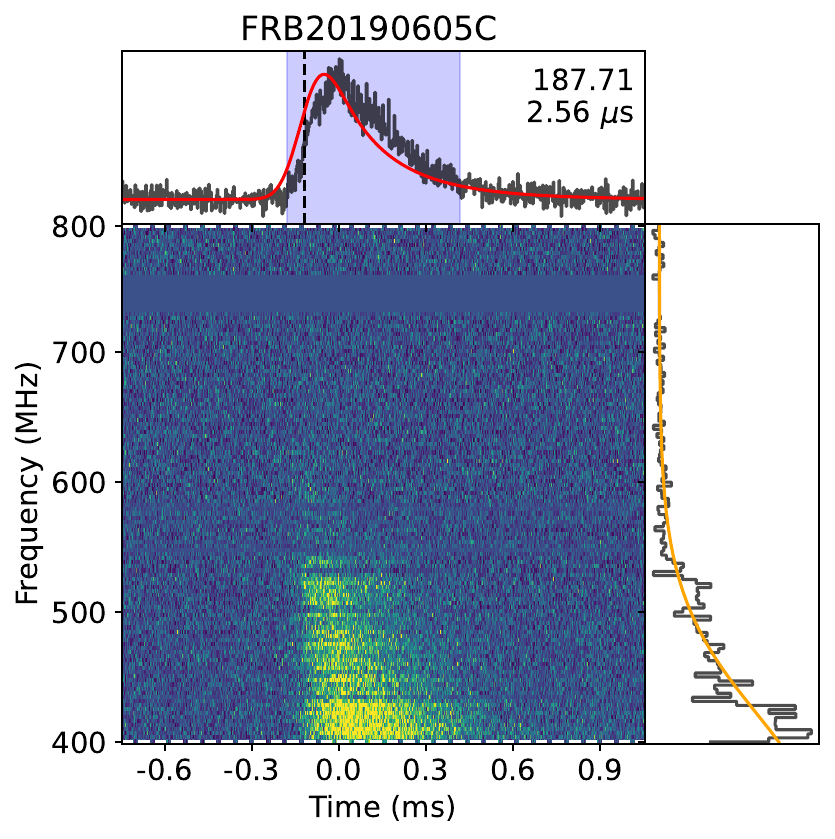}{0.18\textwidth}{}
          \fig{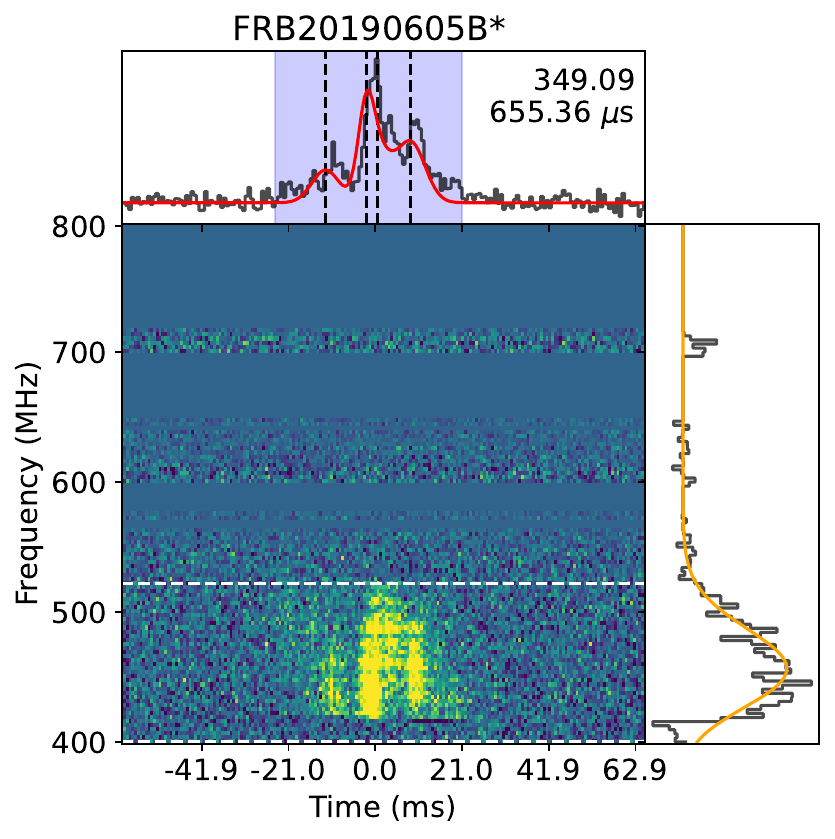}{0.18\textwidth}{}
          }
\gridline{\fig{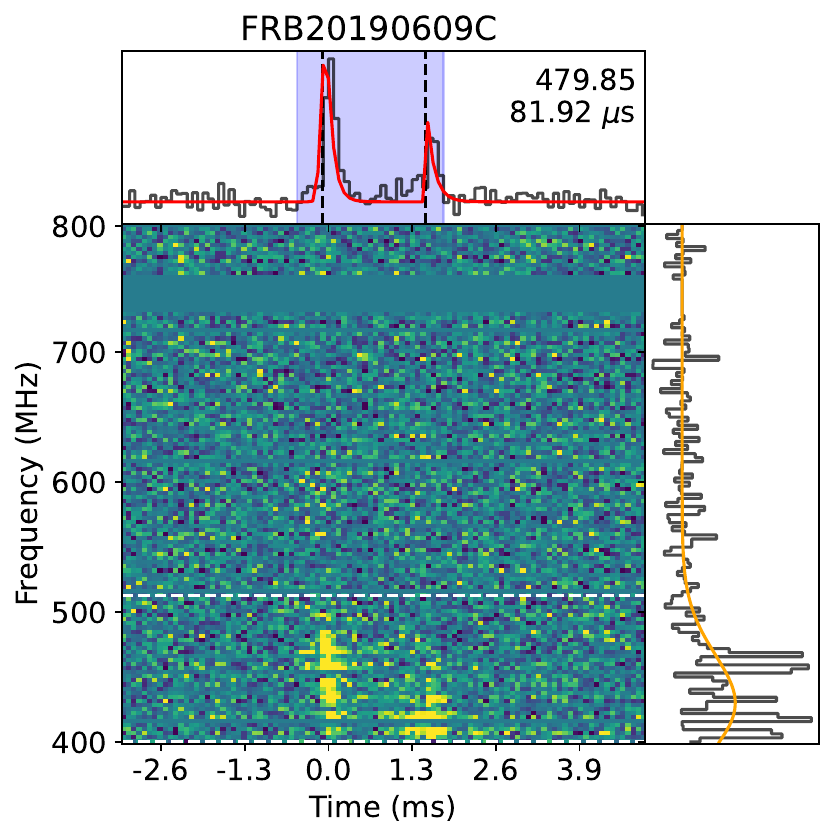}{0.18\textwidth}{}
          \fig{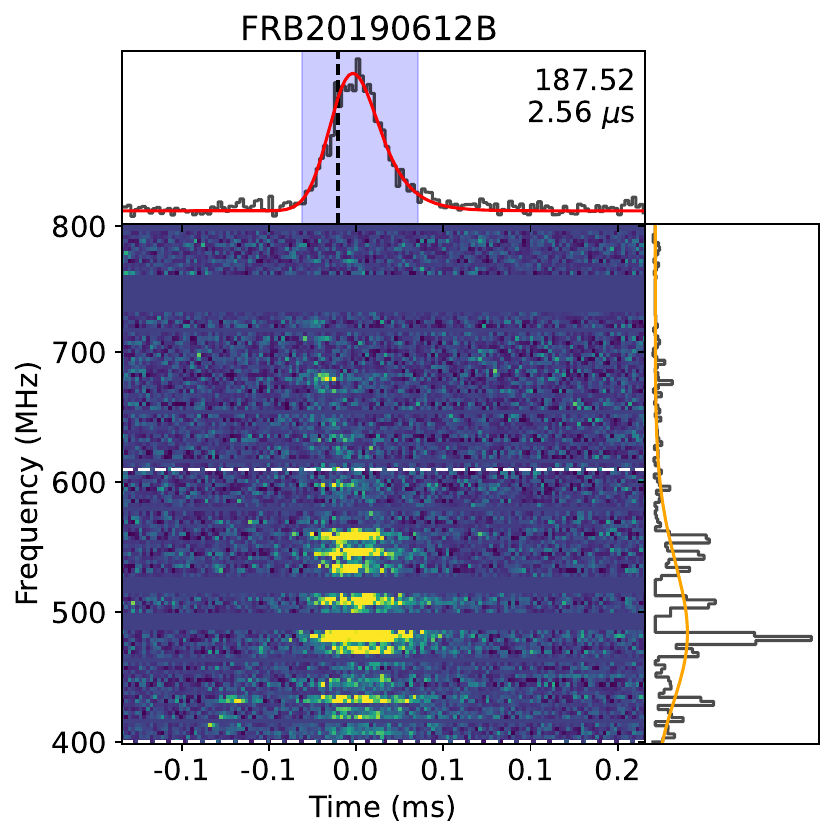}{0.18\textwidth}{}
          \fig{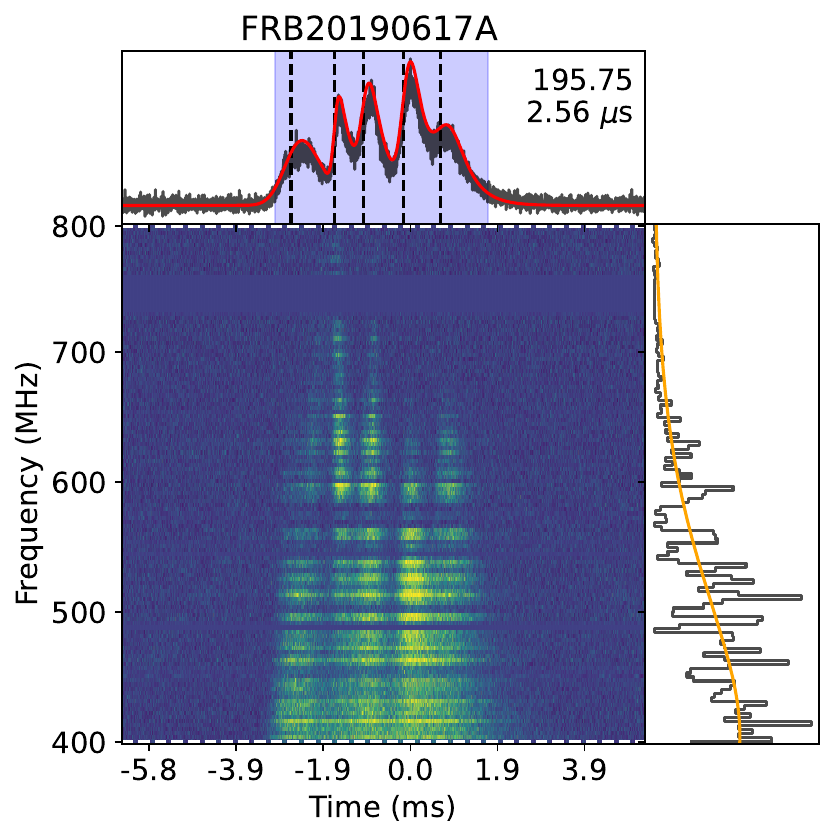}{0.18\textwidth}{}
          \fig{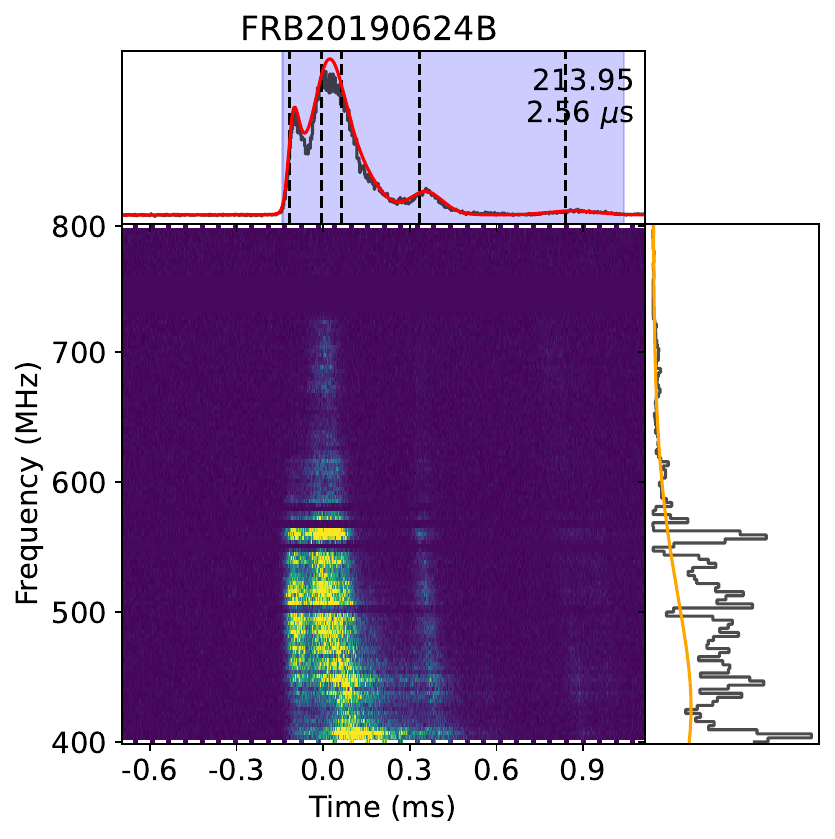}{0.18\textwidth}{}
          \fig{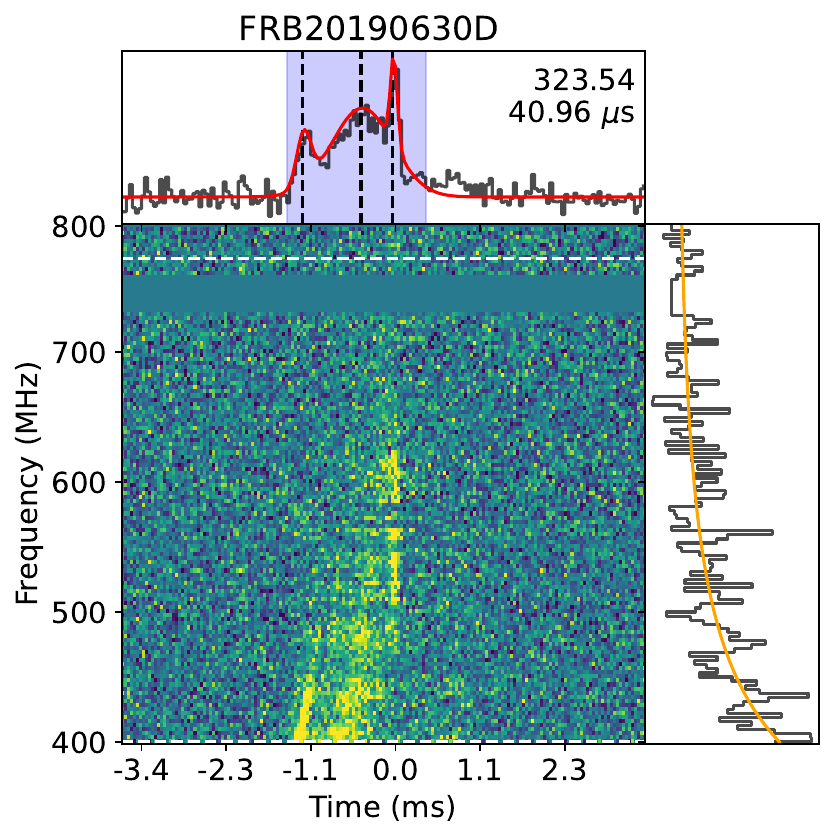}{0.18\textwidth}{}
          }
    \caption{A representative sub-sample of our bursts. Each subfigure displays the burst's time series in the top panel, with the best-fit shown by a red line. The shaded blue region shows the burst duration. Beneath this, dynamic spectra are depicted, alongside the frequency distribution of power, with the best-fit spectrum indicated by an orange line. The title of each sub-figure corresponds to its TNS name (* are repeaters). In the top right corner of the time series panel, the DM to which it has been dedispersed is provided, along with its time resolution. The number of channels utilized is 128. White dashed lines delineate the extent of bandwidth. Waterfall plots for all the baseband bursts have been presented in \cite{aaa+23_basecat}.}\label{fig:burst_wfalls}
\end{figure*}

\begin{ThreePartTable}
\renewcommand\TPTminimum{\textwidth}
\begin{TableNotes}
\item\tablenotetext{a}{Optimized \texttt{fitburst} DM. A single DM is determined for all sub-components. For bursts for which the best fit is given by the profile, the structure maximizing DM is reported instead. }

\tablenotetext{b}{Temporal downsampling factor used to determine the burst properties. The equivalent time resolution is 2.56 $\upmu$s $\times$ downsampling factor (see Section \ref{sec:Morphology_pipe}).}

\tablenotetext{c}{Number of components identified in the burst at the given time resolution.}

\tablenotetext{d}{Width of the burst as measured by \texttt{fitburst}. If multiple components are present, we provide the minimum width from all the components along with the maximum width from all components. Intrinsic widths for all the components will be available in the live table. Multiply by 2.35 to get the FWHM value.}

\tablenotetext{e}{Scattering time as determined using \texttt{fitburst}.  Upper limits (shown with $<$) are presented for bursts for which we were unable to measure scattering and the limit is equal to the measured minimum width (Model 2 fits). Scattering measurements which were determined using the profile fits are shown with a $\sim$ symbol (Model 1 fits). See Section \ref{sec:fit} for discussion of the different scattering fitting methods.}

\tablenotetext{f}{Bandwidth of the burst as determined using a running power-law. Bursts for which the entire 400-MHz CHIME/FRB band was not stored are shown with $>$. Bandwidths have not been corrected for the primary beam responses of CHIME/FRB. }

\tablenotetext{g}{Event of the repeating source FRB 20180916B \citep{abb+19c}.}

\tablenotetext{h}{Repeating source \citep{abb+19c}.}

\tablenotetext{i}{Repeating source \citep{fab+20}.}

\tablenotetext{j}{Event of the repeating source FRB 20180908B \citep{fab+20}.}

\tablenotetext{k}{Event of the repeating source FRB 20190222A \citep{abb+19c}.}

\tablenotetext{l}{Event of the repeating source FRB 20190604A \citep{fab+20}.}

\tablenotetext{m}{Event of the repeating source FRB 20180814A \citep{abb+19b}.}

\tablenotetext{n}{Total duration of the burst in ms determined using the time between 10\% fluence from the rise of the first and 90\% fluence at tdecay of the last component.}

\end{TableNotes}
\begin{longtable}[c]{l c c c c c c c}
\caption{Measured morphological properties for 137 bursts from the first CHIME/FRB baseband catalog. A live table of all the Baseband catalog measurements used in this paper can be found here (insert link on acceptance).} \label{tab:burst}
\endfirsthead
\caption{\textit{continued}} \endhead
\hline
\\[-2ex]
TNS Name & DM\tablenotemark{a} & Downsample\tablenotemark{b} & Num. Comp.\tablenotemark{c} & Width\tablenotemark{d} & Scattering\tablenotemark{e} & Bandwidth\tablenotemark{f} & Duration\tablenotemark{n} \\
 & (pc cm$^{-3}$) &  &  & (ms) & (ms) & (MHz) & (ms) \\
\\[-2ex]
\hline
\\[-2ex]
FRB 20181209A & 328.59(1) & 1 & 1 & 0.048(2) & 0.114(4) & 400.0 & 0.58\\ 
FRB 20181213A & 678.69(1) & 8 & 3 & 0.048(6), 0.078(4) & 0.17(5) & 400.0 & 2.7\\ 
FRB 20181214C & 632.827(7) & 32 & 2 & 0.067(6), 0.38(3) & 1.6(2) & 400.0 & 6.9\\ 
FRB 20181215B & 494.044(6) & 1 & 4 & 0.0138(2), 0.196(6) & $\sim$ 0.01 & 400.0 & 0.63\\ 
FRB 20181219C & 647.76(4) & 128 & 1 & 0.61(3) & 1.1(1) & 400.0 & 10.0\\ 
FRB 20181220A & 209.517(8) & 8 & 1 & 0.074(4) & 0.21(1) & 400.0 & 1.3\\ 
FRB 20181221A & 316.29(5) & 64 & 1 & 0.47(2) & 1.38(6) & 133.3 & 7.9\\ 
FRB 20181221B & 1394.84(1) & 2 & 1 & 0.220(4) & 0.275(9) & $>$230.7 & 1.6\\ 
FRB 20181222E & 327.949(5) & 32 & 3 & 0.29(1), 0.63(3) & 0.84(4) & 400.0 & 41.0\\ 
FRB 20181223C & 112.49(1) & 64 & 1 & 0.58(2) & $<$ 0.58 & 400.0 & 3.1\\ 
FRB 20181224E & 581.84(1) & 8 & 5 & 0.05(1), 0.40(2) & $\sim$ 0.07 & 400.0 & 7.4\\ 
FRB 20181225A\tablenotemark{g} & 348.80(4) & 64 & 2 & 0.23(5), 0.41(6) & 0.90(8) & 107.9 & 10.0\\ 
FRB 20181226A\tablenotemark{g} & 348.80(1) & 64 & 2 & 0.21(1), 0.84(5) & 0.53(2) & 177.8 & 6.9\\ 
FRB 20181226D & 385.335(5) & 2 & 4 & 0.014(2), 0.054(1) & 0.126(3) & 400.0 & 0.96\\ 
FRB 20181226E & 308.78(1) & 32 & 1 & 0.29(1) & 0.29(3) & $>$154.8 & 3.1\\ 
FRB 20181228B & 568.508(8) & 256 & 1 & 0.35(4) & 1.1(2) & 76.2 & 9.2\\ 
FRB 20181229A & 955.43(2) & 128 & 1 & 0.18(3) & 2.3(1) & 400.0 & 12.0\\ 
FRB 20181231A & 1376.8(3) & 256 & 1 & 0.5(1) & 1.6(7) & $>$199.8 & 8.5\\ 
FRB 20181231B & 197.358(9) & 2 & 1 & 0.039(5) & 1.34(2) & 400.0 & 2.1\\ 
FRB 20181231C & 556.11(3) & 128 & 1 & 0.25(3) & 0.23(8) & $>$45.4 & 3.9\\ 
FRB 20190102A & 699.0(4) & 32 & 1 & 1.16(4) & 1.03(3) & $>$73.9 & 11.0\\ 
FRB 20190102B & 367.08(4) & 32 & 1 & 0.32(1) & 0.55(3) & 400.0 & 3.5\\ 
FRB 20190103C & 1349.0(1) & 256 & 1 & 1.41(8) & 1.6(2) & $>$140.0 & 15.0\\ 
FRB 20190106B & 316.531(2) & 4 & 5 & 0.0146(3), 0.23(2) & $<$ 0.01 & $>$376.5 & 1.4\\ 
FRB 20190110A & 472.786(3) & 2 & 2 & 0.019(3), 0.026(2) & 0.18(2) & $>$290.5 & 1.4\\ 
FRB 20190110C & 222.05(2) & 256 & 1 & 0.63(3) & $<$ 0.63 & 76.2 & 4.6\\ 
FRB 20190111B & 1336.86(1) & 4 & 1 & 0.130(4) & 0.190(9) & $>$238.5 & 1.1\\ 
FRB 20190115B & 748.19(3) & 64 & 3 & 0.32(7), 0.55(6) & $<$ 0.32 & 400.0 & 5.9\\ 
FRB 20190116A\tablenotemark{h} & 445.4(6) & 256 & 1 & 5.3(3) & $<$ 5.31 & 298.4 & 19.0\\ 
FRB 20190117A\tablenotemark{i} & 393.08(5) & 32 & 2 & 0.62(1), 2(0) & 0.80(5) & 139.7 & 10.0\\ 
FRB 20190118A & 225.097(5) & 4 & 2 & 0.0494(4), 0.1287(9) & 0.26(2) & 400.0 & 1.8\\ 
FRB 20190121A & 425.30(3) & 32 & 1 & 0.75(2) & 4.01(7) & 400.0 & 10.0\\ 
FRB 20190122C & 690.032(8) & 8 & 7 & 0.099(7), 5.8(5) & $\sim$ 0.27 & 400.0 & 28.0\\ 
FRB 20190124B & 441.5(2) & 256 & 1 & 1.6(1) & 5.4(5) & 400.0 & 21.0\\ 
FRB 20190124F & 254.799(4) & 8 & 6 & 0.02(2), 0.117(4) & $\sim$ 0.18 & 400.0 & 2.8\\ 
FRB 20190130B & 988.76(1) & 8 & 1 & 0.085(5) & 0.56(1) & 171.4 & 3.8\\ 
FRB 20190131E & 279.794(6) & 2 & 1 & 0.072(2) & 0.118(5) & 400.0 & 0.63\\ 
FRB 20190201B & 749.08(2) & 64 & 1 & 0.39(2) & 0.62(5) & 400.0 & 5.9\\ 
FRB 20190202B & 464.839(4) & 8 & 2 & 0.083(3), 0.10(2) & 0.53(1) & $>$329.6 & 1.5\\ 
FRB 20190203A & 420.550(7) & 32 & 1 & 0.57(2) & 0.82(5) & 400.0 & 7.4\\ 
FRB 20190204B & 1464.835(9) & 128 & 1 & 0.13(2) & 1.00(9) & $>$220.1 & 5.9\\ 
FRB 20190206A & 188.273(9) & 64 & 2 & 0.4(2), 1.13(3) & 1.88(9) & 400.0 & 11.0\\ 
FRB 20190208C & 238.322(5) & 8 & 1 & 0.028(2) & 0.103(6) & $>$319.1 & 0.98\\ 
FRB 20190210B & 624.24(1) & 4 & 2 & 0.155(5), 0.33(1) & 0.157(5) & $>$129.8 & 2.4\\ 
FRB 20190212B & 600.181(3) & 4 & 3 & 0.087(4), 0.69(3) & $<$ 0.09 & 400.0 & 1.7\\ 
FRB 20190212C & 1015.7(7) & 256 & 1 & 2.8(2) & 12(2) & $>$319.1 & 44.0\\ 
FRB 20190213D & 1346.8(4) & 256 & 1 & 2.3(1) & $<$ 2.33 & $>$78.6 & 10.0\\ 
FRB 20190214C & 532.84(2) & 64 & 1 & 0.49(3) & 1.8(1) & 400.0 & 7.5\\ 
FRB 20190217A & 798.14(4) & 256 & 1 & 0.43(5) & 3.9(4) & 400.0 & 15.0\\ 
FRB 20190224C & 497.12(2) & 128 & 1 & 0.18(3) & 8.1(8) & 400.0 & 19.0\\ 
FRB 20190224D & 752.892(6) & 4 & 5 & 0.033(9), 0.53(3) & $\sim$ 0.02 & 400.0 & 1.6\\ 
FRB 20190226A & 601.538(7) & 4 & 1 & 0.053(3) & 0.159(7) & 400.0 & 0.67\\ 
FRB 20190227A & 394.037(8) & 16 & 5 & 0.0505(6), 0.81(1) & 0.29(2) & $>$251.8 & 8.7\\ 
FRB 20190301A\tablenotemark{k} & 459.78(4) & 256 & 3 & 0.50(3), 3.3(3) & $<$ 0.5 & 133.3 & 14.0\\ 
FRB 20190303B & 193.441(5) & 2 & 3 & 0.077(5), 0.46(2) & 1.303(5) & 400.0 & 6.8\\ 
FRB 20190304A & 483.516(8) & 8 & 1 & 0.057(6) & 0.82(2) & 400.0 & 2.8\\ 
FRB 20190304B & 469.92(2) & 64 & 1 & 0.11(2) & 0.56(4) & $>$158.7 & 5.7\\ 
FRB 20190320A & 614.0(1) & 256 & 1 & 0.85(5) & 1.6(2) & 400.0 & 9.2\\ 
FRB 20190320B & 489.501(8) & 4 & 5 & 0.0085(5), 0.103(8) & $\sim$ 0.14 & 400.0 & 1.8\\ 
FRB 20190320E & 299.07(2) & 32 & 1 & 0.24(1) & 1.89(9) & 400.0 & 5.7\\ 
FRB 20190322A & 1060.27(9) & 256 & 1 & 1.0(1) & 15(1) & $>$327.7 & 35.0\\ 
FRB 20190323B & 789.527(7) & 1 & 3 & 0.0056(3), 0.022(1) & $\sim$ 0.23 & 400.0 & 0.69\\ 
FRB 20190327A & 346.579(7) & 2 & 2 & 0.060(5), 0.09(3) & 0.44(2) & 400.0 & 1.2\\ 
FRB 20190405B & 1113.71(7) & 128 & 1 & 0.51(4) & 3.1(2) & $>$282.3 & 10.0\\ 
FRB 20190410B & 642.152(8) & 8 & 1 & 0.110(5) & 0.21(1) & $>$198.6 & 1.8\\ 
FRB 20190411B & 1229.42(1) & 64 & 3 & 0.32(5), 0.5(2) & $<$ 0.32 & $>$228.0 & 4.6\\ 
FRB 20190411C & 233.714(8) & 1 & 11 & 0.019(2), 0.249(9) & $\sim$ 0.01 & 400.0 & 2.6\\ 
FRB 20190412A & 364.56(1) & 16 & 3 & 0.049(6), 0.22(1) & 0.96(3) & 400.0 & 4.4\\ 
FRB 20190417C & 320.266(4) & 1 & 2 & 0.022(3), 0.04679(9) & $\sim$ 0.14 & 400.0 & 0.77\\ 
FRB 20190418A & 184.476(9) & 64 & 1 & 0.59(3) & $<$ 0.59 & 133.3 & 2.5\\ 
FRB 20190419B & 165.13(2) & 8 & 2 & 0.12(6), 0.37(5) & $\sim$ 0.37 & 400.0 & 1.9\\ 
FRB 20190420B & 846.846(9) & 256 & 1 & 0.8(1) & 3(1) & $>$114.6 & 20.0\\ 
FRB 20190423A & 242.600(8) & 1 & 2 & 0.287(1), 0.553(7) & $\sim$ 2.35 & $>$294.4 & 8.5\\ 
FRB 20190423D & 496(1) & 256 & 1 & 1.8(1) & 6.6(4) & 400.0 & 20.0\\ 
FRB 20190425A & 128.14(1) & 4 & 5 & 0.007(1), 0.07718(8) & $\sim$ 0.03 & 400.0 & 0.73\\ 
FRB 20190425B & 1031.63(1) & 4 & 2 & 0.135(4), 0.17(1) & 0.188(8) & $>$253.8 & 1.7\\ 
FRB 20190427A & 455.77(1) & 32 & 1 & 0.093(9) & 0.99(6) & 400.0 & 2.8\\ 
FRB 20190430C & 400.4(3) & 32 & 1 & 0.83(2) & $<$ 0.83 & 196.8 & 3.4\\ 
FRB 20190501B & 783.974(4) & 8 & 2 & 0.077(6), 0.118(6) & 0.44(1) & 400.0 & 4.1\\ 
FRB 20190502A & 625.74(1) & 16 & 1 & 0.78(3) & 0.93(9) & 400.0 & 5.7\\ 
FRB 20190502B & 918.5(2) & 256 & 1 & 1.95(9) & $<$ 1.95 & $>$61.8 & 9.8\\ 
FRB 20190502C & 396.875(9) & 4 & 7 & 0.0169(9), 0.126(5) & 0.045(2) & 400.0 & 2.2\\ 
FRB 20190517C & 335.49(6) & 8 & 1 & 0.249(5) & 0.25(1) & 400.0 & 2.1\\ 
FRB 20190518C & 443.955(7) & 4 & 2 & 0.06(1), 0.12(1) & 0.43(2) & 400.0 & 1.4\\ 
FRB 20190519E & 693.619(7) & 8 & 1 & 0.013(2) & 0.040(3) & $>$186.1 & 0.37\\ 
FRB 20190519G & 430.0(5) & 256 & 2 & 1.9(3), 7.8(4) & $<$ 1.89 & 400.0 & 28.0\\ 
FRB 20190519H & 1170.878(6) & 1 & 2 & 0.037(1), 0.068(2) & $\sim$ 0.12 & $>$107.1 & 0.54\\ 
FRB 20190604G & 232.998(7) & 32 & 6 & 0.05(2), 1.5(4) & $\sim$ 0.18 & 400.0 & 7.5\\ 
FRB 20190605C & 187.710(5) & 1 & 1 & 0.062(1) & 0.064(2) & 400.0 & 0.62\\ 
FRB 20190605A\tablenotemark{g} & 349.8(5) & 128 & 1 & 2.46(6) & $<$ 2.46 & 196.8 & 10.0\\ 
FRB 20190605B\tablenotemark{g} & 349.09(6) & 256 & 4 & 1.92(8), 3.5(3) & $<$ 1.92 & $>$123.1 & 45.0\\ 
FRB 20190606A\tablenotemark{l} & 552.58(4) & 256 & 1 & 1.27(9) & $<$ 1.27 & 228.6 & 5.9\\ 
FRB 20190606B & 277.50(3) & 128 & 1 & 0.33(4) & 4.8(4) & 400.0 & 14.0\\ 
FRB 20190607B & 289.267(2) & 128 & 1 & 0.02(1) & 1.6(1) & 400.0 & 4.3\\ 
FRB 20190608A & 722.13(1) & 16 & 1 & 0.154(9) & 0.11(2) & 400.0 & 0.78\\ 
FRB 20190609A & 316.681(4) & 16 & 2 & 0.146(5), 1.0(1) & 0.40(2) & 400.0 & 10.0\\ 
FRB 20190609B & 292.149(7) & 4 & 3 & 0.069(6), 0.249(6) & 0.115(3) & 400.0 & 2.1\\ 
FRB 20190609C & 479.855(5) & 32 & 2 & 0.014(1), 0.052(2) & 0.030(3) & 114.3 & 2.3\\ 
FRB 20190609D & 511.56(2) & 32 & 1 & 0.19(2) & 0.74(5) & 400.0 & 3.7\\ 
FRB 20190611A\tablenotemark{m} & 190(2) & 256 & 2 & 1.31(8), 1.3(1) & 5.4(3) & 165.1 & 26.0\\ 
FRB 20190612B & 187.524(7) & 1 & 1 & 0.0141(3) & $\sim$ 0.02 & 209.5 & 0.082\\ 
FRB 20190613A & 714.71(3) & 128 & 2 & 0.09(5), 0.55(4) & $\sim$ 2.07 & 400.0 & 12.0\\ 
FRB 20190613B & 285.088(5) & 1 & 2 & 0.015(4), 0.0243(5) & $\sim$ 0.1 & 400.0 & 0.39\\ 
FRB 20190614A & 1063.909(7) & 16 & 1 & 0.15(1) & 0.55(4) & 400.0 & 2.6\\ 
FRB 20190614C & 589.1(2) & 256 & 1 & 2.8(2) & $<$ 2.78 & 190.5 & 7.9\\ 
FRB 20190616A & 212.513(5) & 8 & 2 & 0.082(4), 0.37(2) & $<$ 0.08 & 400.0 & 0.72\\ 
FRB 20190617A & 195.749(6) & 1 & 5 & 0.077(2), 0.324(4) & $\sim$ 0.38 & 400.0 & 4.8\\ 
FRB 20190617B & 272.73(7) & 32 & 1 & 0.65(3) & 2.1(1) & 399.8 & 8.4\\ 
FRB 20190617C & 639.03(4) & 256 & 1 & 1.2(1) & 6.8(7) & 400.0 & 25.0\\ 
FRB 20190618A & 228.922(6) & 2 & 3 & 0.032(6), 0.12(2) & 0.108(2) & 400.0 & 1.8\\ 
FRB 20190619A & 899.83(1) & 32 & 2 & 0.17(1), 0.22(2) & $<$ 0.17 & 400.0 & 1.6\\ 
FRB 20190619B & 270.552(4) & 16 & 2 & 0.141(9), 0.146(9) & $<$ 0.14 & 400.0 & 1.3\\ 
FRB 20190619C & 488.072(3) & 16 & 3 & 0.03(1), 0.14(1) & $\sim$ 0.18 & 400.0 & 2.2\\ 
FRB 20190619D & 378.5(3) & 64 & 1 & 1.3(2) & 7.8(4) & 400.0 & 14.0\\ 
FRB 20190621A\tablenotemark{j} & 195.9(2) & 256 & 1 & 3.2(2) & $<$ 3.15 & 57.1 & 11.0\\ 
FRB 20190621B & 1061.17(3) & 256 & 1 & 0.91(7) & $<$ 0.91 & 400.0 & 3.9\\ 
FRB 20190621C & 570.342(7) & 1 & 1 & 0.0365(9) & 0.035(3) & 400.0 & 0.24\\ 
FRB 20190621D & 647.32(4) & 64 & 1 & 0.24(2) & 2.2(1) & 400.0 & 7.5\\ 
FRB 20190622A & 1122.803(9) & 16 & 1 & 0.031(5) & 0.13(1) & 400.0 & 0.74\\ 
FRB 20190623A & 1082.17(1) & 64 & 1 & 0.164(9) & $<$ 0.16 & 400.0 & 1.1\\ 
FRB 20190623C & 1049.88(2) & 128 & 1 & 0.43(4) & 0.8(2) & 400.0 & 4.6\\ 
FRB 20190624A & 973.8(1) & 256 & 1 & 0.97(6) & 1.7(2) & 400.0 & 13.0\\ 
FRB 20190624B & 213.947(8) & 1 & 5 & 0.015(2), 0.086(8) & $\sim$ 0.03 & 400.0 & 1.2\\ 
FRB 20190625E\tablenotemark{m} & 188.57(5) & 256 & 3 & 1.1(1), 1.6(1) & $<$ 1.11 & 69.8 & 35.0\\ 
FRB 20190626A\tablenotemark{m} & 192.2(6) & 256 & 1 & 8.3(4) & $<$ 8.28 & 63.5 & 26.0\\ 
FRB 20190627A & 404.3(1) & 64 & 1 & 0.17(2) & $\sim$ 0.28 & $>$80.5 & 1.5\\ 
FRB 20190627C & 968.50(1) & 8 & 2 & 0.20(1), 1(0) & 1.20(3) & 400.0 & 3.2\\ 
FRB 20190628A & 745.792(9) & 64 & 1 & 0.29(2) & $<$ 0.29 & 400.0 & 1.5\\ 
FRB 20190628B & 408.03(2) & 128 & 1 & 0.76(5) & $<$ 0.76 & 400.0 & 4.3\\ 
FRB 20190629A & 503.60(3) & 256 & 1 & 0.36(6) & 1.4(3) & 400.0 & 5.9\\ 
FRB 20190630B & 652.0(3) & 16 & 1 & 1.49(4) & 13.0(2) & 400.0 & 22.0\\ 
FRB 20190630C & 1660.22(1) & 32 & 1 & 0.100(9) & 0.31(2) & $>$139.6 & 3.9\\ 
FRB 20190630D & 323.540(3) & 16 & 3 & 0.044(4), 0.41(2) & $\sim$ 0.02 & 374.6 & 1.9\\ 
FRB 20190701A & 637.088(9) & 16 & 1 & 0.124(7) & $<$ 0.12 & 400.0 & 0.74\\ 
FRB 20190701B & 749.096(8) & 8 & 2 & 0.028(2), 0.31(3) & 0.120(6) & 400.0 & 1.4\\ 
FRB 20190701C & 973.81(1) & 128 & 1 & 0.24(2) & 0.41(6) & 400.0 & 3.9\\ 
FRB 20190701D & 933.39(3) & 64 & 1 & 0.43(3) & 2.14(8) & 400.0 & 14.0\\ 
\hline
\insertTableNotes
\\[-2ex]
\end{longtable}\label{table}
\end{ThreePartTable}
\vspace*{1cm}

\section{Additional correlation plots}

We present here correlation plots of remaining properties shown in in Table \ref{Tab:corr} and show results from the MADFM analysis as discussed in Section \ref{sec:corr}.

\setcounter{figure}{0}
\renewcommand{\thefigure}{B\arabic{figure}}

\begin{figure*}
\centering
\gridline{\fig{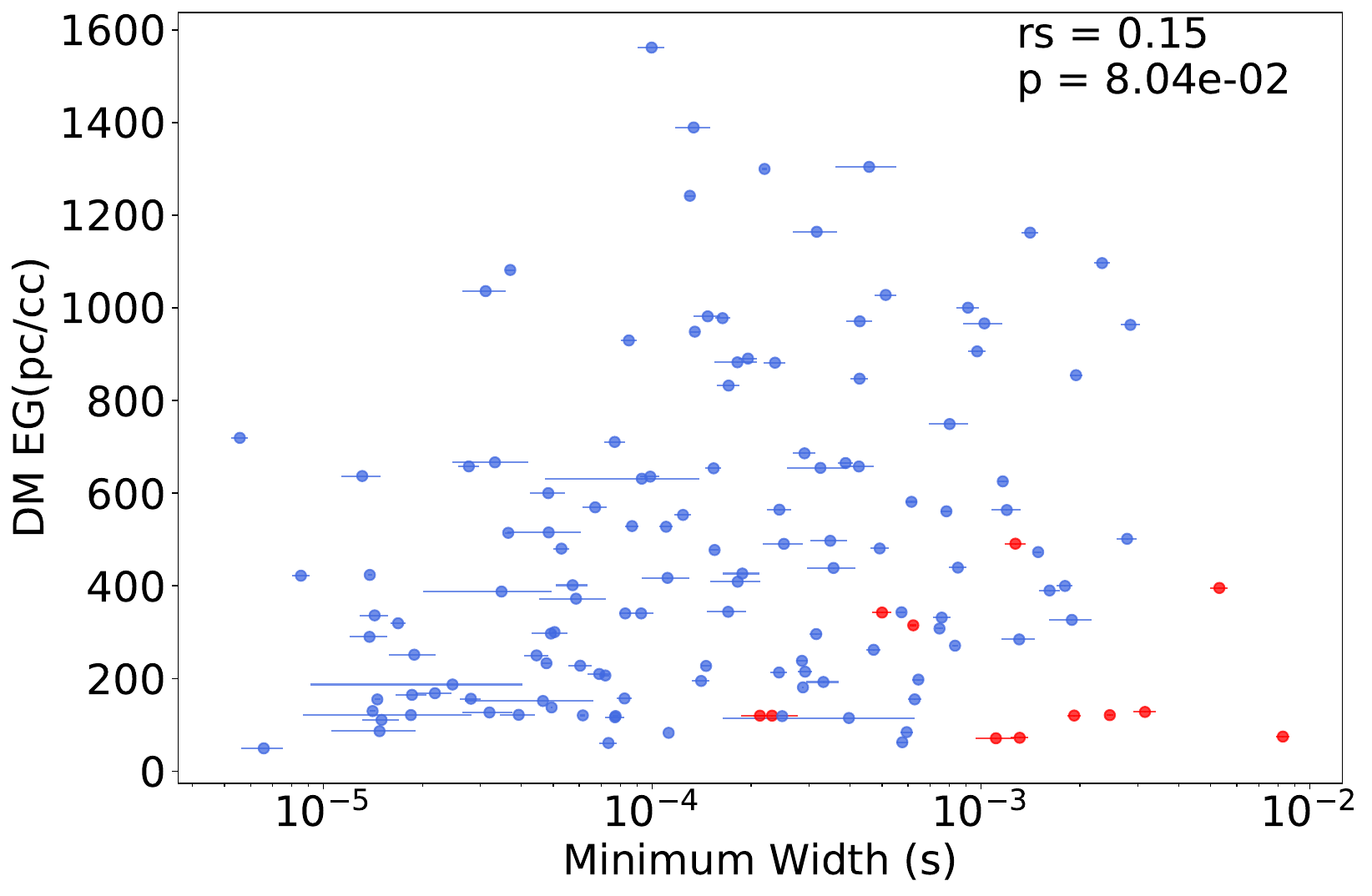}{0.32\textwidth}{}
          \fig{FLuence_vs_scattering.pdf}{0.32\textwidth}{}
          \fig{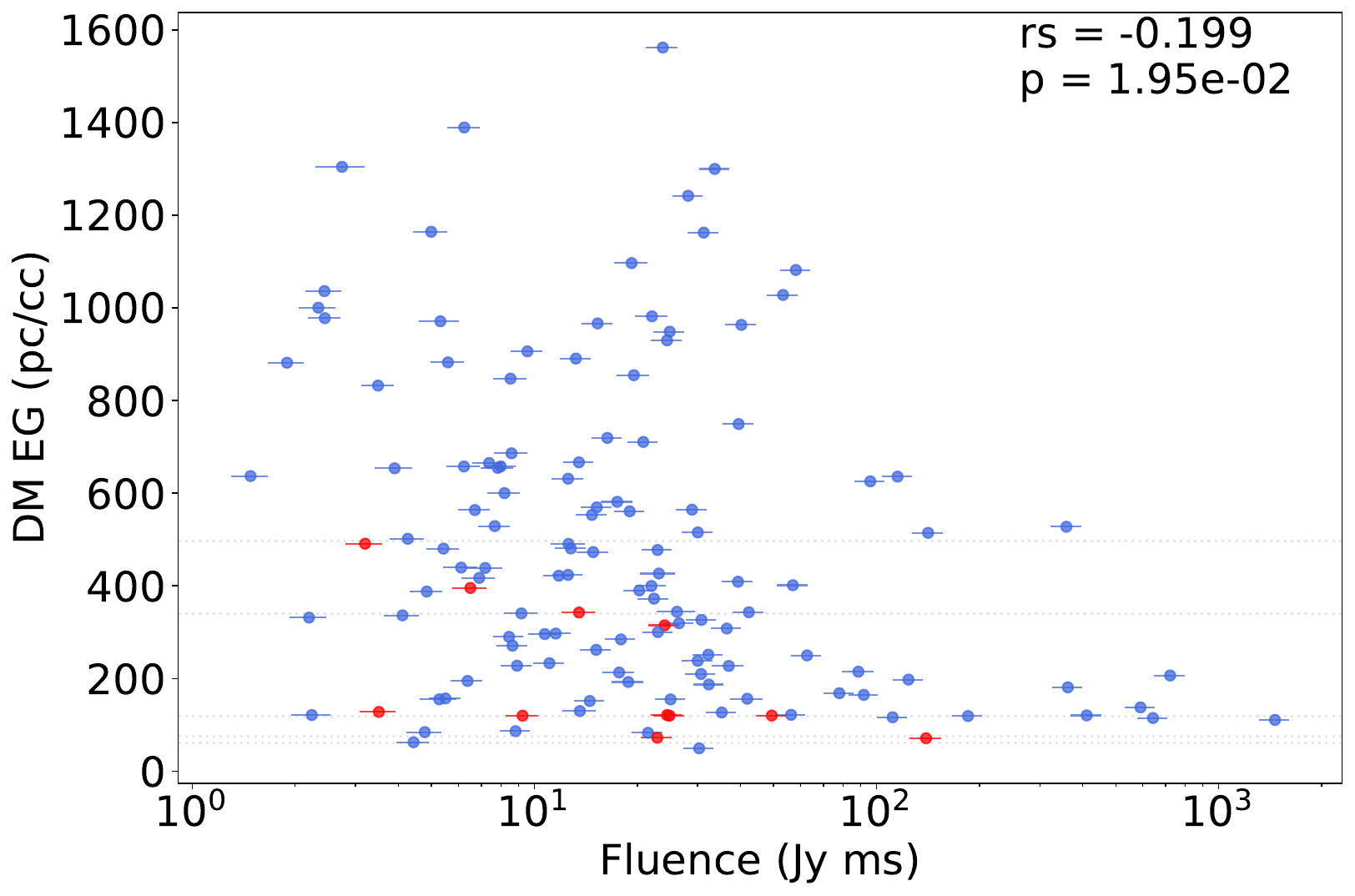}{0.32\textwidth}{}
          }
    \caption{Additional Correlation Plots. Left Panel: Extragalactic DM vs. minimum intrinsic width—no significant correlation is observed, indicating that FRBs can exhibit a broad range of intrinsic widths. Middle Panel: Scattering vs. Fluence—no significant correlation is found. Various symbols represent different scattering estimates as detailed in Figure \ref{fig:corr_DM_scat}. Right Panel: Extragalactic DM vs. Fluence—a very mild, but not statistically significant, correlation is observed. Notably, true fluence estimates for high DM sources and sources missing channels are unavailable due to incomplete channel storage (see Section \ref{sec:bias}). Dashed lines indicate sidelobe events where the primary beam is poorly characterized, so fluences are not reported for these events. This suggests that intrinsic luminosities of bursts from an FRB source can vary significantly, as seen in certain repeating sources. Spearman rank correlation coefficients (rs) and p-values are displayed in the top right corner of each plot. Red points represent repeaters, and blue points denote one-off FRBs.}\label{fig:additional_corr}
\end{figure*}

\begin{figure*}
\centering
\gridline{\fig{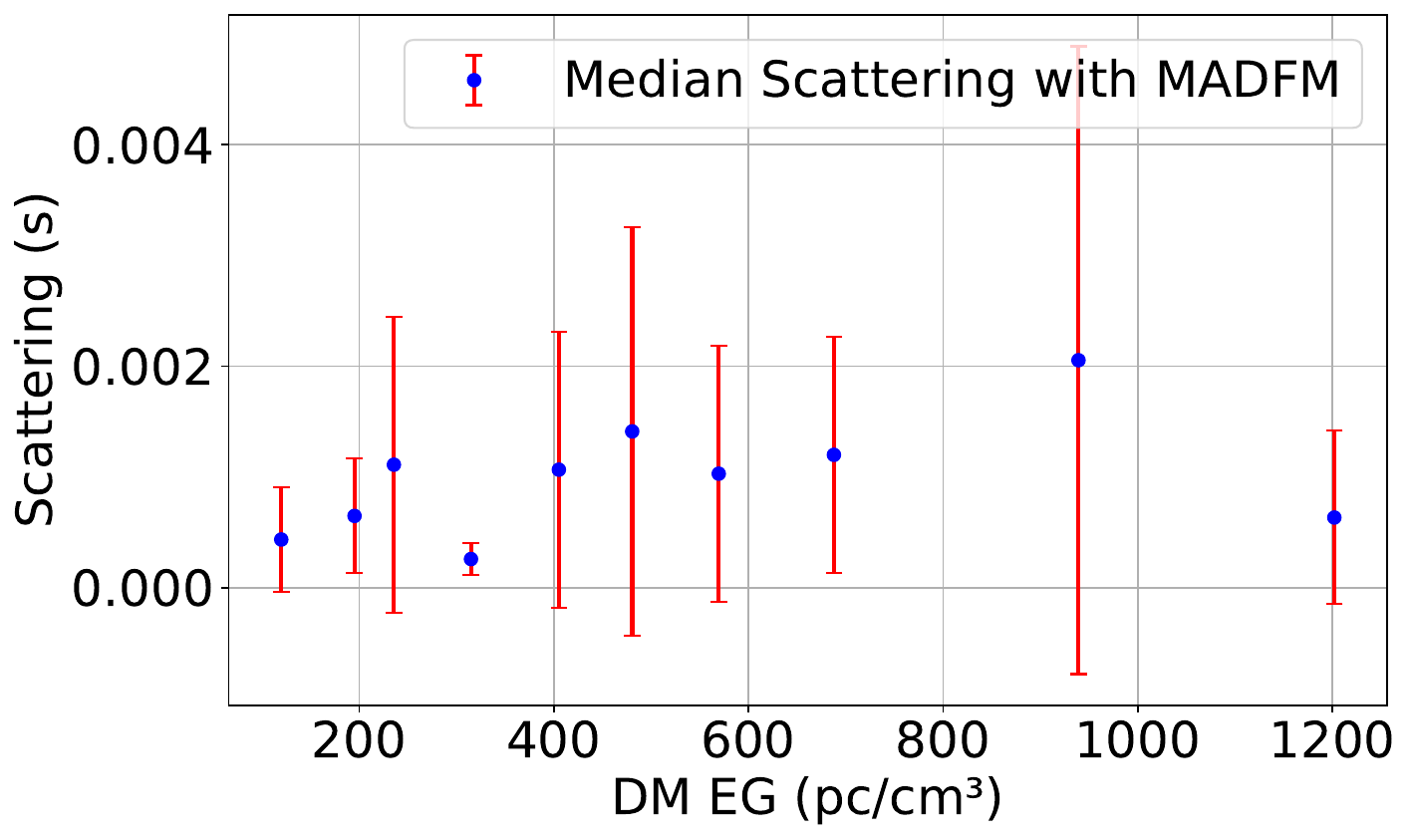}{0.32\textwidth}{}
          \fig{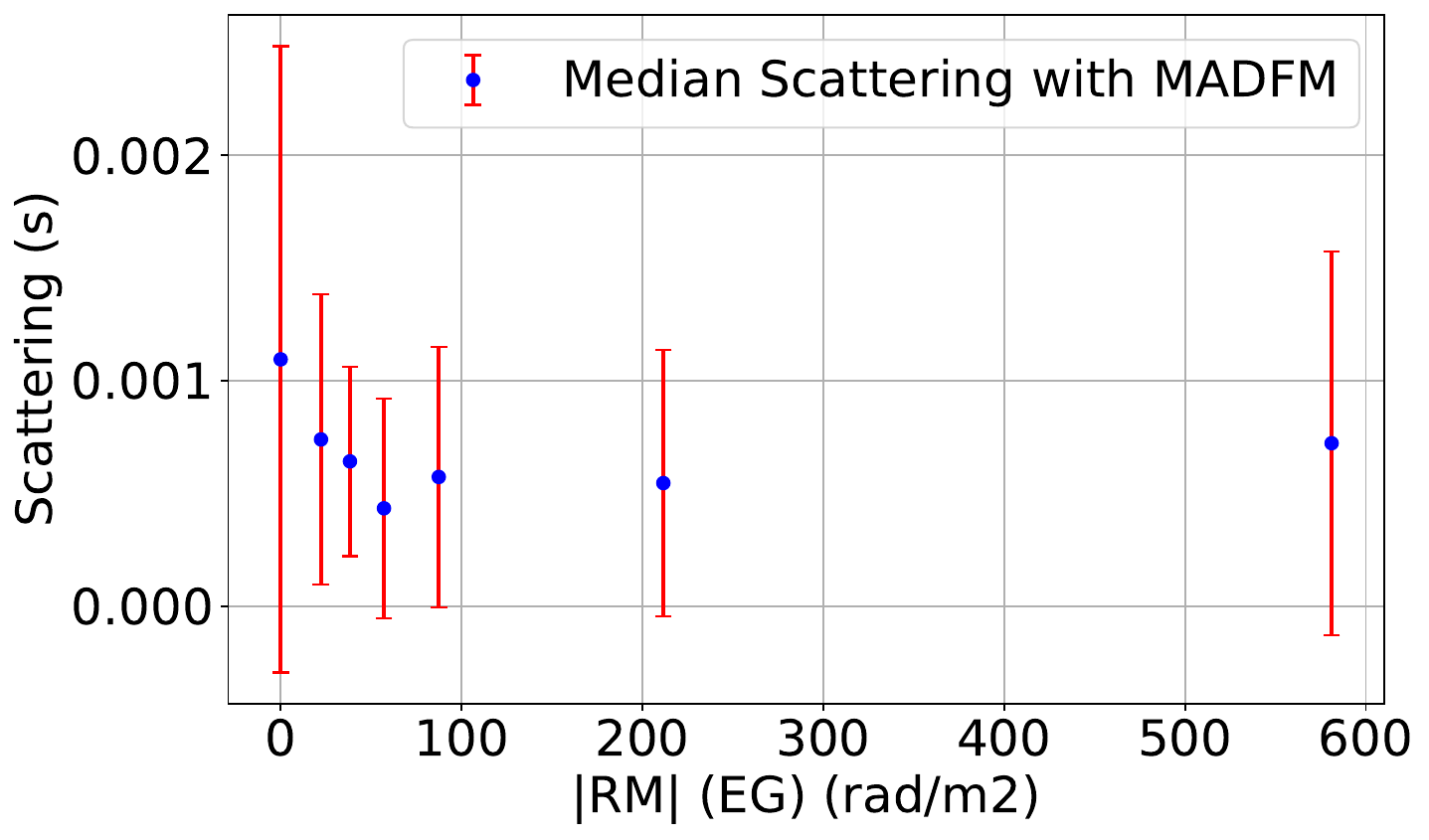}{0.32\textwidth}{}
          \fig{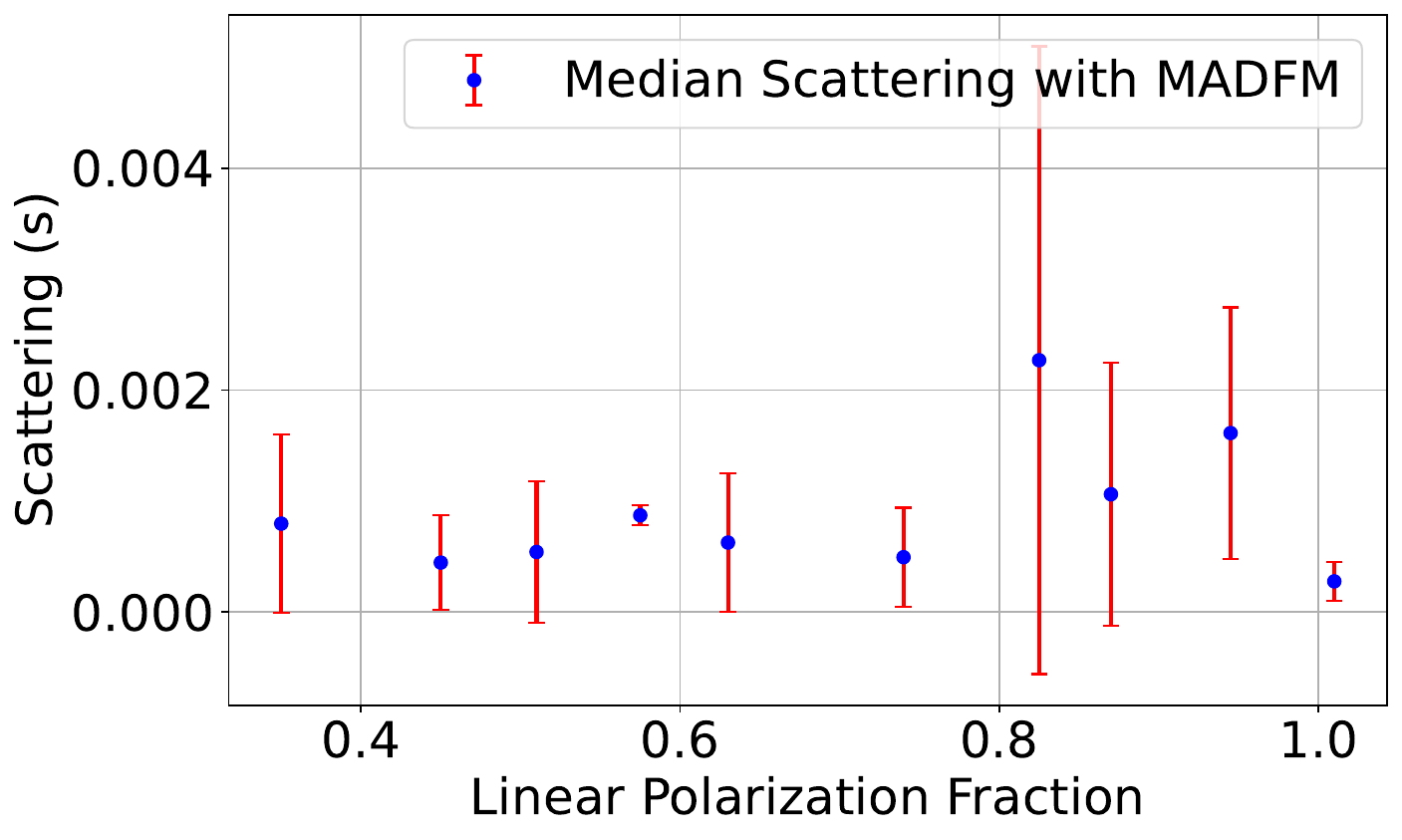}{0.32\textwidth}{}
          }
    \caption{To further quantify the absence of correlations, as depicted in Figures \ref{fig:corr_DM_scat}, \ref{fig:corr_RM_scat}, and \ref{fig:corr_polfrac_scat}, we computed the median absolute deviation from the median (MADFM) for the scattering timescale and its dependence on the associated properties. The blue markers represent the median values calculated from subsets of 10 data points, while the error bars reflect the magnitude of the mean absolute deviation (MAD). In this analysis, only the measured data points (depicted as filled circles in the referenced figures) were considered. The trend of the data is well-described by a horizontal line, reinforcing the absence of any significant correlation within our dataset.} \label{fig:madfm_corr}
\end{figure*}

\bibliography{frbrefs,in_prep,R3_baseband}{}
\bibliographystyle{aasjournal}

\end{document}